\documentclass[aps,prfluids,preprint,superscriptaddress,tightenlines]{revtex4-2}
\usepackage{multirow,makecell,colortbl,xcolor,tikz,siunitx,parskip,amssymb,hyperref}
\usetikzlibrary{shapes,shapes.misc}
\usetikzlibrary{plotmarks,calc}
\makeatletter
\AtBeginDocument{%
  \@ifclassloaded{revtex4-2}{%
    \ifx\CT@everycr\everycr
      \newtoks\everycr
    \fi
    \CT@everycr{\noalign{\global\let\CT@row@color\relax}\the\everycr}%
  }{}%
}
\makeatother

\definecolor{c1}{RGB}{229,0,172}
\definecolor{c2}{RGB}{229,0,221}
\definecolor{c3}{RGB}{184,0,229}
\definecolor{c4}{RGB}{130,0,229}
\definecolor{c5}{RGB}{81,0,229}
\definecolor{c6}{RGB}{27,0,229}
\definecolor{c7}{RGB}{0,26,229}
\definecolor{c8}{RGB}{0,75,229}
\definecolor{c9}{RGB}{0,128,229}
\definecolor{c10}{RGB}{0,182,229}
\definecolor{c11}{RGB}{0,229,227}
\definecolor{c12}{RGB}{136,229,0}
\definecolor{c13}{RGB}{189,229,0}
\definecolor{c14}{RGB}{229,220,0}
\definecolor{c15}{RGB}{229,167,0}
\definecolor{c16_SR}{RGB}{229,113,0}
\definecolor{c16_AF}{RGB}{229,64,0}
\definecolor{c17_SR}{RGB}{229,11,0}
\definecolor{c17_AF}{RGB}{229,0,36}
\definecolor{c144}{HTML}{FC87F7}
\definecolor{c256}{HTML}{E600DD}
\definecolor{c384}{HTML}{81007C}
\definecolor{c64}{HTML}{FECBFC}

\DeclareSIUnit\bpm{bpm}

\newcommand{\LA}{\mathrm{LA}}
\newcommand{\LAA}{\mathrm{LAA}}
\newcommand{\PV}{\mathrm{PV}}
\newcommand{\MV}{\mathrm{MV}}

\DeclareRobustCommand{\lineCoarse}{\raisebox{2pt}{\tikz{\draw[c144,very thick, dash pattern= on 3pt off 2pt on 0pt off 2pt ,line cap = round](0,0) -- (4.5mm,0);}}}
\DeclareRobustCommand{\lineVeryCoarse}{\raisebox{2pt}{\tikz{\draw[c64,very thick, dashed,line cap = round](0,0) -- (4mm,0);}}}
\DeclareRobustCommand{\lineNom}{\raisebox{2pt}{\tikz{\draw[c256,very thick        ](0,0) -- (5mm,0);}}}
\DeclareRobustCommand{\lineFine}{\raisebox{2pt}{\tikz{\draw[c384,very thick, dash pattern= on 0pt off 2pt, line cap = round](0,0) -- (4.5mm,0);}}}
\DeclareRobustCommand{\lineOne}{\raisebox{2pt}{\tikz{\draw[c1,very thick, dash pattern= on 3pt off 2pt on 0pt off 2pt, line cap = round](0,0) -- (4.75mm,0);}}}
\DeclareRobustCommand{\lineFive}{\raisebox{2pt}{\tikz{\draw[c5,very thick,dash pattern= on 0pt off 2pt, line cap = round](0,0) -- (5mm,0);}}}
\DeclareRobustCommand{\lineEleven}{\raisebox{2pt}{\tikz{\draw[c11,very thick, dashed, line cap = round](0,0) -- (5mm,0);}}}
\DeclareRobustCommand{\lineTwelve}{\raisebox{2pt}{\tikz{\draw[c12,very thick](0,0) -- (5mm,0);}}}

\DeclareRobustCommand{\linegray}{ \raisebox{2pt}{\tikz{\draw[very thick,gray, dashed, line cap = round](0,0) -- (5.2mm,0);}}}
\DeclareRobustCommand{\line}{ \raisebox{2pt}{\tikz{\draw[very thick](0,0) -- (5mm,0);}}}
\DeclareRobustCommand{\dashed}{\raisebox{2pt}{\tikz{\draw[very thick, dashed, line cap = round](0,0) -- (5.2mm,0);}}}

\DeclareRobustCommand{\filled}{ \tikz{\draw[draw=none,fill=black!20]  (0,0) rectangle ++ (.4,.2) ;}}

\DeclareRobustCommand{\stream}{ \tikz{\draw[draw=c5,-latex]  (0,0) -- ++(.3,.25) ;}}
\DeclareRobustCommand{\motion}{ \tikz{\draw[draw=red,-latex,dash pattern=on 2pt off 2pt]  (0,0) -- ++(.3,.25) ;}}

\makeatletter
\renewcommand{\frontmatter@affiliationfont}{%
  \linespread{1}%
  \fontsize{10pt}{14pt}\selectfont
  \itshape
  \setlength{\parskip}{0pt}%
}
\makeatother

\hypersetup{
    colorlinks=true,
    linkcolor=blue,
    filecolor=blue,      
    urlcolor=blue,
    citecolor=blue,
}

\usepackage{glossary/myglossaries}
\usepackage{glossary-inline}
\begin{document}

\title{
Fluid exchange and renewal in the left atrial appendage
}

\author{Y.~St\"ocker}
\affiliation{Institute of Fluid Mechanics and Heat Transfer, TU Wien, 1060 Vienna, Austria}
\author{M.~Guerrero-Hurtado} 
\affiliation{ Department of Aerospace Engineering, Universidad Carlos III de Madrid, Legan\'es, Spain}
\author{A.~Gonzalo}
\affiliation{Department of Mechanical Engineering, University of Washington, Seattle, WA, USA}
\author{S.~Sharda} 
\affiliation{Department of Mechanical Engineering, University of Washington, Seattle, WA, USA} 
\author{E.~Dur{\'a}n} 
\affiliation{Department of Mechanical, Thermal and Fluids Engineering, Universidad de M\'alaga, M\'alaga, Spain}
\author{P.~Martinez-Legazpi}
\affiliation{Department of Mathematical Physics and Fluids, Universidad Nacional de Educaci\'on a Distancia, Spain}
\affiliation{CIBERCV, Madrid, Spain}
\author{A.~M.~Kahn} 
\affiliation{Division of Cardiovascular Medicine, University of California San Diego, La Jolla, CA, USA}
\author{E.~McVeigh}
\affiliation{Division of Cardiovascular Medicine, University of California San Diego, La Jolla, CA, USA}
\affiliation{Department of Bioengineering, University of California San Diego, La Jolla, CA, USA}
\affiliation{Department of Radiology, University of California San Diego, La Jolla, CA, USA}
\author{J.~Bermejo}
\affiliation{CIBERCV, Madrid, Spain}
\affiliation{Hospital General Universitario Gregorio Mara\~n\'on, Madrid, Spain}
\affiliation{Instituto de Investigaci\'on Sanitaria Gregorio Mara\~n\'on, Madrid, Spain}
\affiliation{Facultad de Medicina, Universidad Complutense de Madrid, Madrid, Spain}

\author{O.~Flores}
\affiliation{ Department of Aerospace Engineering, Universidad Carlos III de Madrid, Legan\'es, Spain}
\author{J.~C.~del~{\'A}lamo}
\affiliation{Department of Mechanical Engineering, University of Washington, Seattle, WA, USA}
\affiliation{Center for Cardiovascular Biology, University of Washington, Seattle, WA, USA}
\affiliation{Division of Cardiology, University of Washington, Seattle, WA, USA}
\author{M.~Garcia-Villalba}
\affiliation{Institute of Fluid Mechanics and Heat Transfer, TU Wien, 1060 Vienna, Austria}
\date{\today}

\begin{abstract}
Fluid exchange through the opening of a cavity does not uniquely determine
how the cavity is washed out.
We investigate this phenomenon in the left atrial appendage (LAA),
a deforming side cavity of the heart, using simulations of
17 patient-specific atria with wall motion reconstructed from
time-resolved computed tomography.
Fluid exchange through the LAA ostium is decomposed into a
component driven by LAA volume change and
a component due to simultaneous bidirectional flow, with no net volume transport.
Washout of a scalar initially confined to the LAA is compared
with predictions from two well-mixed washout models
and assessed alongside residence-time distributions.
Volume-driven transport accounts for approximately 24--78\%
of total exchange across the simulated atria.
Relative washout efficiencies are approximately 26--53\%
with respect to the inflow-driven washout model that assumes scalar-free inflow.
A scalar-balance decomposition identifies two contributions
to this discrepancy: re-ingestion of previously expelled scalar
and preferential export of scalar-poor fluid.
Both effects reduce the effective washout rate in every configuration.
Residence-time distributions reveal coexisting distinct younger and older fluid
regions, and a regional analysis of one appendage localizes
the older population predominantly in its distal region.
These findings show that substantial exchange across the LAA ostium
coexists with prolonged distal retention.
Effective renewal depends on how much fluid crosses
the ostium and which fluid is exchanged.
\end{abstract}
\maketitle
\hrule height 0.4pt
\smallskip
\begingroup
  \linespread{1}\footnotesize
  \renewcommand*{\glossarysection}[2][]{\noindent\textbf{#2:}\space}
  \renewcommand*{\glsinlinedescformat}[3]{,\space#1}
  \renewcommand*{\glspostinline}{.}
  \printglossary[type=\acronymtype,style=inline,title={Abbreviations}]
  \par
\endgroup

\newpage
\section{Introduction}

\begin{figure}[h!tb]
    \centering
    \includegraphics[width=\linewidth]{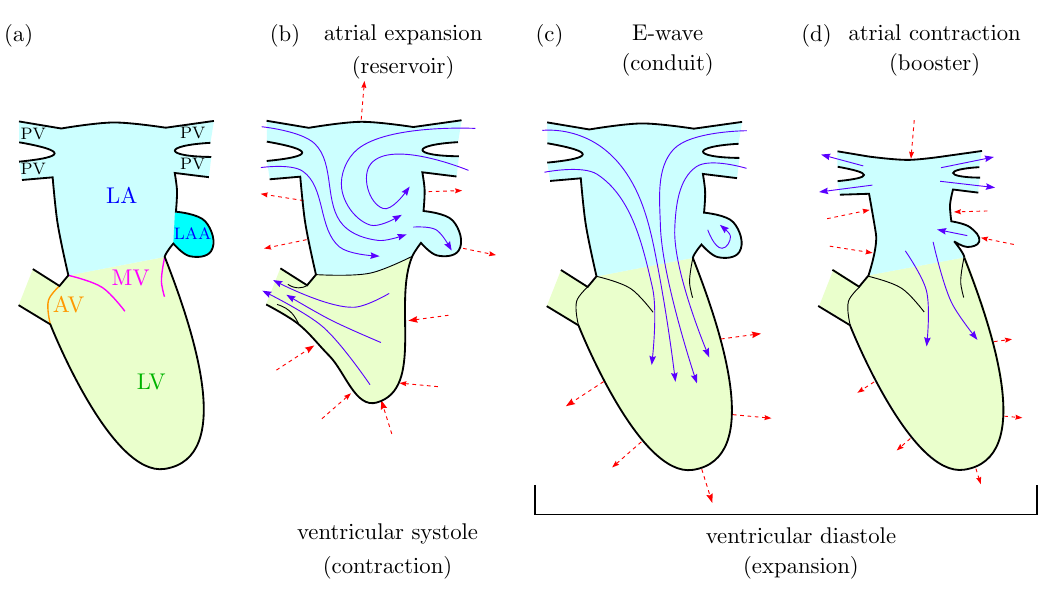}
    \caption{
    Schematic of left-heart anatomy and atrial function
    during the cardiac cycle.
    (a)~Anatomical structures.
    (b)~Atrial expansion during ventricular systole (reservoir phase).
    (c)~Early ventricular filling (E-wave; conduit phase).
    (d)~Active atrial contraction during late ventricular filling
    (A-wave; booster-pump phase).
    Solid arrows (\stream) indicate fluid flow, and dashed arrows
    (\motion) indicate wall motion.
    PV: pulmonary vein; LA: left atrium; LAA: left atrial appendage; MV: mitral valve; AV: aortic valve; LV: left ventricle.
    }
    \label{fig:sketch}
\end{figure}

The \gls*{LA} is one of the four cardiac chambers and connects
the pulmonary circulation to the \gls*{LV} (Fig.~\ref{fig:sketch}a).  
It receives oxygenated
blood from the lungs through the \glspl*{PV} and delivers it
to the ventricle through the \gls*{MV}.
Over the cardiac cycle, it acts successively as a
reservoir during ventricular systole, 
as a conduit during early ventricular filling, and 
as a booster pump during late ventricular filling 
(Fig.~\ref{fig:sketch}b--d).
%
%
This cyclic filling and emptying is accompanied by distinct
three-dimensional atrial flow patterns that are
driven jointly by the PV inflow jets, the transmitral outflow
and the deformation of the chamber wall.
In \gls*{SR}, the early diastolic transmitral filling wave is
termed the E-wave, while the later wave generated by atrial
contraction is termed the A-wave.

Extending laterally from the atrial body is the \gls*{LAA}, a smaller pouch-shaped cavity
that communicates with the \gls*{LA} through a single opening called the ostium.
Because the appendage lies outside the main flow path from the PVs to the MV, 
and its volume returns to approximately the same value every cardiac cycle, its contents can only be renewed by exchange across the ostium.
This exchange is of particular interest because impaired blood
renewal can create conditions favorable to LAA thrombus formation. 
Thrombi
that embolize from the \gls*{LAA} can cause ischemic stroke or transient
ischemic attacks~\cite{goldman1999,stoddard2003}.
Loss of coordinated atrial contraction can impair blood renewal, as occurs in
\gls*{AF}, the most common sustained cardiac arrhythmia~\cite{VanGelder2024}.
In subjects with this condition, reduced \gls*{LAA} emptying
velocities have been associated with spontaneous echocardiographic
contrast and appendage thrombus~\cite{goldman1999,hautmann2023}.
Among subjects with existing atrial thrombi, lower velocities
have also been associated with subsequent ischemic stroke or
systemic embolism~\cite{okada2024}.
Understanding how the LAA exchanges fluid with the atrial body 
is thus a fluid mechanics problem with direct clinical relevance. 

%


The relationship between flow organization and blood transport
has been examined extensively in the left ventricle~\cite{Eriksson2013,Hendabadi2013}.
Classical studies of vortex-ring formation~\cite{Gharib1998} and their application
to diastolic filling have linked fundamental flow phenomena to
cardiac function~\cite{Gharib2006,Charonko2013,MartinezLegazpi2014},
while studies of
ventricular flow asymmetry have examined its role in reducing
energy dissipation~\cite{Pedrizzetti2005}.
Laboratory and numerical 
studies in idealized and anatomically realistic ventricles have
examined how geometry, wall motion and mitral-valve dynamics
organize the filling flow~\cite{Domenichini2005,Querzoli2010,Le2012,Meschini2018}.
Complementary investigations have related flow organization and
ventricular anatomy to mixing, washout and regional retention,
using Lagrangian analyses, passive scalars and residence-time
measures~\cite{Seo2013,Vedula2016,Espa2012,Badas2016,
Rossini2016,DiLabbio2018,DiLabbio2022}.
Several studies provide mechanistic and clinical evidence of 
the association between long residence times or shallow inflow jets and 
LV mural thrombosis~\cite{Seo2016,Harfi2017,Duus2021,MartinezLegazpi2018}. 
More recent work demonstrates that LV stasis measured by echocardiographic vector 
flow mapping predicts brain thromboembolism~\cite{RodriguezGonzalez2024,RodriguezGonzalez2025}.

More broadly, cardiac flow modeling has also advanced through image-based
reconstruction and multiphysics models that couple blood flow,
tissue mechanics and
electrophysiology~\cite{Mittal2016,Viola2020,Viola2023PRF}.
In a recent perspective article, \textcite{Verzicco2022} reviewed 
cardiac hemodynamics together with the 
electrophysiology and tissue mechanics that drive it.
His discussion of physiological and pathological flows centers on the ventricle and its 
valves, while the atrial appendage is mentioned mainly as a preferential site of clot 
formation.
%

%
Compared with the LV, the left atrium has received considerably less 
attention from fluid mechanicians.
Dedicated contributions include investigations of transitional turbulence effects on atrial
flow~\cite{Zingaro2021}, and of how geometry, blood rheology,
coagulation and circulatory support affect atrial and appendage
hemodynamics~\cite{Chen2023,Wang2024,Bouchnita2024,Yang2025,Langer2026}.
The atrium has also been resolved within larger cardiac models.
Some of these directly examine atrial flow organization,
chamber dynamics and wall shear stress~\cite{Chnafa2014,Viola2020},
while others include the atrium but emphasize ventricular
flow~\cite{Viola2023PRF,Guglietta2026}.
These studies often consider idealized geometries, population-averaged anatomical 
models that omit the LAA, or small cohorts, and few focus on blood 
inside the appendage itself, even if approximately 90\% of atrial thrombi in patients with AF 
originate in this small chamber~\cite{Blackshear1996}.

Most studies of atrial flow are found outside of the literature primarily 
devoted to fluid mechanics.
Early simulations of an idealized atrium with an attached
appendage examined the effects of atrial contraction using
fluid--structure interaction~\cite{zhang2008}.
Subsequent frameworks incorporated patient-specific anatomy and
wall motion derived from \gls*{4D-CT}~\cite{Vedula2015,Otani2016,Masci2020,Villalba2021}.
%
Comparisons of moving and rigid-wall simulations have shown that
\gls*{LA} wall motion, including appendage deformation,
alters \gls*{LAA} flow patterns even in subjects with
impaired atrial function~\cite{Villalba2021,
DuenasPamplona2021,Kjeldsberg2024}.
Wall motion also affects blood renewal within the appendage,
although moving-wall simulations did not consistently yield
faster renewal than their rigid-wall
counterparts~\cite{Villalba2021,DuenasPamplona2021}.
%
Interaction with the surrounding atrial flow also influences appendage
renewal. 
The importance of this interaction is evident in studies
that varied \gls*{PV} orientation, flow split and atrial or
appendage geometry~\cite{GarciaIsla2018,DuenasPamplona2022,Duran2023,Rodriguez2024,Rodriguez2025}.
\textcite{Duran2023} found that changing
the \gls*{PV} flow split affected \gls*{LAA} renewal, with greater sensitivity
in their subjects with impaired atrial function,
and proposed a role for secondary flow near the ostium.
\textcite{Rodriguez2024,Rodriguez2025} further showed that changes in 
\gls*{PV} flow split
and orientation alter the position of the dominant atrial vortex
and thereby affect blood stasis within the appendage.
Together, these studies show that appendage renewal depends on
both its deformation and its coupling to atrial-body flow.

Classical exchange flow problems provide a useful framework for
examining this connection.
From a fluid mechanics standpoint, the \gls*{LAA} is a
deforming side cavity that exchanges fluid with an unsteady
main flow through a single opening.
In open cavities and river groyne fields, fluid transfer
between a side cavity and the main stream 
is driven by eddies in the shear
layer that spans the opening~\cite{Uijttewaal2001,Chang2006}.
When a passive scalar is released inside such a cavity, its
mean concentration decays nearly exponentially.
Treating the cavity as a well-mixed volume, this washout
can be approximated by a first-order dead-zone model 
with a single exchange coefficient~\cite{Uijttewaal2001,Chang2006}.
In tidal estuaries, by contrast, exchange is driven by the
periodic variation of the water volume. Assuming that this 
volume is locally well mixed at high tide,
\textcite{Ketchum1951} defined an exchange ratio as the fraction of 
the high-tide volume removed during the ebb
to estimate water renewal by successive tides.

Fluid exchange through the LAA ostium is driven by
both cyclic changes in appendage volume and
interactions with the atrial flow.
However, exchange across the ostium does not necessarily
produce effective renewal throughout the appendage.
Following the washout of a passive scalar initially
confined to the appendage,
%
\textcite{Otani2016} found similar appendage
washout histories in two atria with substantially different
structure and function.
\textcite{Bosi2018} compared residual virtual contrast across
four representative appendage morphologies and observed retained scalar
at their tips.
\textcite{sanatkhani2021,sanatkhani2023} reported substantial
variation in clearance within conventional \gls*{LAA} shape groups and persistent
tracer retention at the tips.
Spatially resolved transport has also been assessed in vivo.
\textcite{Severance2026} used dynamic contrast \gls*{4D-CT} to quantify
LAA contrast transport in 93 subjects, demonstrating that patients 
imaged during AF episodes 
showed slower transport than patients imaged during normal sinus 
rhythm. However, they also reported substantial variability among patients imaged during AF, 
suggesting that coordinated LAA contraction is not the only factor affecting fluid exchange.

Blood residence time, or blood age, provides a complementary description of this
transport. Here, residence time measures the elapsed time
since entry into the \gls*{LA} through the \glspl*{PV}.
Unlike washout of a scalar initialized in the \gls*{LAA}, this metric
includes time spent in the atrial body before appendage entry and
during repeated \gls*{LA}--\gls*{LAA} exchange.
Previous studies have reported
elevated residence times toward the appendage tips and distinct younger and older fluid
parcels within the \gls*{LAA}~\cite{Villalba2021,DuenasPamplona2021,
DuenasPamplona2022,Stoecker2025}.
These observations show that renewal in the \gls*{LAA} is heterogeneous and
involves more than one timescale.
Weak circulation in remote regions may contribute to this pattern,
as suggested by the secondary recirculations found in deep
driven cavities~\cite{Shankar2000}.
Renewal can also be reduced by the return of previously ejected
fluid, as demonstrated for tidal embayments~\cite{Sanford1992}.
However, the connection between 
bulk washout, local retention and exchange through the ostium
remains unclear.
Transit models developed for the LV~\cite{Eriksson2013,MartinezLegazpi2025} are 
difficult to extrapolate to the LAA, because the ostium simultaneously
acts as an inlet and outlet, which
adds significant complexity. 

\newpage
The present study addresses three questions.
First, how much do changes in \gls*{LAA} volume contribute
to fluid exchange through the ostium, and how does this
contribution vary over the cardiac cycle and between subjects?
Second, how effectively does the total exchange through the ostium
remove fluid initially present in the appendage, and which mechanisms cause 
departures from well-mixed washout predictions?
Third, what do residence-time distributions reveal about the multi-cycle
and spatially heterogeneous renewal that bulk washout rates 
do not capture?

To answer these questions, we use multi-cycle \gls*{CFD} simulations driven by \gls*{4D-CT}-derived
anatomy and wall motion of 17 subjects in \gls*{SR} or \gls*{AF}.
We decompose total ostium exchange into a volume-driven
contribution, termed \emph{active}, and a remainder representing
simultaneous inflow and outflow, termed \emph{passive}.
To interpret the relationship between fluid exchange and
washout, we compare scalar washout with predictions from
simple well-mixed washout models.
Residence-time distributions and a regional analysis of one appendage
relate bulk transport measures to the spatial distribution of retained
fluid.
The appendage emptying fraction ranges from approximately 
13 to 65\%.
Based on the ostium's equivalent diameter and the peak bulk
relative velocity across 
the ostium, the Reynolds number ranges from 185 to 957.

The manuscript is organized as follows.
Section~\ref{sec:methodology} describes the imaging-derived models,
the numerical method, and the transport measures.
Section~\ref{sec:results} first examines phase-dependent flow and
the contributions to ostium exchange, then compares scalar washout
with washout predictions and analyzes residence-time distributions.
The main conclusions are presented in
Section~\ref{sec:conclusions}.

\section{Methodology}
\label{sec:methodology}

\subsection{Cohort and Imaging}
Patient-specific \gls*{LA} geometries and wall motion were obtained from time-resolved three-dimensional X-ray computed tomography (\gls*{4D-CT}).
The study comprised 17 subjects.
Subjects 1--13 corresponded to the cohort described in Chapters 5 and 6
of \textcite{guerrero2024thesis}.

Imaging data were acquired at the National Institutes of Health,
the University of California San Diego,
and Hospital General Universitario Gregorio Mara\~n\'on as reported in~\cite{guerrero2024thesis},
with a spatial resolution ranging from 0.32 to \SI{0.62}{mm} in-plane and from 0.45 to \SI{1}{mm}
in the through-plane direction.
Images were obtained at intervals of 5--10\% of the cardiac cycle.
Snapshots of the anatomies are displayed in \autoref{fig:cases}, while
anatomical and functional parameters, \gls*{AF} history, and rhythm during
imaging are reported in \autoref{tab:cases}.

{\renewcommand{\arraystretch}{0.1}
\begin{figure}[h!tb]
    \setlength{\tabcolsep}{0pt} 
    \centering
\begin{tabular}{ccccc}
    1 & 2 & 3 & 4 & 5 \\
    \includegraphics[scale=0.12, trim = 150 250 200 150, clip]{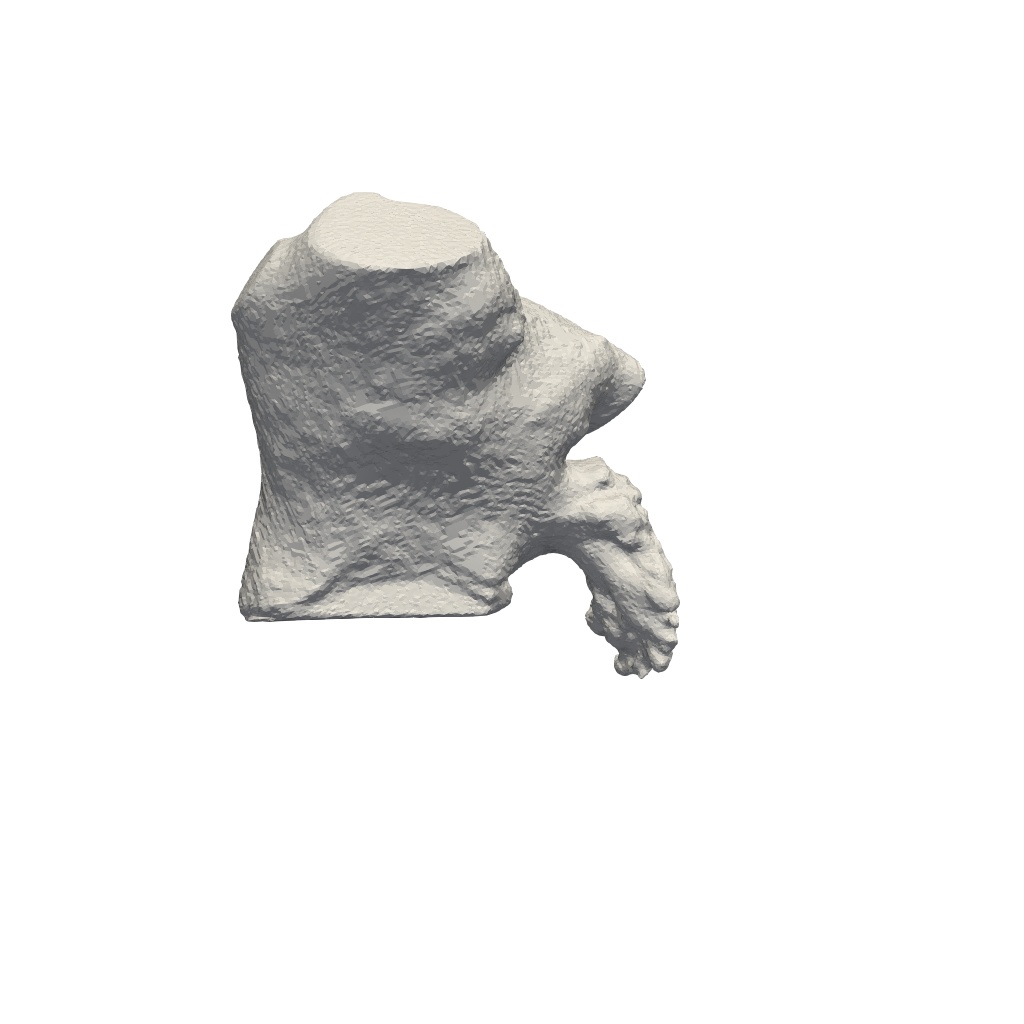} &
    \includegraphics[scale=0.12, trim = 150 150 200 250, clip]{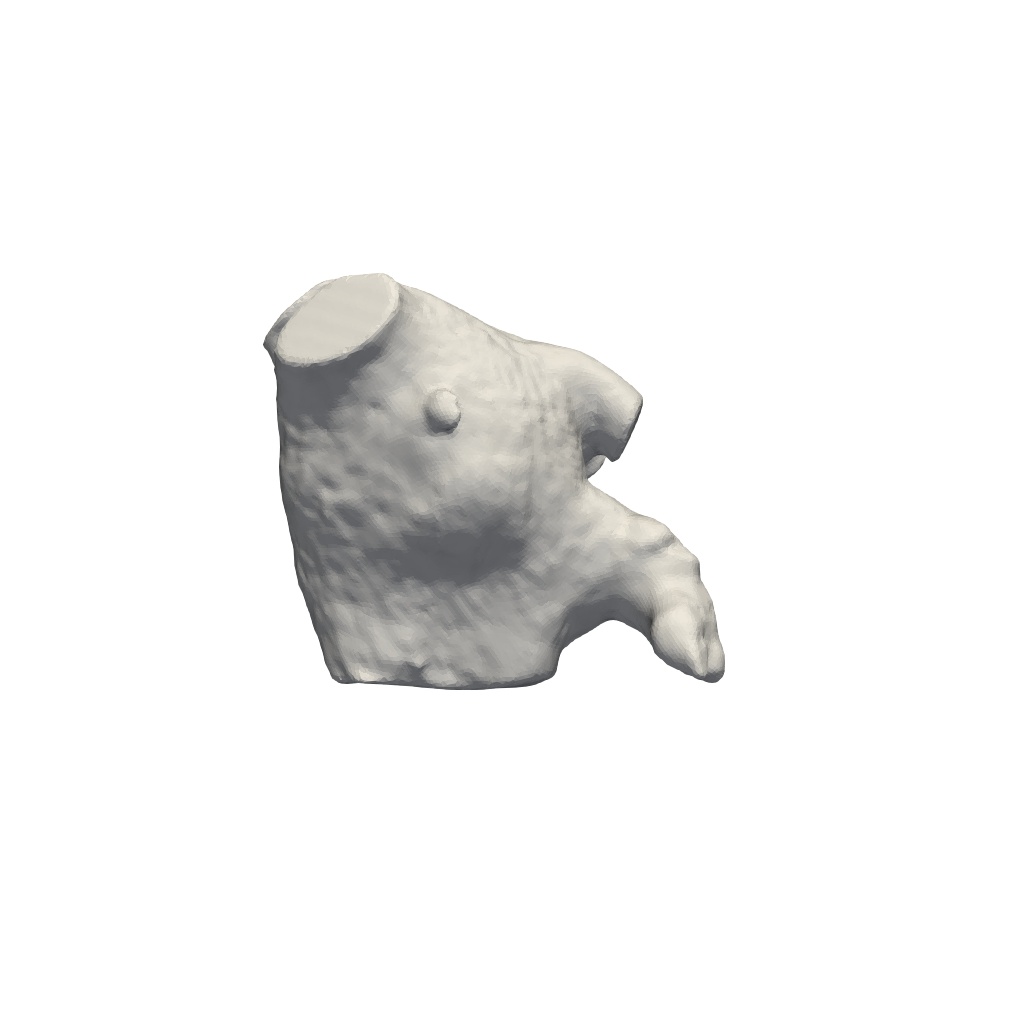}&
    \includegraphics[scale=0.12, trim = 150 150 200 250, clip]{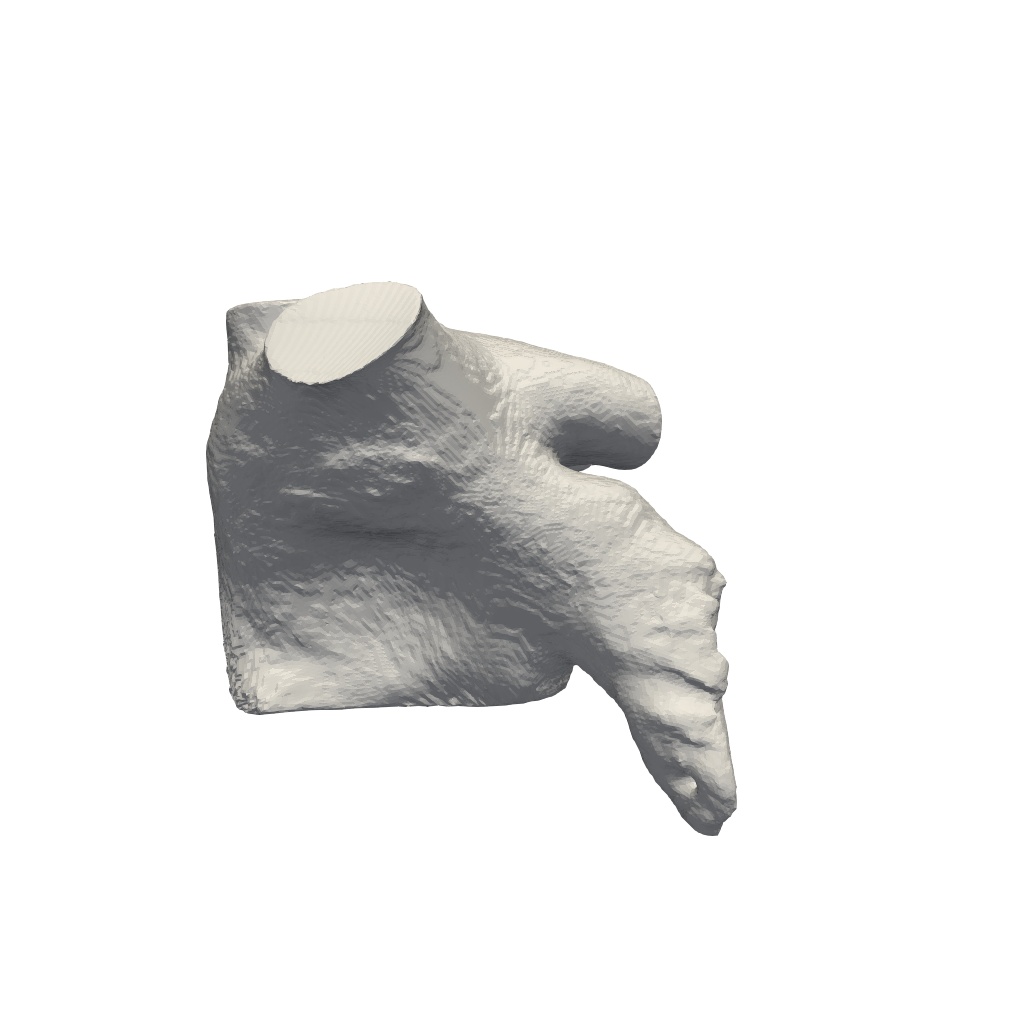}&
    \includegraphics[scale=0.12, trim = 150 150 200 200, clip]{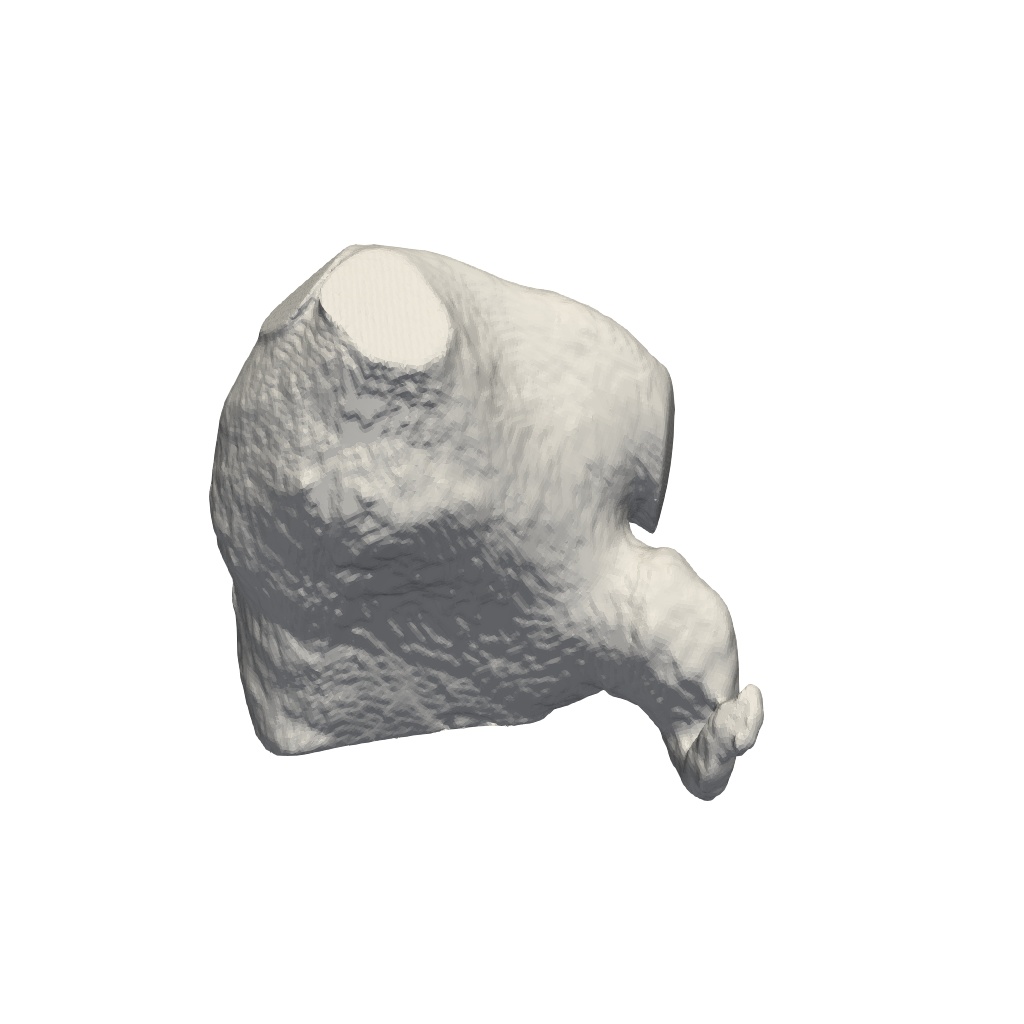}&
    \includegraphics[scale=0.12, trim = 150 150 200 200, clip]{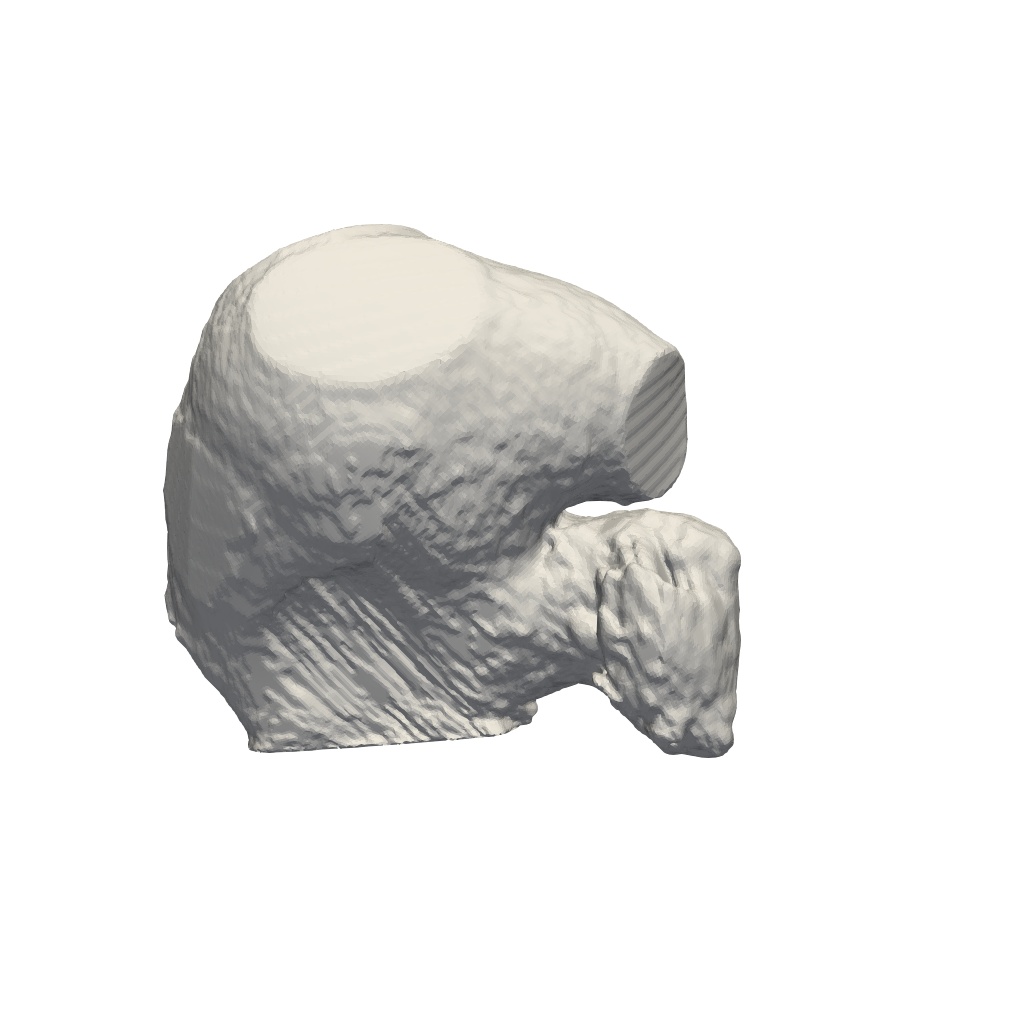} \\
    6 & 7 & 8 & 9 & 10 \\
    \includegraphics[scale=0.12, trim = 150 150 200 80, clip]{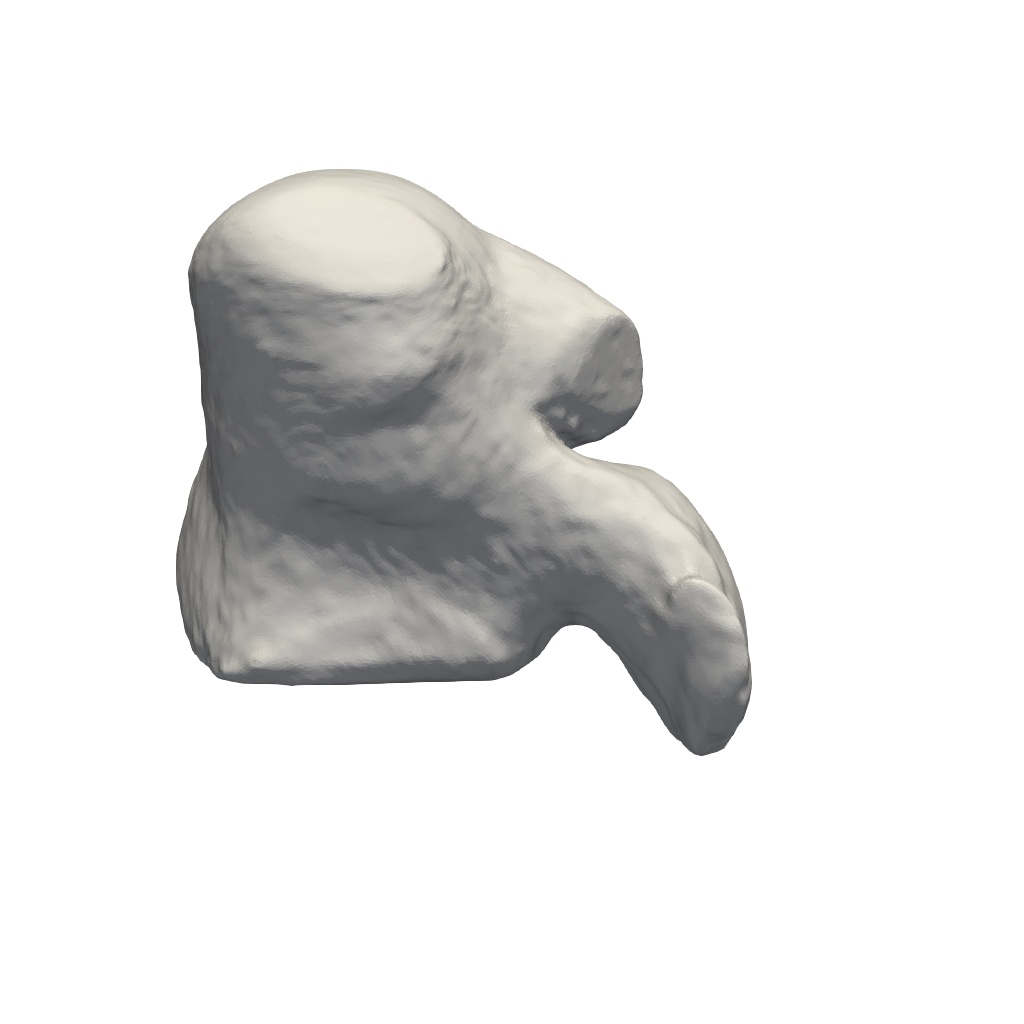}&
    \includegraphics[scale=0.12, trim = 150 150 200 80, clip]{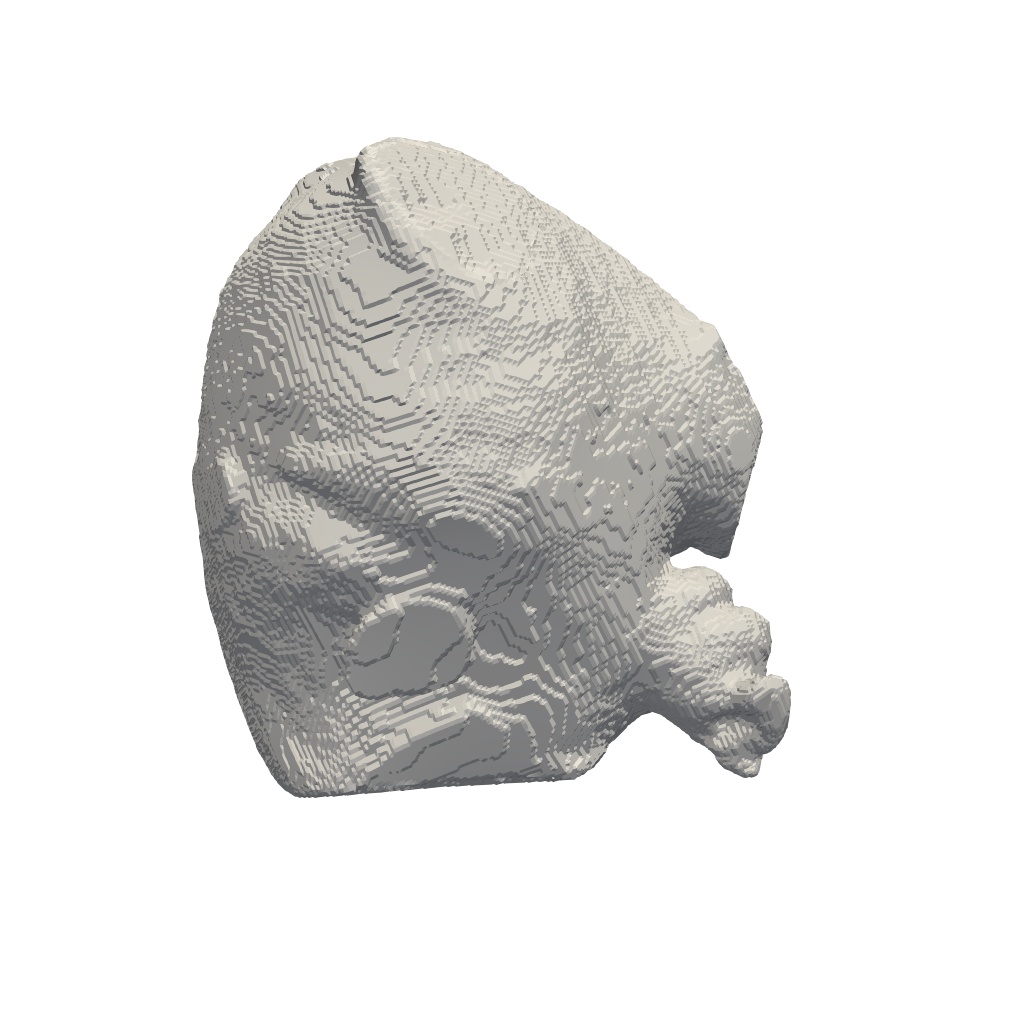}&
    \includegraphics[scale=0.12, trim = 150 150 200 80, clip]{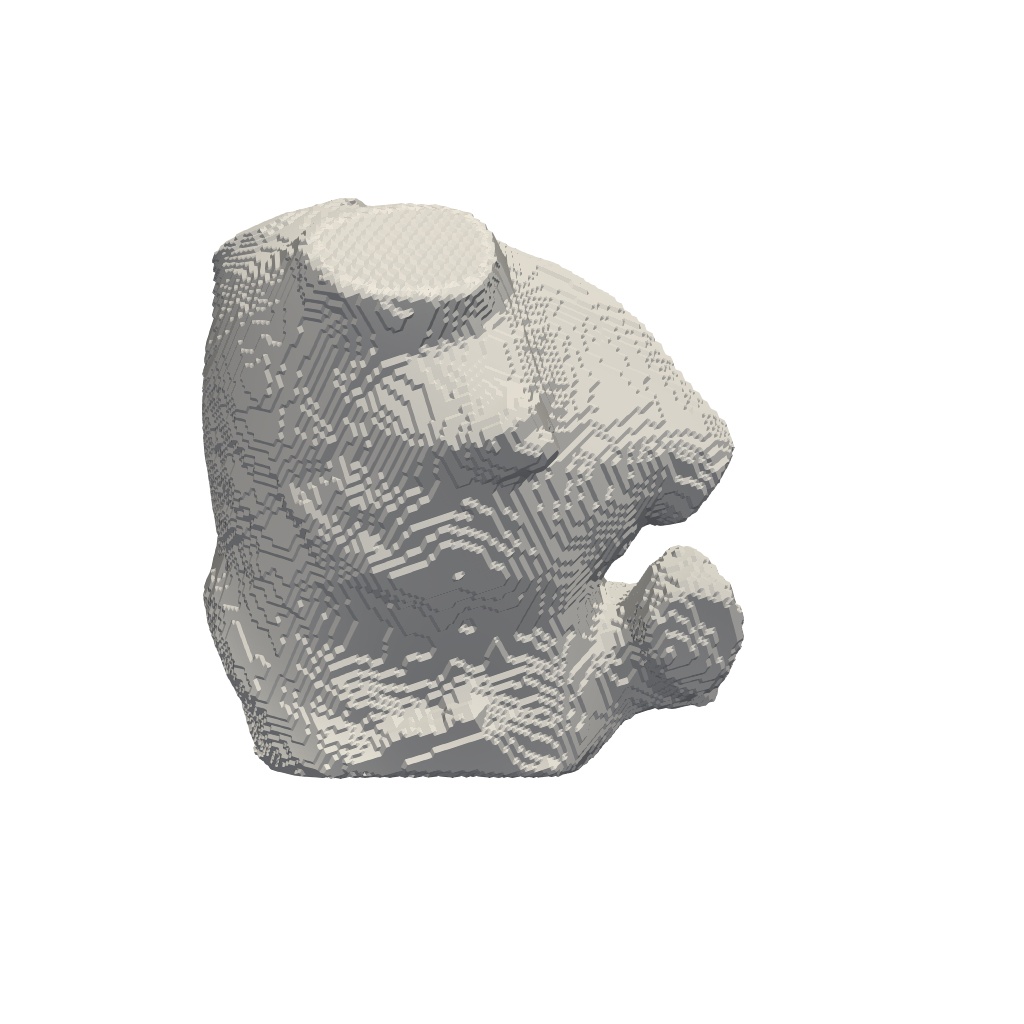}&
    \includegraphics[scale=0.12, trim = 150 200 200 30, clip]{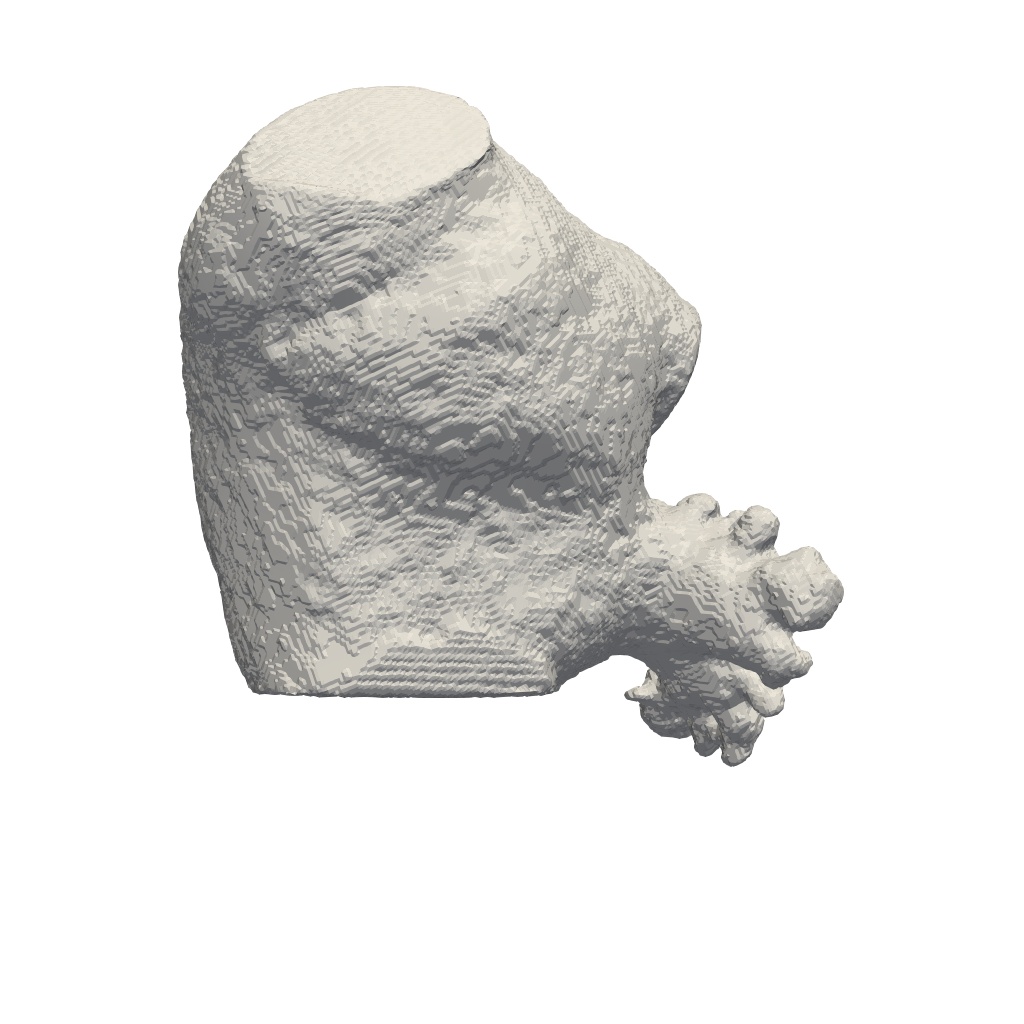}&
    \includegraphics[scale=0.12, trim = 100 200 200 30, clip]{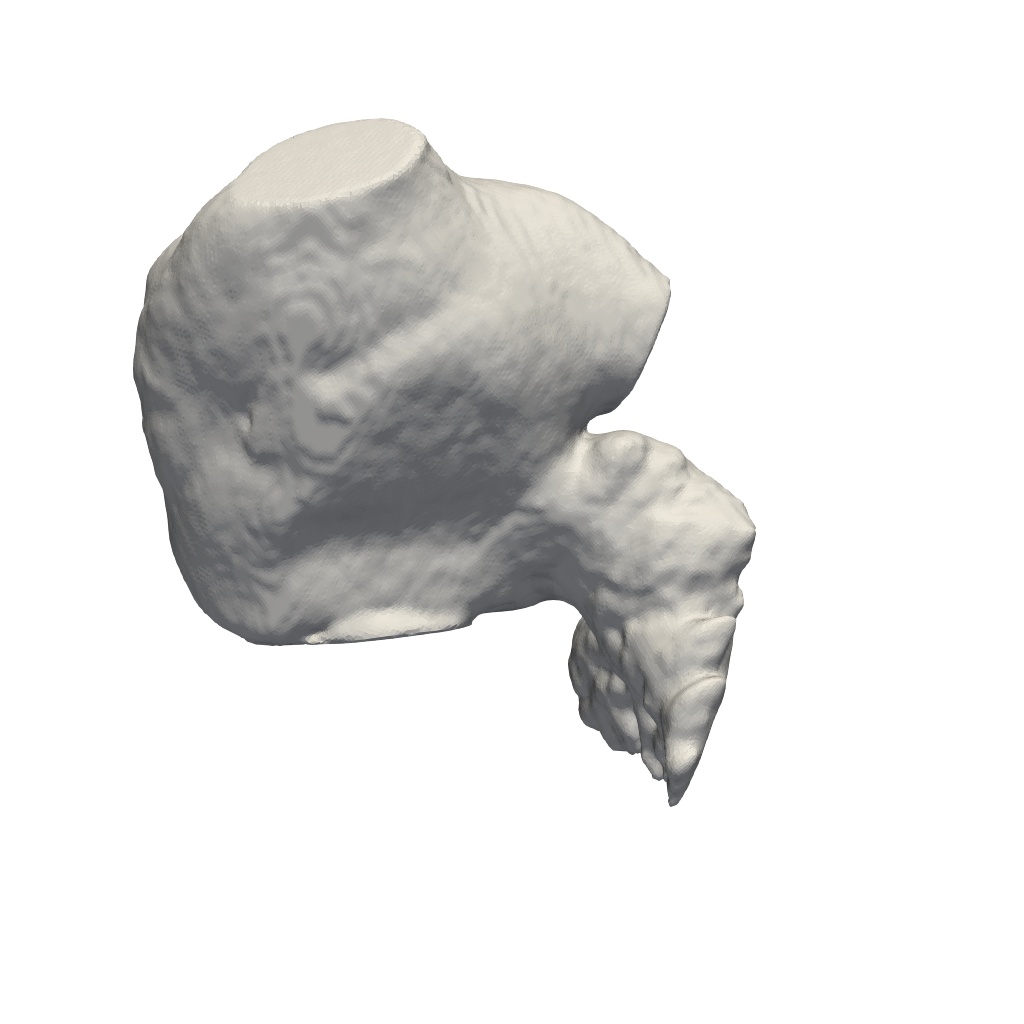} \\
    11 & 12 & 13 & 14 & 15 \\
    \includegraphics[scale=0.12, trim = 150 250 200 80, clip]{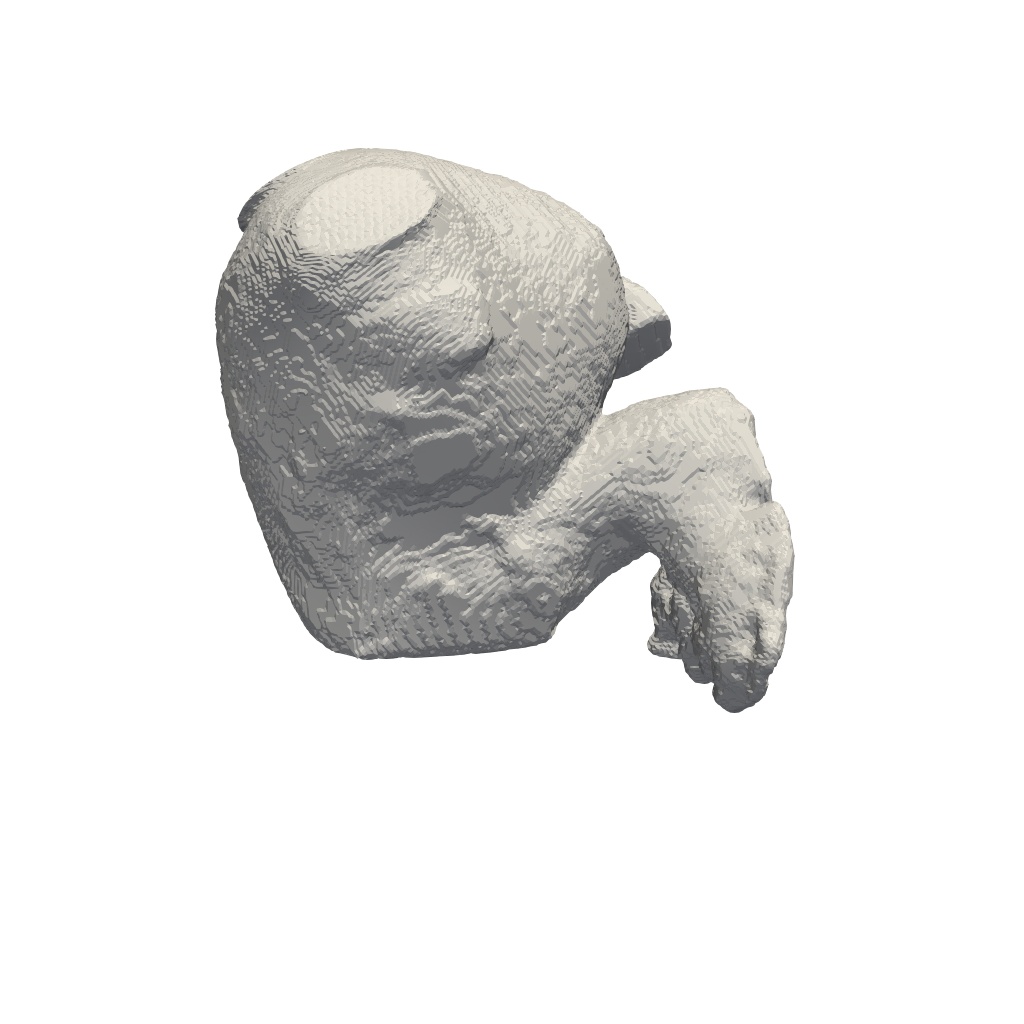}&
    \includegraphics[scale=0.12, trim = 150 200 200 130, clip]{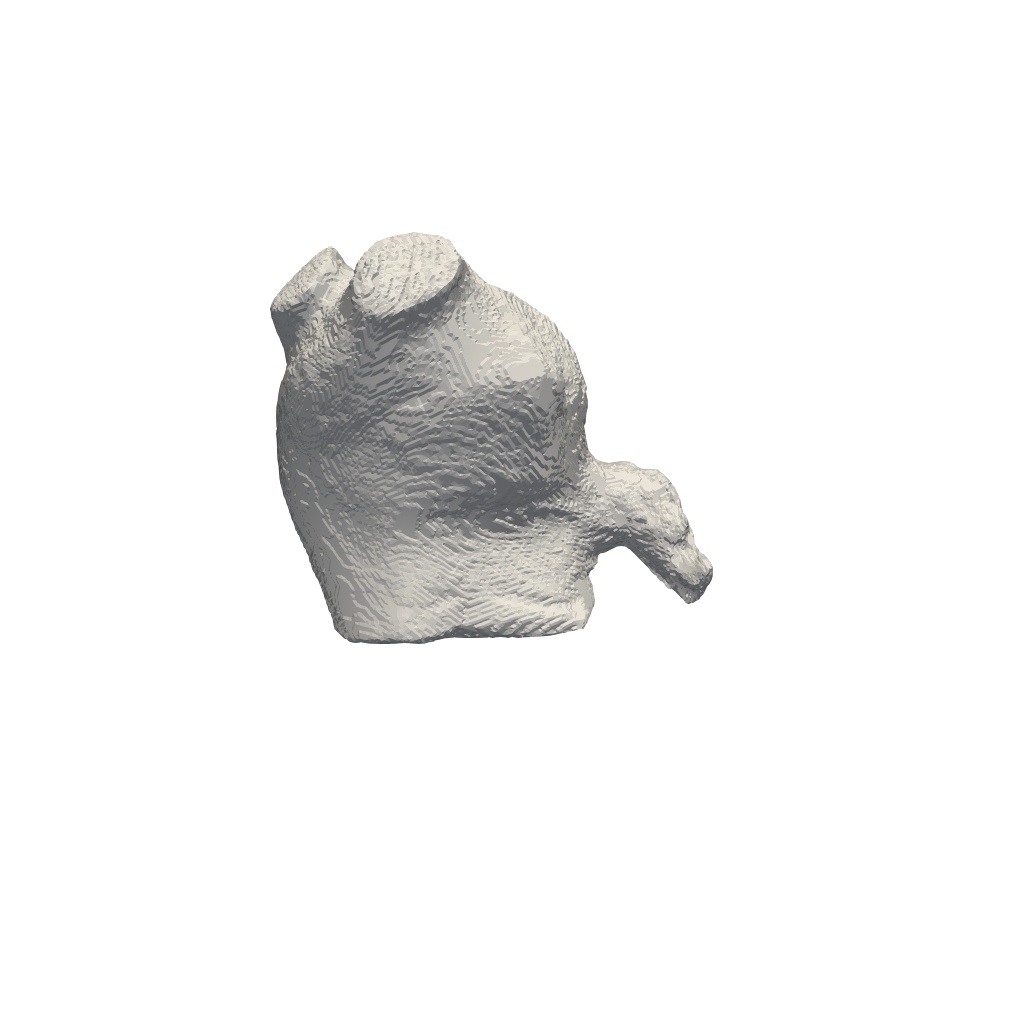}&
    \includegraphics[scale=0.12, trim = 150 150 200 180, clip]{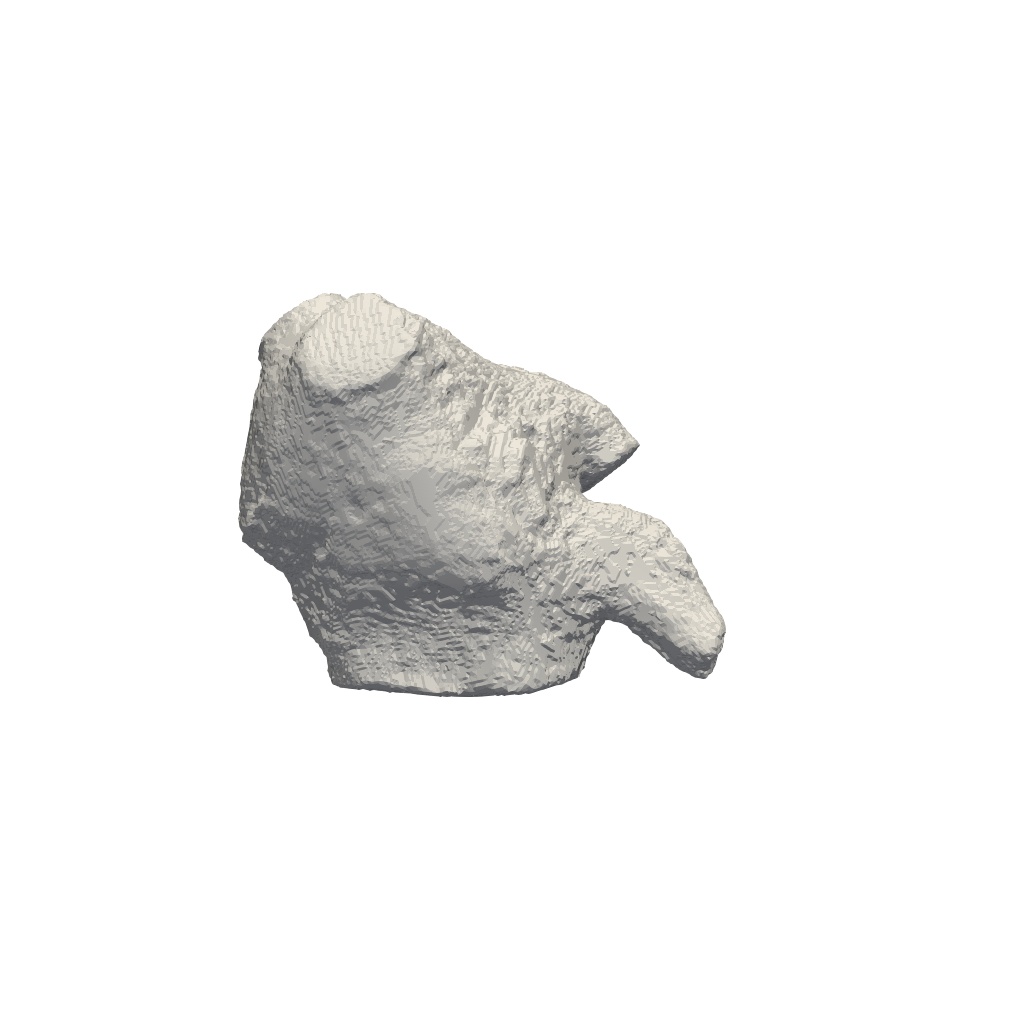}&
    \includegraphics[scale=0.12, trim = 50 200 125 80, clip]{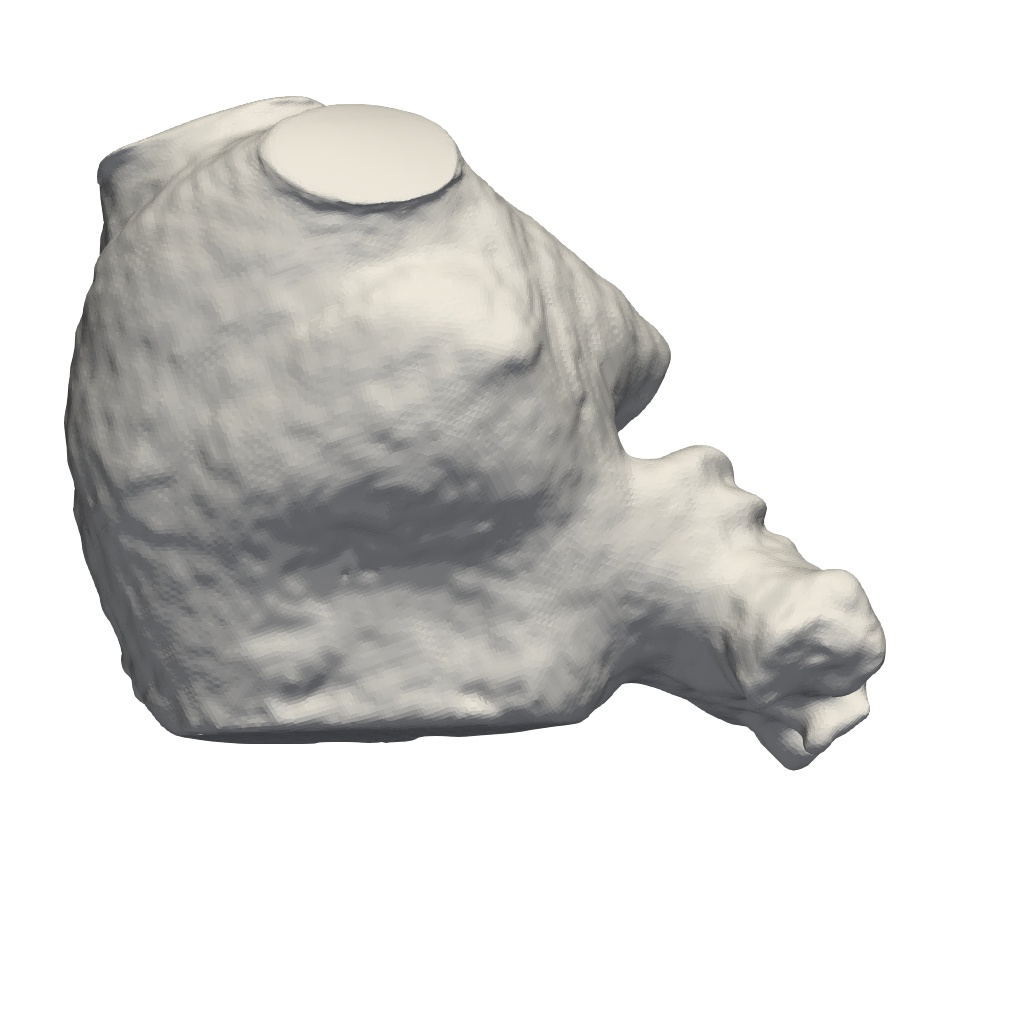}&
    \includegraphics[scale=0.12, trim = 100 200 150 150, clip]{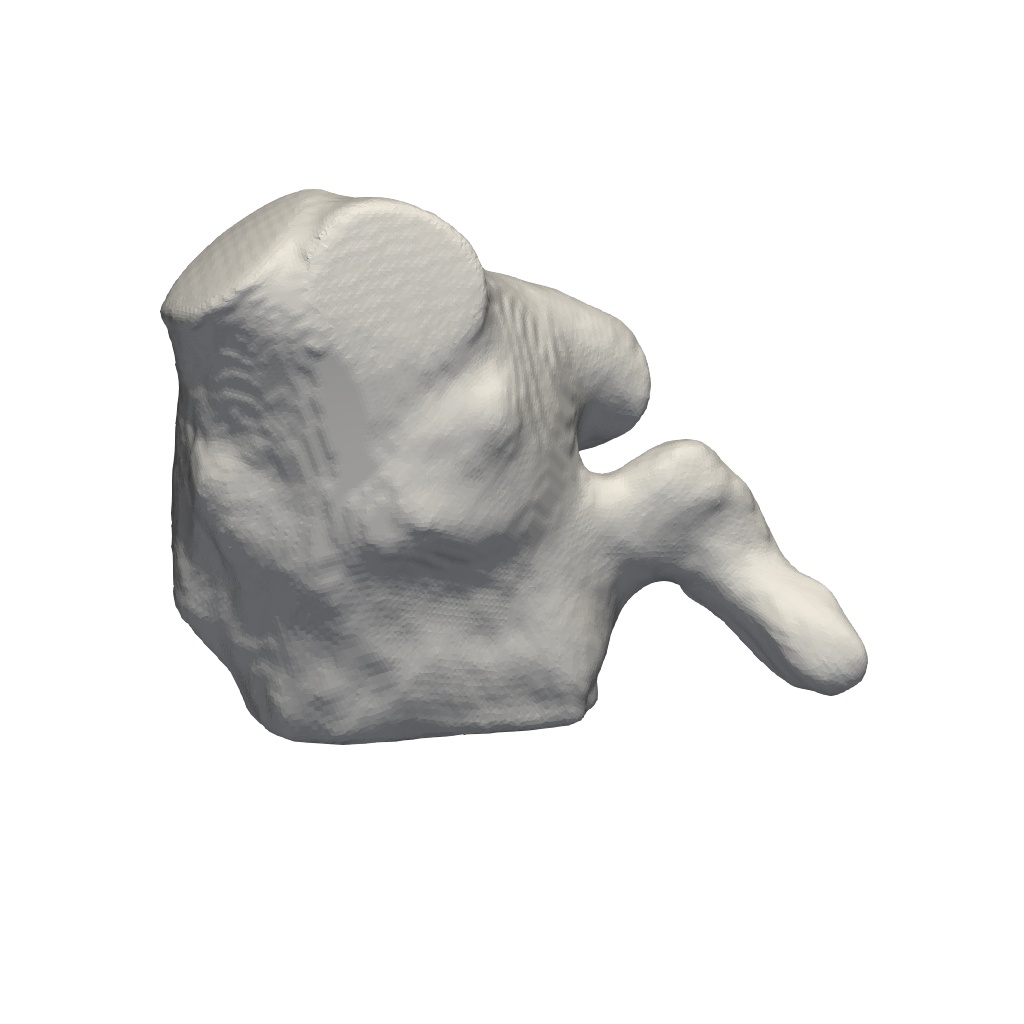} 
    \\
    16 & 17 \\
    \includegraphics[scale=0.12, trim = 150 330 200 0, clip]{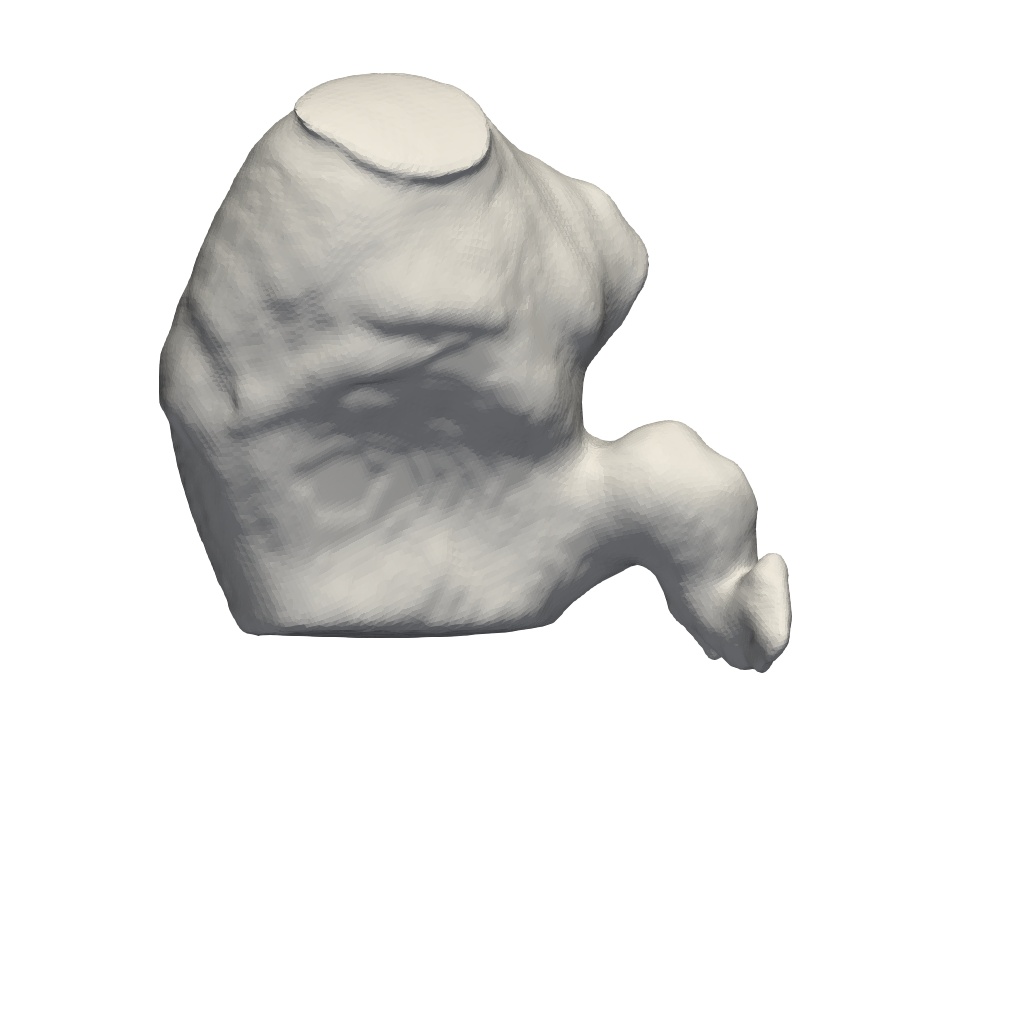}&
    \includegraphics[scale=0.12, trim = 100 250 200 80, clip]{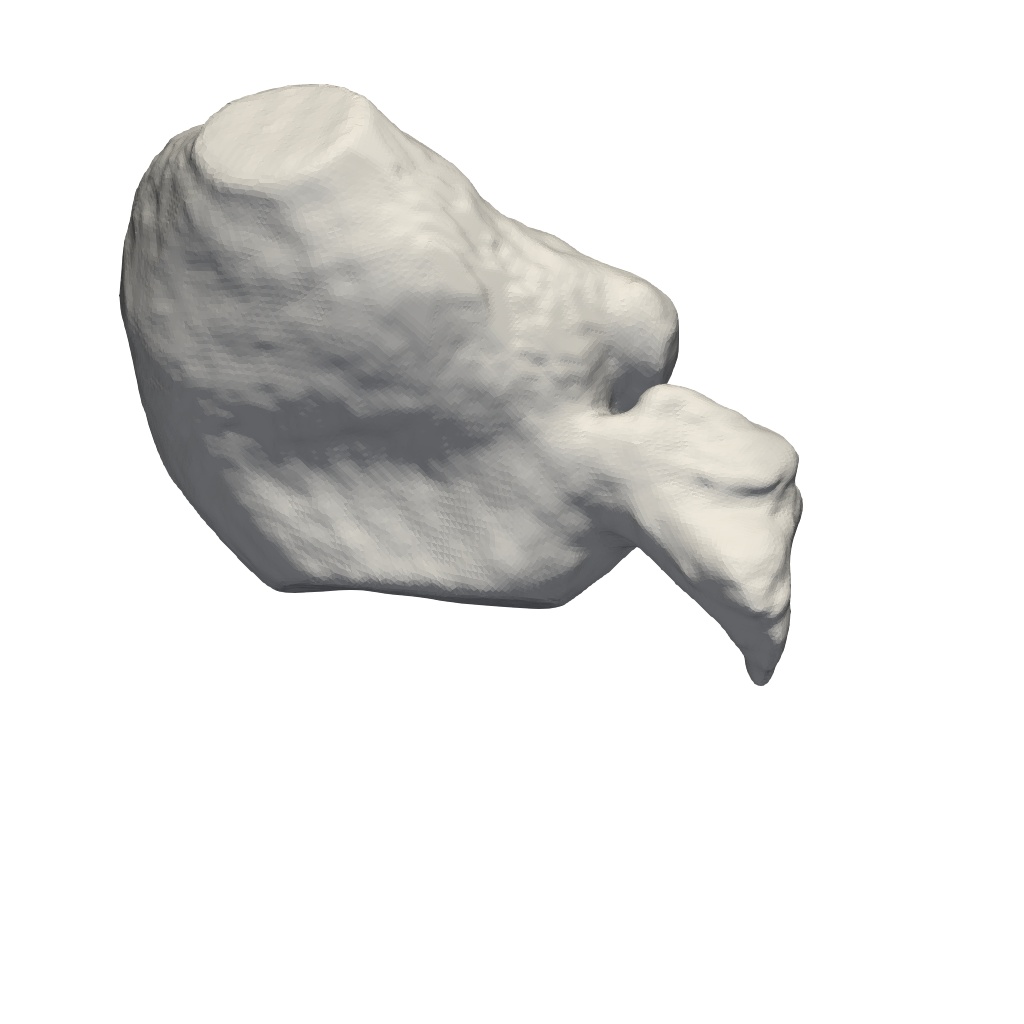}
\end{tabular}
\caption{Patient-specific atrial anatomies shown at the same physical scale.}
\label{fig:cases}
\end{figure}}

\begin{table}[h!tb]
    \centering
    \begin{tabular}{|c|c|c|c|c|c|c|c|}
    \hline
    \multirow{2}{*}{\makecell{subject}}
    & \multicolumn{2}{c|}{$\overline V$ [mL]}&  \multicolumn{2}{c|}{EF [\%]} & \multirow{2}{*}{\makecell{LVSV \\ {[mL]}}} & \multirow{2}{*}{\makecell{AF \\ history}} & \multirow{2}{*}{rhythm} \\
    & LA & LAA
    & LA & LAA & & & \\ \hline
    1 
        & 88 & 6.4 & 42.4 & 53.0 & 80.2 & no & SR \\
    2 
        & 69 & 4.9 & 46.3 & 50.0 & 86.6 & no & SR \\
    3 
        & 113 & 14.3 & 39.9 & 43.1 & 98.6 & no & SR \\
    \rowcolor{gray!20}4 
        & 144 & 10.7 & 23.3 & 21.3 & 72.8 & yes & AF \\
    \rowcolor{gray!20}5 
        & 155 & 15.5 & 9.3 & 20.5 & 76.7 & yes & AF \\
    \rowcolor{gray!20}6 
        & 178 & 22.0 & 23.8 & 20.1 & 138.5 & yes & AF \\
    \rowcolor{gray!20}7 
        & 226 & 5.5 & 12.3 & 23.8 & 86.0 & yes & AF \\
    \rowcolor{gray!20}8 
        & 175 & 6.2 & 12.7 & 29.7 & 109.2 & yes & AF \\
    \rowcolor{gray!20}9 
        & 191 & 14.1 & 11.9 & 13.4 & 109.8 & yes & AF \\
    10 
        & 135 & 17.6 & 26.0 & 32.0 & 90.5 & yes & SR \\
    \rowcolor{gray!20}11 
        & 148 & 15.8 & 13.2 & 28.0 & 76.3 & yes & AF \\
    12 
        & 84 & 3.2 & 44.4 & 65.0 & 109.8 & no & SR \\
    13 
        & 96 & 3.8 & 42.7 & 51.5 & 104.0 & no & SR \\ 
    \rowcolor{gray!20}14 
        & 191 & 17.2 & 15.3 & 17.5 & 93.8 & yes & AF \\
    15 
        & 112 & 7.3 & 22.2 & 41.8 & 90.2 & yes & SR \\ 
    \rowcolor{gray!20}16 
        & 93 & 7.8 & 17.2 & 37.1 & 63.2 & yes & AF \\
    17 
        &  125 & 11.5 & 30.1 & 41.6 & 120.0 & yes & SR \\ \hline
\end{tabular}
\caption{Geometry, function and clinical descriptors. Gray rows highlight subjects imaged during AF. $\overline V$ denotes cycle-averaged chamber volume; EF denotes global emptying fraction, defined as $ \mathrm{EF}=\frac{V_{\max}-V_{\min}} {V_{\max}}$, 
where $V_{\max}$ and $V_{\min}$ are the maximum
and minimum chamber volumes during the cardiac cycle; LVSV  (left ventricular stroke volume) denotes the volume ejected by the \gls*{LV}
during the cycle.}
\label{tab:cases}
\end{table}

\subsection{Geometry and Wall-Motion Reconstruction}
For subjects 1--13, the \gls*{4D-CT} images were segmented in {ITK-SNAP}~\cite{itksnap2006}
using a semi-automatic active contour method.
The \gls*{PV} inlet planes, \gls*{MV} annulus plane, and \gls*{LAA} ostium were
manually delineated at each imaging frame to obtain the \gls*{LA}, \gls*{LAA}
and \gls*{LV} volumes.
Subjects 14--17 were segmented using
nnU-Net~\cite{isensee2021nnu},which was trained using 145 manually segmented 
3D frames from 25 patients with CFD-ready LA, LV, LAA and PV labels.
The resulting segmentations were then corrected manually.
Artificial caps spanning the \gls*{PV} inlet planes and the \gls*{MV} annulus
delimited the reconstructed atrial cavity at its openings, which were used to impose boundary conditions as described below.
Subsequently, surface triangulations were generated,
smoothed and resampled using iso2mesh~\cite{fang2009tetrahedral}.
The frame with the largest \gls*{LA} volume served as the reference
for registering the resulting surface point clouds with the
Coherent Point Drift algorithm~\cite{myronenko2010CPD}.
Finally, periodic wall motion was reconstructed from the registered surfaces by
interpolating the vertex coordinates in time using Fourier series.
A fixed cardiac cycle period of $T=\SI{1}{s}$, corresponding to a heart rate of \SI{60}{bpm},
was used for all simulations.
Two \glspl*{ROI} were defined: the \gls*{LA} \gls*{ROI}, $\Omega_{LA}$, comprised the entire
atrial cavity, including the \gls*{LAA}, whereas the \gls*{LAA} \gls*{ROI}, $\Omega_{LAA}$, comprised
only the appendage cavity.

\newpage
\subsection{Numerical Methods and Computational Setup}
Blood was modeled as an incompressible Newtonian fluid
with a kinematic viscosity of $\nu=\SI{0.04}{cm^2/s}$. The Navier--Stokes
equations were solved with the in-house solver TUCANGPU~\cite{Guerrero2025a}
on a staggered Cartesian grid using second-order central finite differences,
while time integration was performed with a low-storage, three-stage,
semi-implicit Runge--Kutta scheme.

Unless otherwise stated, the computational domain was a triply periodic
cube with an edge length of \SI{13}{cm}
and 256 uniformly distributed points per direction,
giving a grid spacing of $\Delta x\approx\SI{0.51}{mm}$.
A constant time step of $\Delta t=\SI{50}{\us}$ was used,
with the \gls*{CFL}
number remaining below 0.3 throughout all simulations.
Sensitivity to spatial resolution was assessed for subject~2
using grids of $64^3$, $144^3$, $256^3$, and $384^3$ points
(Appendix~\ref{app:grid_refinement}).
The two finest grids yielded similar mean residence times, with a
remaining difference of 3.6\% in the mean washout rate.

The moving \gls*{LA} surface was embedded in this fixed computational domain
and represented by an \gls*{IBM}~\cite{Uhlmann2005}. In this method,
the no-slip condition is imposed through localized volumetric forcing in the vicinity of marker points located on the \gls*{LA} surface.
The marker points were defined as the centroids of the triangular surface elements,
and their motion was prescribed from the imaging-derived kinematic reconstruction.
To reduce residual leakage across the immersed boundary,
a second layer of marker points was added by shifting duplicates
of the original marker points outward
by one grid spacing $\Delta x$ along the local surface normals.

The transmitral outflow rate $Q_{\MV}$ was obtained from the time-dependent \gls*{LV}
volume reconstructed from \gls*{4D-CT}.
For this purpose, the \gls*{LV} volume was Fourier-interpolated and differentiated in time,
with backflow removed by retaining only positive values:
\begin{equation}
 Q_{\MV}(t)=\max\left(\frac{\mathrm{d}V_{\mathrm{LV}}}{\mathrm{d}t},0\right).
 \label{eq:mvflow}
\end{equation}
The total \gls*{PV} inflow rate was then determined from mass conservation in the \gls*{LA}
\begin{equation}
 Q_{\PV}=\frac{\mathrm{d}V_{\LA}}{\mathrm{d}t}+Q_{\MV},
 \label{eq:pvbalance}
\end{equation}
and divided among the \glspl*{PV} to obtain equal inlet velocities in all veins.

This velocity was imposed as a plug profile 
in buffer regions upstream of the \gls*{PV} inlet surfaces
using immersed-boundary forcing that relaxed the local fluid velocity
toward the target velocity with a relaxation timescale of
$\tau_{\mathrm{IBM}}=10\Delta t=\SI{0.5}{ms}$~\cite{Villalba2021}.
The mitral outlet was represented as a no-slip surface when the valve was closed ($Q_\MV=0$) and as an open boundary, i.e. no \gls{IBM} forcing was applied,
when it was open ($Q_\mathrm{MV}>0$).

After initialization with zero velocity, 
all simulations were run for 20 cardiac cycles to reach a quasi-periodic state.
The subsequent eight cycles ($20\leq t/T \leq 28$) were used for analysis and sampled at 40 equidistant instants per cycle.
For the phase-resolved fluid-flow comparisons, values at the same
cardiac phase were averaged across these eight cycles to
describe the recurring phase dependence while
reducing cycle-to-cycle fluctuations.
For a quantity $q$, the phase average was defined as
\begin{equation}
    \langle q\rangle(\phi)
    =\frac{1}{8}\sum_{j=20}^{27}q\big((j+\phi)T\big),
    \qquad 0\leq\phi<1,
    \label{eq:phase_average}
\end{equation}
where $\phi$ is the fraction of the cardiac period.

\subsection{Appendage Ostium and Fluid Exchange}
To evaluate fluid exchange between the \gls*{LAA} and the \gls*{LA} body,
a moving surface spanning the \gls*{LAA} ostium was defined. 
The ostium rim was delineated manually 
during segmentation.
To construct the cap, the ostium rim vertices were projected onto a fitted plane
and the enclosed region was triangulated.
Each interior vertex was defined as a
weighted average of the rim vertices using the 
mean value coordinate method~\cite{Floater2003}
\begin{equation}
    \vec{x}_i(t)=\sum_j w_{ij}\vec{x}_j(t),
    \qquad \sum_j w_{ij}=1.
\end{equation}
These weights were 
held fixed throughout the cycle and applied to the three-dimensional rim positions and velocities,
such that the cap followed the local motion and deformation of the ostium rim.
Fluid velocities were interpolated from their staggered grid locations to 
the centroid of each cap triangle using the same interpolation strategy as
used in the \gls*{IBM} implementation in TUCANGPU.

The relative normal velocity through the ostium surface was obtained as
\begin{equation}
 \gls{w}(\vec{x},t)=\left[\gls{uf}(\vec{x},t)-\gls{uA}(\vec{x},t)\right]\cdot\gls{n}(\vec{x},t),
 \qquad \vec{x}\in \gls{ostium-surface}(t),
 \label{eq:w}
\end{equation}
where $\gls{ostium-surface}(t)$ is the moving ostium surface with
local velocity \gls{uA}, $\gls{uf}$ is the fluid velocity,
and \gls{n} is the unit normal pointing from
the \gls{LAA} into the \gls{LA} body.
Based on this, the instantaneous \gls*{LAA} inflow and outflow rates were calculated as
\begin{align}
 \gls{Qf;in;CFD}(t)&=\int_{\gls{ostium-surface}(t)}\max(-\gls{w}(\vec{x},t),0)\,\mathrm{d}A, \label{eq:Qin} \\
 \gls{Qf;out;CFD}(t)&=\int_{\gls{ostium-surface}(t)}\max(\gls{w}(\vec{x},t),0)\,\mathrm{d}A.
\end{align}
Additionally, we defined the \emph{total} ostial exchange rate as the
total flux magnitude crossing the ostium, counting both crossing directions,
\begin{equation}
 \gls{Qf;total}(t)=\gls{Qf;in;CFD}(t)+\gls{Qf;out;CFD}(t).
\end{equation}

Part of this total flow rate can be directly attributed to the
\gls{LAA} volume change. This volume-driven contribution is termed \emph{active} exchange
and determined as
\begin{equation}
 \gls{Qf;active}(t)=\left| \frac{\mathrm{d}\gls{VLAA}}{\mathrm{d}t}\right|.
\end{equation}
The remaining contribution, termed \emph{passive} exchange,
\begin{equation}
 \gls{Qf;passive}(t)=\gls{Qf;total}(t)-\gls{Qf;active}(t)
 = 2\,\min\!\big(\gls{Qf;in;CFD}(t), \gls{Qf;out;CFD}(t)\big) \;\geq 0,
 \label{eq:passive}
\end{equation}
quantifies simultaneous inflow and outflow beyond the net
transport required by appendage volume change.
It vanishes when the ostium carries flow in only one direction.

Integrating the total flow rate over one cardiac cycle gives the
total exchanged volume per cycle
\begin{equation}
    \gls{Vf;total} = \int_{t_n}^{t_n+T} \gls{Qf;total}(t)\,\mathrm{d}t = \gls{Vf;active} +\gls{Vf;passive},
\end{equation}
with $\gls{Vf;active}$ and $\gls{Vf;passive}$ defined analogously.
Here, $t_n$ is the beginning of the cycle.

To enable comparison of exchange quantities
across appendages of different sizes,
volumes were normalized by the time-averaged \gls{LAA} volume \gls{VLAA;mean}
and flow rates by $\gls{VLAA;mean}/T$,
\begin{equation}
 \widehat V=\frac{V}{\gls{VLAA;mean}},\qquad
 \widehat Q=\frac{QT}{\gls{VLAA;mean}}.
\end{equation}

\newpage
\subsection{Regional Analysis}
\begin{figure}[h!tb]    
    \centering
    \includegraphics[width=.2\linewidth]{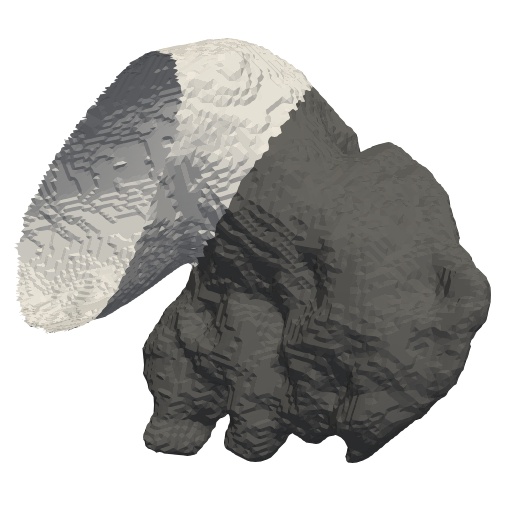}
    \caption{
    Separation of the appendage of subject~11 into proximal (lighter) and distal (darker) regions of similar volumes.
    }
    \label{fig:LAAcut}
\end{figure}

Regions near the ostium are termed \emph{proximal}, while
regions farther into the appendage, toward its tips, are
termed \emph{distal}.
For subject~11, the \gls*{LAA} was divided into
proximal and distal regions of similar volume as depicted in \autoref{fig:LAAcut}.
To locate the cutting plane, the distance of each voxel from the \gls*{LAA}
tip was determined at $t/T=0$, and a cutting plane was fitted to
separate the regions.
The resulting surface regions were tracked throughout the prescribed wall motion.

\subsection{Scalar Washout and Well-Mixed Washout Models}
To investigate the washout of fluid from the \gls{LAA},
we analyzed the transport of a passive scalar that was initialized at $t_0=20\,T$
with a uniform concentration $c=1$ inside the \gls{LAA}
and $c=0$ everywhere else.
The scalar concentration was evolved over time by solving the transport equation
\begin{equation}
     \frac{\partial c}{\partial t}+\gls{uf}\cdot\nabla c=0,
\end{equation}
using a third-order \gls*{WENO} scheme~\cite{Shu2020} for spatial discretization
to limit spurious oscillations near sharp gradients in the
absence of an explicit diffusion term.

The instantaneous mean concentration within the \gls{LAA} was then defined as
\begin{equation}
 \gls{c}(t)=\frac{1}{V_{\LAA}(t)}
 \int_{\Omega_{\LAA}(t)}c(\vec{x},t)\,\mathrm{d}V.
 \label{eq:cmean}
\end{equation}

Two reduced models were used to assess whether the observed washout
could be explained by \gls{LAA} pumping, that is, through volume change,
or by the actual flow crossing the ostium, as determined from the \gls{CFD} simulations.
Both models assumed that
(1) the fluid within the \gls{LAA} is perfectly mixed,
that is, it has a uniform concentration \gls{c;model};
(2) fluid entering the \gls{LAA} has zero concentration;
(3) fluid leaving the \gls{LAA} has an outflow concentration
equal to the instantaneous mixed concentration \gls{c;model}.
Consequently, the model-predicted concentration of scalar
cannot increase and
only decreases when fluid enters the \gls{LAA}.
With $\gls{c;model}(t_0)=1$, the resulting concentration was
\begin{equation}
 \gls{c;model}(t)=\exp\left[-\int_{t_0}^{t}
 \frac{Q_{\mathrm{in}}(\tau)}{\gls{VLAA}(\tau)}\,\mathrm{d}\tau\right],
 \label{eq:mix}
\end{equation}
where $Q_{\mathrm{in}}$ is the inflow rate assumed by a model.
The derivation of this expression is provided
in Appendix~\ref{app:perfect_mixing}.
For the volume-driven model,
inflow was attributed solely to appendage expansion,
\begin{equation}
 Q_{\mathrm{in}}(t)=\max\left( \frac{\mathrm{d}\gls{VLAA}}{\mathrm{d}t},0\right),
\end{equation}
which neglects passive exchange.
In the inflow-driven model,
the CFD-derived inflow
rate (see Eq.~\ref{eq:Qin}) was used instead.

An effective washout rate was evaluated from the decrease in mean
\gls{LAA} concentration over each cardiac cycle beginning at $t_n=t_0+nT$,
\begin{equation}
 k_n=-\frac{1}{T}\ln\!\left[\frac{\gls{c}(t_n+T)}{\gls{c}(t_n)}\right],
  \label{eq:cycle_decay}
\end{equation}
where $n$ counts cycles since initialization.
This rate was averaged over the
seven cycles following the first cycle after initialization.
The first cycle was excluded to reduce sensitivity to the
initial washout transient, which can differ substantially from subsequent cycles.
We denote the mean washout rate by $k=\frac{1}{7}\sum_{n=1}^{7}k_n$ and
use the same averaging interval for the model rates and
the scalar-balance correction terms below.
Concentrations were evaluated at the beginning of successive
cycles to characterize decay separately from within-cycle
fluctuations.

The corresponding mean washout rates of the 
volume-driven and inflow-driven models
are denoted by $k_V$ and $k_Q$, respectively.
The relative washout efficiency was then defined as
$R_m=k/k_m$, with $m\in\{V,Q\}$.
This ratio compares CFD-derived washout with the corresponding
washout-model prediction.
Differences between model-predicted and CFD-derived concentration
may reflect spatially nonuniform concentrations over finite-time 
mixing and re-ingestion of previously expelled scalar,
which the fresh-inflow assumption excludes.

We used a cycle-integrated scalar balance
to quantify the effects of scalar re-ingestion
and of differences between the outflow concentration
and the instantaneous LAA mean:
\begin{equation}
 k_n=\underbrace{\frac{1}{T} \int_{t_n}^{t_n+T}
 \frac{\gls{Qf;in;CFD}} {\gls{VLAA}}\,\mathrm{d}t
 }_{k_{Q,n}}
 \underbrace{-\frac{1}{T}\int_{t_n}^{t_n+T}
 \frac{\gls{Qf;in;CFD}}  {\gls{VLAA}}
 \frac{\gls{c}_\mathrm{in}} {\gls{c}}\,\mathrm{d}t
 }_{\gls{alpha}_n}
 +\underbrace{\frac{1}{T} \int_{t_n}^{t_n+T}
 \frac{\gls{Qf;out;CFD}} {\gls{VLAA}}
 \left( \frac{\gls{c}_\mathrm{out}} {\gls{c}} - 1 \right)
 \,\mathrm{d}t
 }_{\gls{beta}_n}.
 \label{eq:washout_decomposition}
\end{equation}
Here, $\gls{c}_\mathrm{in}$ and $\gls{c}_\mathrm{out}$ are the
mean concentrations of incoming and outgoing fluid,
respectively, and $k_{Q,n}$ is the inflow-driven washout
prediction for cycle $n$.
Scalar re-ingestion is quantified by the non-positive term $\gls{alpha}_n$,
which slows down washout.
The term $\gls{beta}_n$ is the outflow concentration correction and
can be either negative or positive, where
$\gls{beta}_n<0$ and $\gls{beta}_n>0$ indicate preferential export of
scalar-poor and scalar-rich fluid, respectively.
Averaging each term over the same seven cycles gives
$k=k_Q+\gls{alpha}+\gls{beta}$.
Details on the derivation are given in
Appendix~\ref{app:washout_decomposition}.

\newpage
\subsection{Residence Time}
Blood residence time $T_R$ was computed alongside the velocity field by
solving the forced transport equation
\begin{equation}
 \frac{\partial T_R}{\partial t}+\gls{uf}\cdot\nabla T_R=1,
 \label{eq:rt}
\end{equation}
which was discretized using the same third-order \gls*{WENO} scheme as the passive scalar.
In the \gls{PV} buffer regions, $T_R$ was
continuously set to zero using the \gls*{IBM} forcing,
such that this quantity measures the accumulated age of blood parcels since
entering the \gls{LA}.

Mean LAA blood residence time was compared with the scalar washout
timescale.
For a steady, constant-volume compartment with an initially
uniform scalar concentration and scalar-free, zero-$T_R$ inflow,
the mean residence time since compartment entry satisfies
\begin{equation}
    \overline{T}_{R}
    = \int_0^\infty
    \frac{\overline{c}(\tau)}
    {\overline{c}(0)}
    \mathrm{d}\tau,
    \label{eq:age_washout_reference}
\end{equation}
where $\tau$ denotes elapsed time since scalar initialization.
For exponential concentration decay, this integral equals $1/k$,
giving $k\overline{T}_{R}=1$.
\section{Results and Discussion}
\label{sec:results}

\subsection{Phase-Dependent Atrial and Appendage Flow}

\begin{figure}[h!tb]
    \centering
    \setlength{\tabcolsep}{0pt} 
    \begin{tabular}{cccccc}
    & \makecell{atrial \\ expansion} & \makecell{peak \\ E-wave} & \makecell{ventricular \\ diastasis} & \makecell{peak atrial \\ contraction rate } \\
    & $t/T = 20.075$ & $t/T = 20.425$ & $t/T = 20.675$ & $t/T = 20.900$\\
    \raisebox{.25cm}[0mm][0mm]{
    \begin{tikzpicture}
        \node[inner sep=0pt,rotate=90, anchor=west](cbar) at (0,0) {\includegraphics[width=2cm,height=0.2cm]{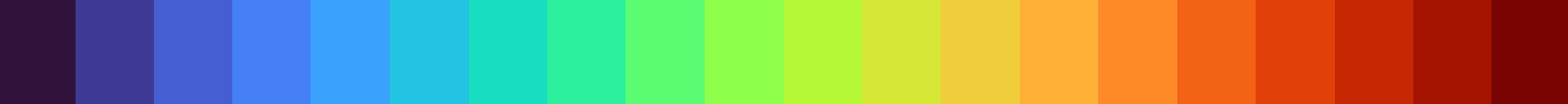}};
        \node(o)[anchor=west] at (-.15,0) {\small $-$};
        \node[anchor=west] at ([yshift=1cm]o.west) {\small $-$};
        \node[anchor=west] at ([yshift=2cm]o.west) {\small $-$};
        \node(o)[anchor=west] at (.25,0) {\small 0};
        \node[anchor=west] at ([yshift=1cm]o.west) {\small 20};
        \node[anchor=west] at ([yshift=2cm]o.west) {\small 40};
        \node[anchor=north west,text centered] at (-.3,-.1)
        {\small $|\gls{uf}|$ (\si{cm/s})};
    \end{tikzpicture}}&
    \includegraphics[scale=.13,trim=170 50 300 250, clip]{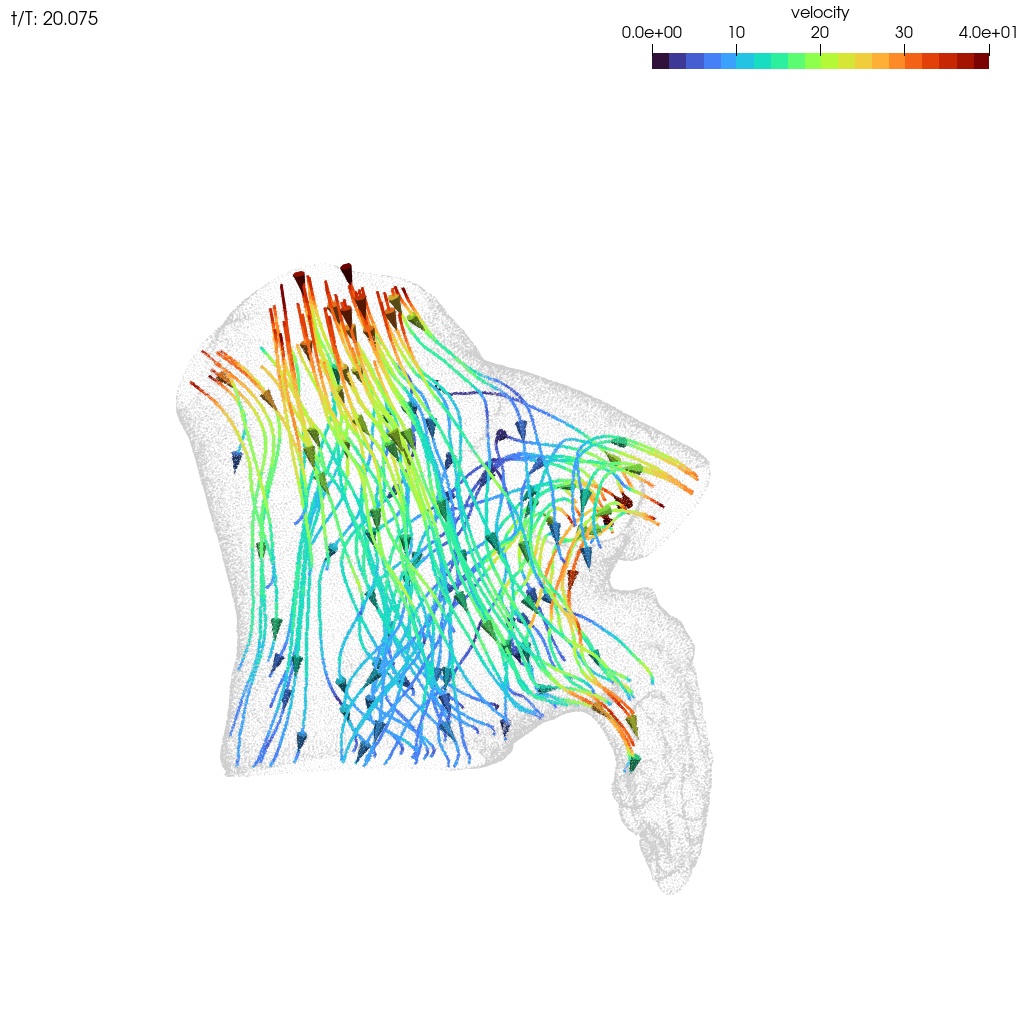}&
    \includegraphics[scale=.13,trim=150 50 280 250, clip]{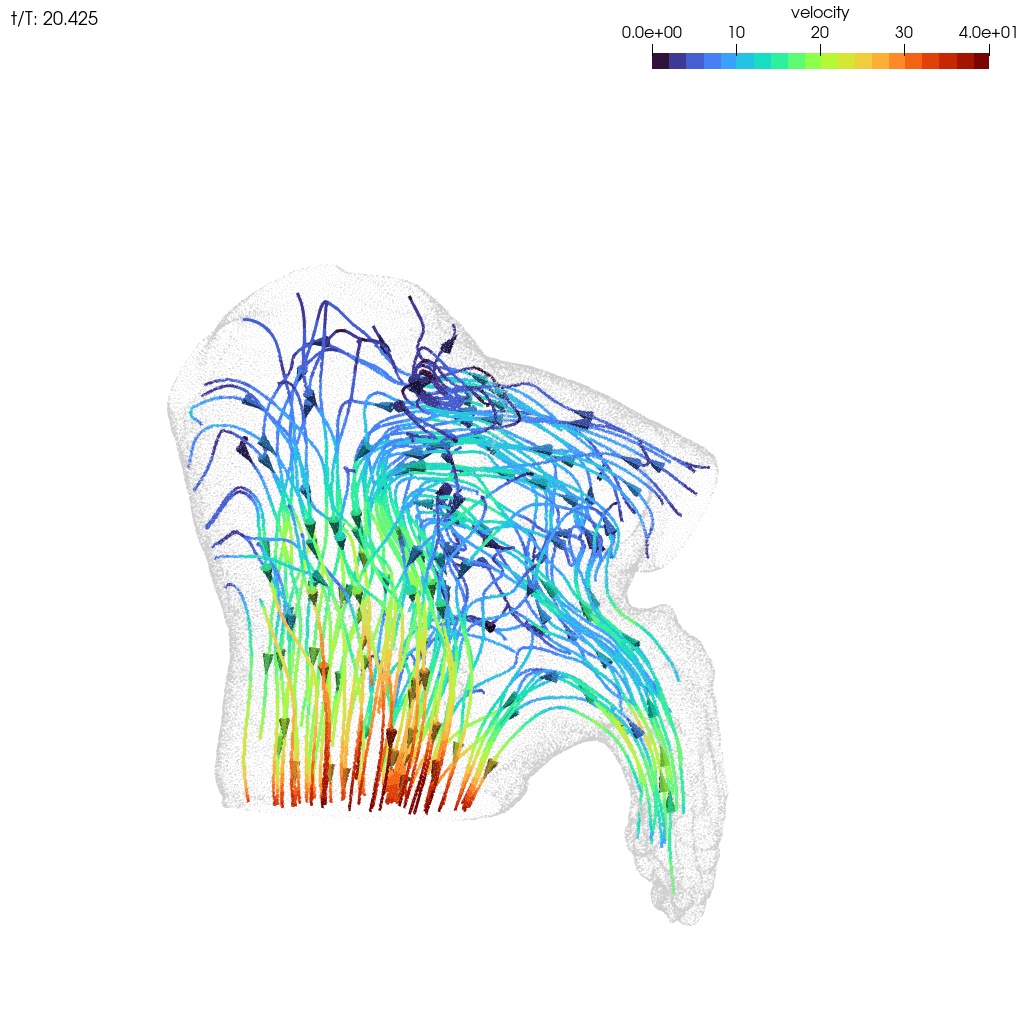} &
    \includegraphics[scale=.13,trim=150 50 280 250, clip]{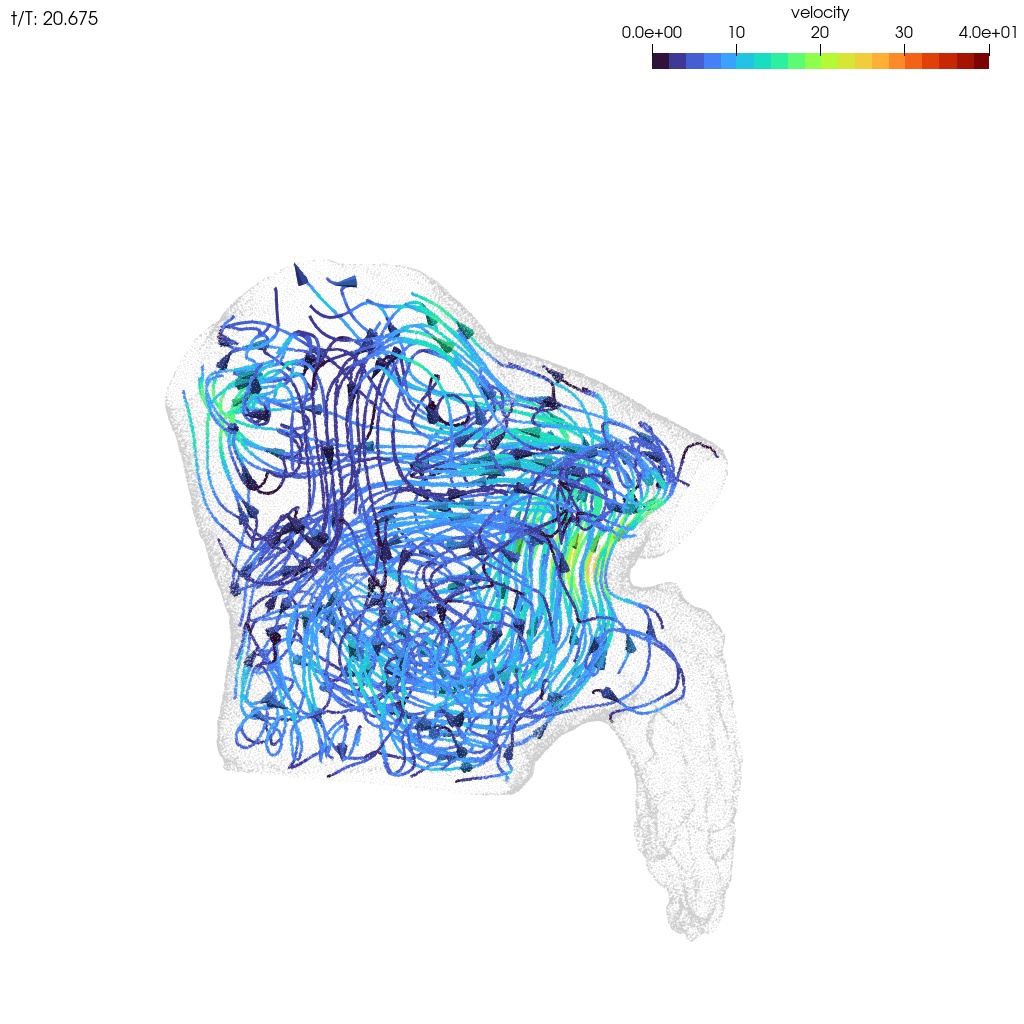} & 
    \includegraphics[scale=.13,trim=170 50 280 250, clip]{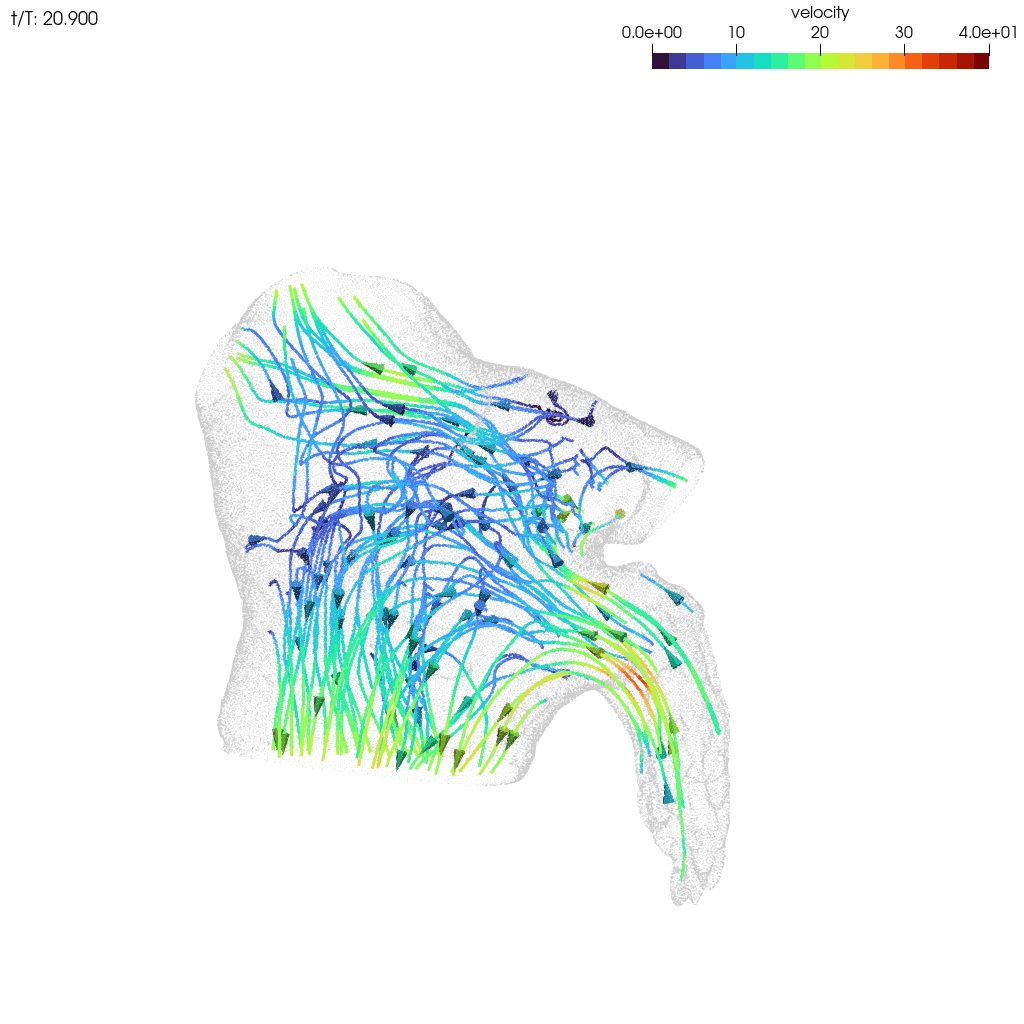} &
    \includegraphics[scale=.4]{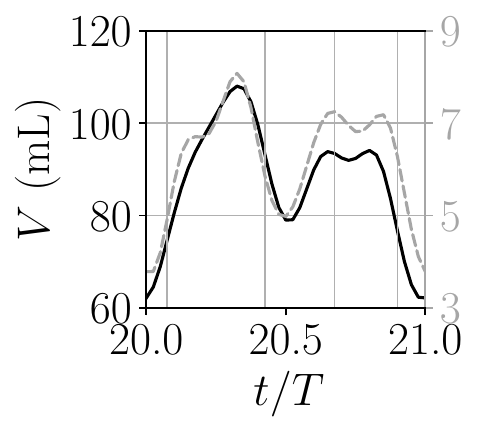} 
    \\
    \begin{tikzpicture}
        \node[inner sep=0pt,rotate=90, anchor=west](cbar) at (0,0) {\includegraphics[width=2cm,height=0.2cm]{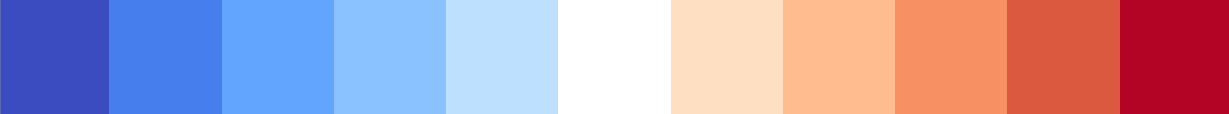}};
        \node(o)[anchor=west] at (-.15,0) {\small $-$};
        \node[anchor=west] at ([yshift=1cm]o.west) {\small $-$};
        \node[anchor=west] at ([yshift=2cm]o.west) {\small $-$};
        \node(o)[anchor=west] at (.25,0) {\small $-20$};
        \node[anchor=west] at ([yshift=1cm]o.west) {\small 0};
        \node[anchor=west] at ([yshift=2cm]o.west) {\small 20};
        \node[anchor=north west,text centered] at (-.3,-.1)
        {\small $\gls{w}$ (\si{cm/s})};
    \end{tikzpicture}
    &
    \includegraphics[scale=.13,trim=200 150 300 150, clip]{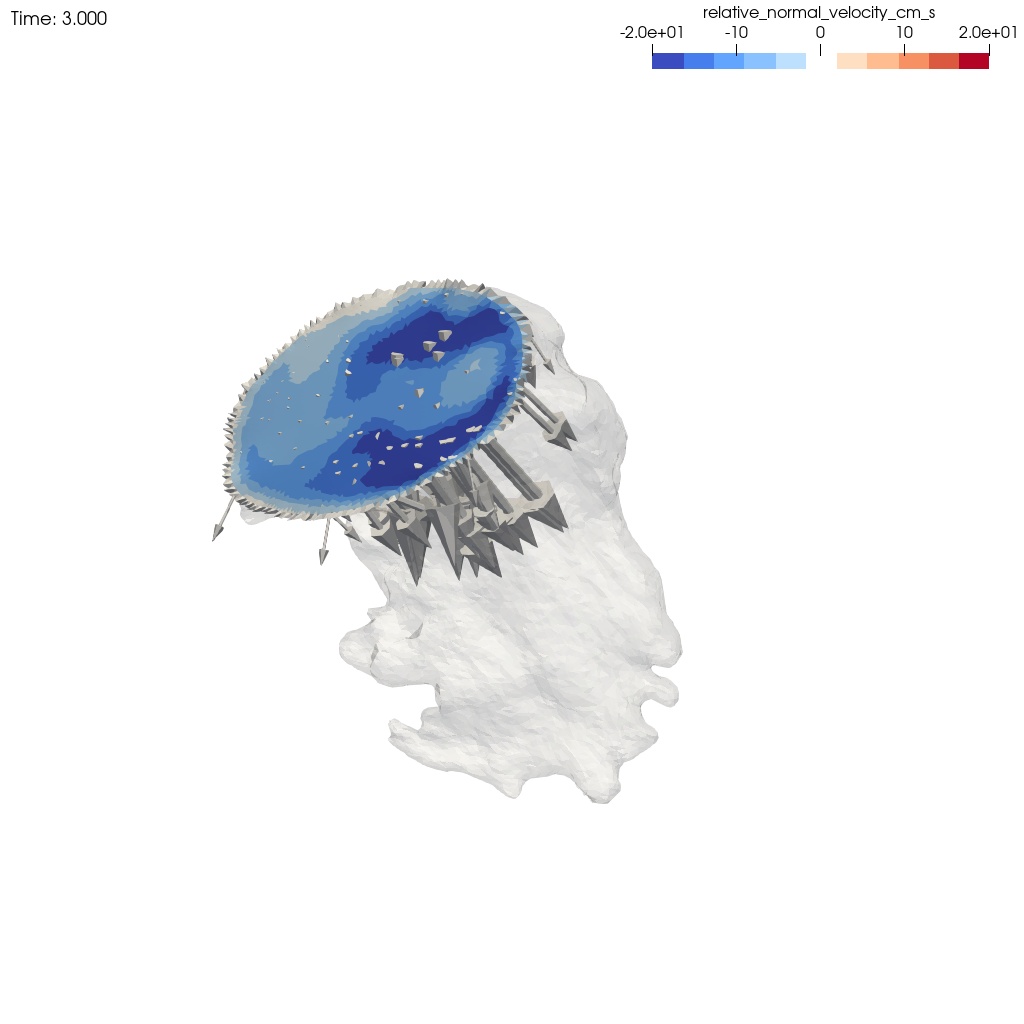}&
    \includegraphics[scale=.13,trim=200 100 300 200, clip]{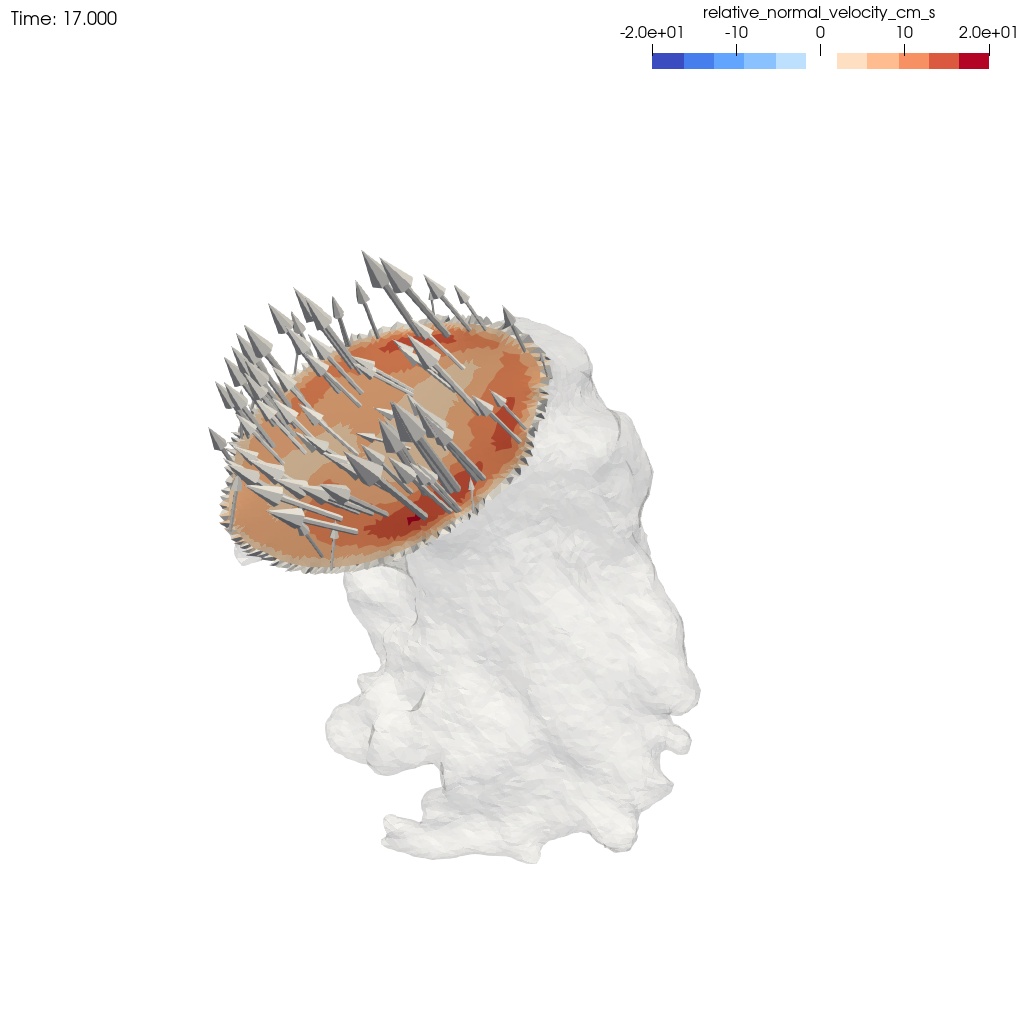} &
    \includegraphics[scale=.13,trim=200 100 280 220, clip]{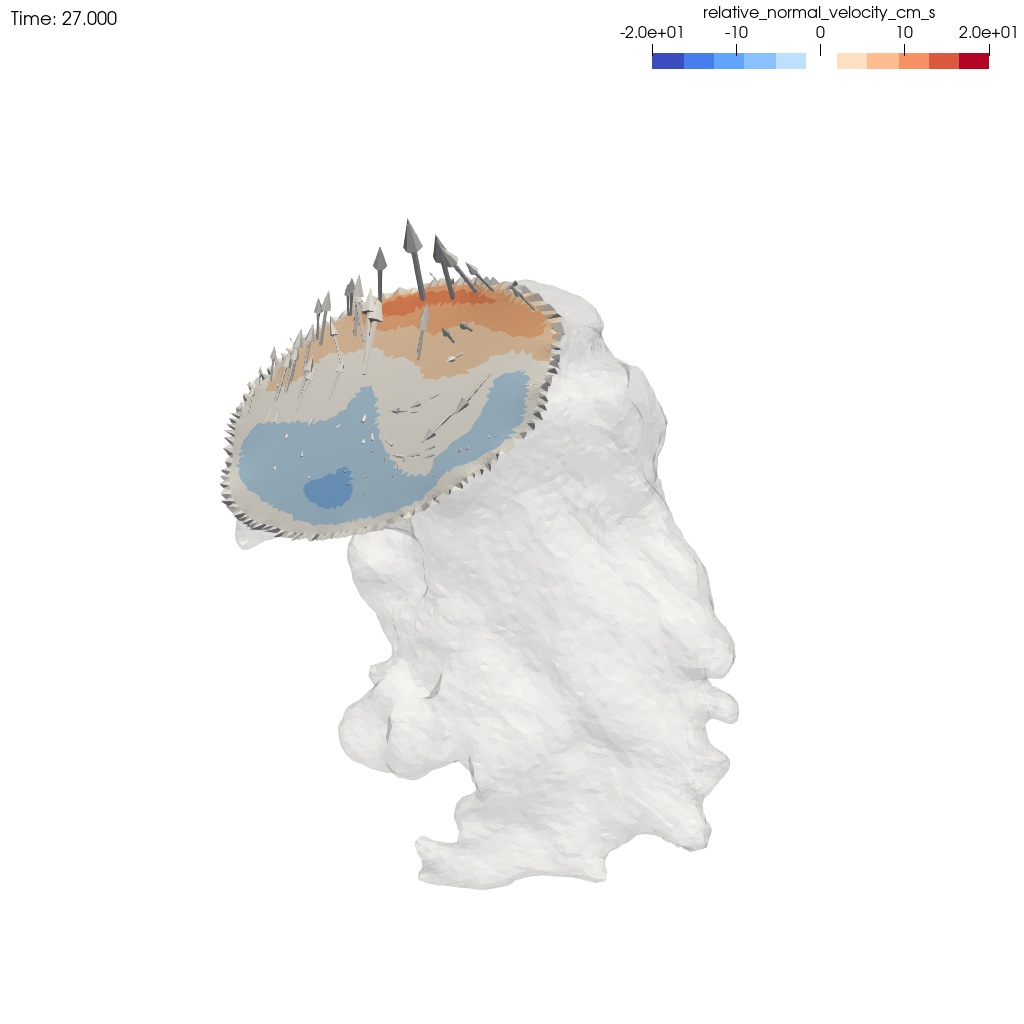} & 
    \includegraphics[scale=.13,trim=200 150 280 120, clip]{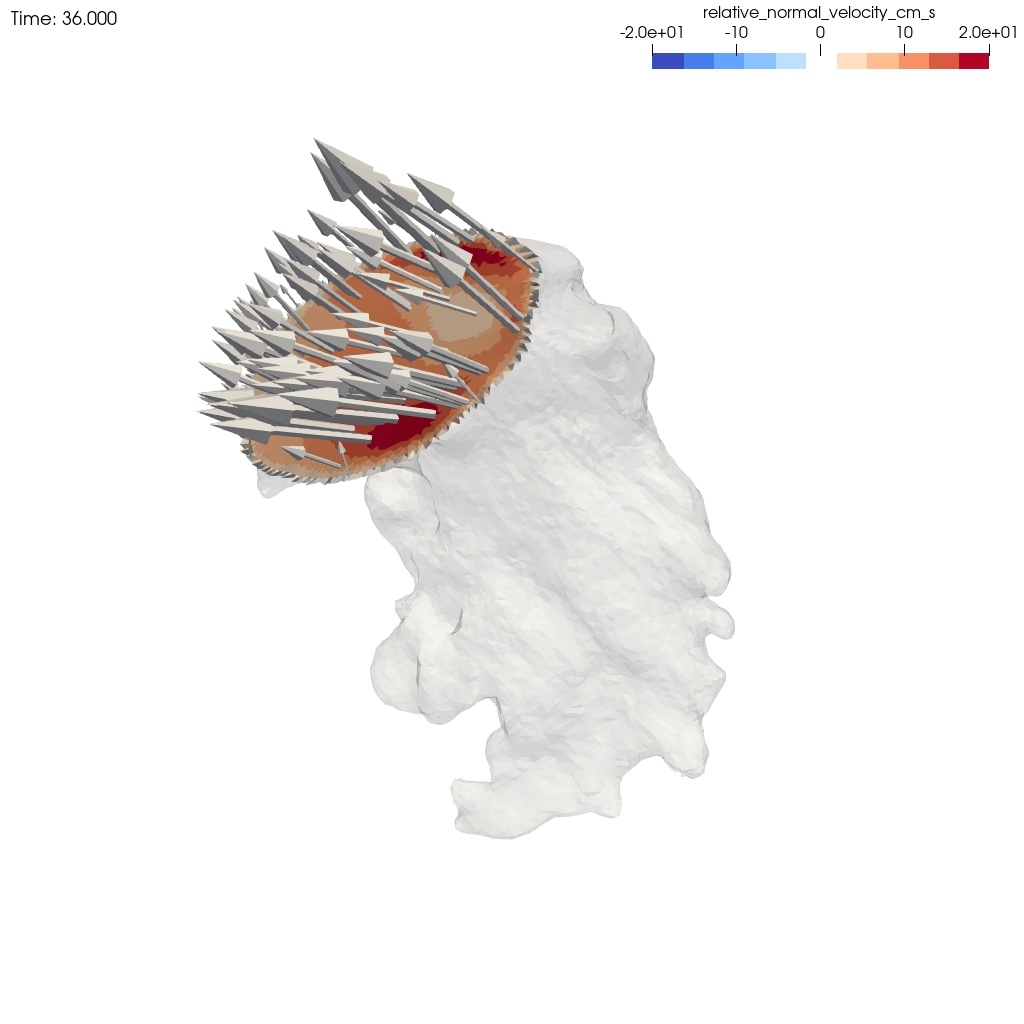} &
    \includegraphics[scale=.4]{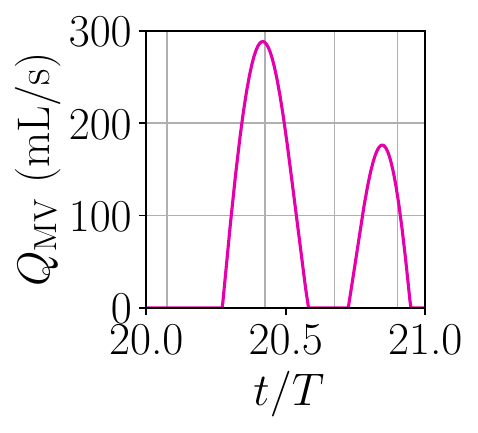}
    \\
    \end{tabular}
    \caption{Phase-dependent flow in subject~1 at the time instants of atrial expansion, 
    peak E-wave, ventricular diastasis, and peak atrial contraction rate. First row: instantaneous 
    streamlines colored by fluid velocity magnitude, with arrows indicating flow direction.
    Second row: the moving ostium surface colored by relative normal velocity \gls{w} 
    (Eq.~\ref{eq:w}). Arrows indicate surface-relative fluid velocity. Panels on the right-hand side
    show\line~LA (left axis) and \linegray~LAA (right axis) volumes and transmitral outflow,
    with vertical lines marking the displayed phases.
    }
    \label{fig:vis}
\end{figure}

The phase dependence of atrial flow and its interaction with the
appendage was examined in subject~1, who did not have arrhythmia 
and was therefore imaged in sinus rhythm.
\autoref{fig:vis} shows instantaneous streamlines in the LA body
and the relative normal velocity across the moving LAA ostium ($w$)
during atrial expansion, peak E-wave, ventricular diastasis,
and at the time instant of maximum atrial contraction rate.
The ostium $w$ maps distinguish fluid entering the LAA ($w<0$, blue)
from fluid leaving it ($w>0$, red), while the panels showing
volume and transmitral-flow histories
provide the temporal context for these spatial patterns.

During atrial expansion, relatively fast streams
entered from the \glspl*{PV} and extended into the atrial cavity toward
the mitral region and into the LAA.
The MV was closed at this instant,
as indicated by the zero transmitral flow rate,
and both the LA and LAA were expanding.
The ostium $w$ was negative over most of the ostium,
consistent with volume-driven appendage filling.
At the E-wave peak, a directed stream
through the \glspl*{PV} toward the open MV dominated the atrial flow,
and the ostium map was predominantly positive.

During ventricular diastasis,
the LA and LAA volumes remained nearly constant.
With the mitral valve closed,
the atrial streamlines formed a more recirculating pattern,
with lower velocities than in the prominent filling and emptying streams.
The ostium exhibited simultaneous inflow and outflow.
Subsequently, active atrial contraction
induced renewed directed flow toward the MV,
accompanied by backflow through the \glspl*{PV},
and a coherent outflow jet from the \gls*{LAA}
due to its contraction.

\newpage
These observations show that volume-driven LAA filling
and emptying coexist with bidirectional exchange over the cycle.
The latter is driven by interactions between the flow
in the atrial body and the appendage.
\textcite{Villalba2021} similarly reported inward and outward
LAA jets with normal wall kinetics, while
subjects with impaired atrial function exhibited cavity-like flow,
and one of their subjects alternated between these patterns over the cycle.
The interaction with atrial-body flow is also in line with the observations of
\textcite{Rodriguez2024}, who examined nine patient-specific AF
geometries and prescribed wall motion through a parametric model.
They found that changing the \gls*{PV} flow split could either improve
or impair LAA renewal, depending on the position of the main
atrial circulatory flow relative to the ostium.
In a subsequent study, they morphed the PV
orientations in two atrial geometries and showed that these
changes shifted the main vortex and altered LAA stasis~\cite{Rodriguez2025}.
The role of atrial body circulatory motions in renewing
the LAA blood pool is highlighted 
by clinical evidence that
lower LA vorticity measured with 4D flow MRI was associated with
a higher risk of embolic brain infarcts~\cite{Spartera2023}.

\newpage
\subsection{Volume-Driven and Bidirectional Ostium Exchange}

Across the 17 subjects, normalized total exchanged volume
through the ostium ranged from 0.94 in subject~14 to 3.67 in subject~12,
while the active, volume-driven contribution accounted for 23.79\% of
total exchange in subject~6 and 78.31\% in subject~1 (Tab.~\ref{tab:results}).
The cohort thus included both predominantly volume-driven exchange
and exchange dominated by passive flow.

Across the cohort, peak bulk absolute relative normal
velocities at the ostium
ranged from 2.8 to \SI{11.7}{cm/s}, giving ostium Reynolds numbers
of 185--957.
Definitions and subject-specific values are provided in
Appendix~\ref{app:ostium_flow_scales}.

\begin{table}[h!tb]
    \centering
    \begin{tabular}{|c|c|c|c|c|c|}
    \hline
    \multirow{2}{*}{\makecell{subject}}
    & \multirow{2}{*}{\makecell{ $\widehat{V}_\mathrm{total}$ \\ {(-)}}}
    & \multirow{2}{*}{\makecell{ $\frac{V_\mathrm{act}}{V_\mathrm{total}}$ \\ {(\%)}}}
    & \multirow{2}{*}{\makecell{ $k$ \\ (\si{1/s})}}
    & \multicolumn{2}{c|}{$\overline T_R$ (s)} \\
    & & & & LA & LAA \\ \hline
    \cellcolor{c1!20}  1 
        & 2.85 & 78.31 & 0.439 & 1.01 & 3.07 \\
    2 
        & 3.31 & 56.79 & 0.541 & 0.73 & 2.01 \\
    3 
        & 3.16 & 47.61 & 0.405 & 1.04 & 2.48 \\
    \rowcolor{gray!20}4 
        & 1.75 & 33.69 & 0.316 & 1.57 & 4.06 \\
    \rowcolor{gray!20}\cellcolor{c5!20}5 
        & 1.61 & 53.42 & 0.432 & 1.57 & 3.56 \\
    \rowcolor{gray!20}6 
        & 2.11 & 23.79 & 0.540 & 1.15 & 2.37 \\
    \rowcolor{gray!20}7 
        & 1.37 & 43.26 & 0.230 & 2.23 & 6.44 \\
    \rowcolor{gray!20}8 
        & 1.96 & 42.28 & 0.285 & 1.56 & 4.81 \\
    \rowcolor{gray!20}9 
        & 1.29 & 25.79 & 0.236 & 1.59 & 5.36 \\
    10 
        & 1.60 & 52.94 & 0.373 & 1.24 & 3.26 \\
    \rowcolor{gray!20}\cellcolor{c11!20}11 
        & 1.29 & 50.34 & 0.200 & 2.03 & 5.82 \\
    \cellcolor{c12!20}12 
        & 3.67 & 65.84 & 0.866 & 0.73 & 1.71 \\
    13 
        & 3.32 & 46.77 & 0.816 & 1.00 & 1.86 \\ 
    \rowcolor{gray!20}14 
        & 0.94 & 43.27 & 0.252 & 1.80 & 6.13 \\
    15 
        & 1.74 & 76.32 & 0.287 & 1.16 & 3.97 \\
    \rowcolor{gray!20}16 
        & 2.04 & 50.71 & 0.369 & 1.24 & 3.76\\
    17 
        & 1.87 & 77.07 & 0.408 & 1.03 & 2.99 \\ \hline
    \end{tabular}
    \caption{Summary of CFD results.
    Normalized total exchanged volume $\widehat{V}_\mathrm{total}$,
    active contribution to total exchange $V_\mathrm{act}/V_\mathrm{total}$,
    effective washout rate $k$,
    and mean blood residence time in the whole LA and LAA.
    Gray rows highlight subjects imaged during AF.
    Colored cells indicate subjects analyzed in detail.
    }
    \label{tab:results}
\end{table}

\begin{figure}[h!tb]
    \centering
    \includegraphics[scale=.5]{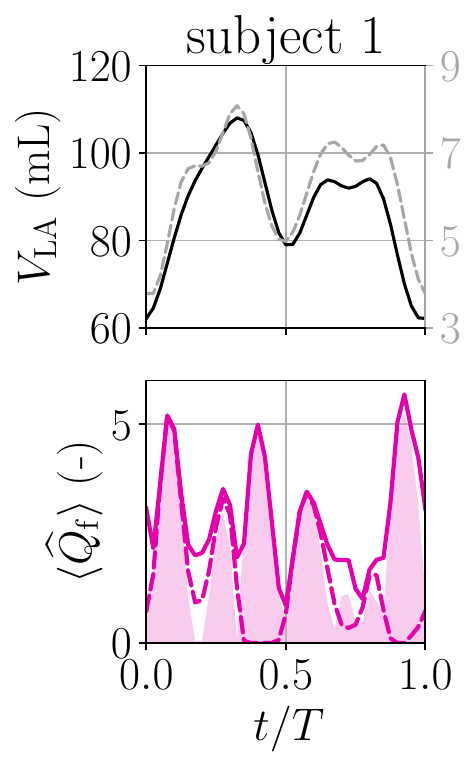}
    \includegraphics[scale=.5]{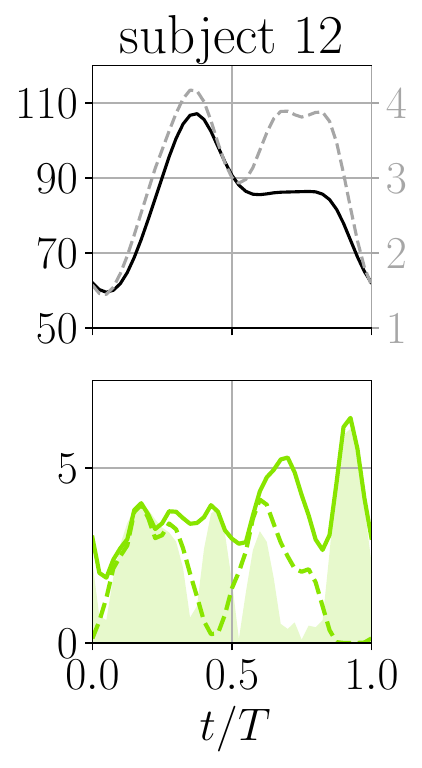}
    \includegraphics[scale=.5]{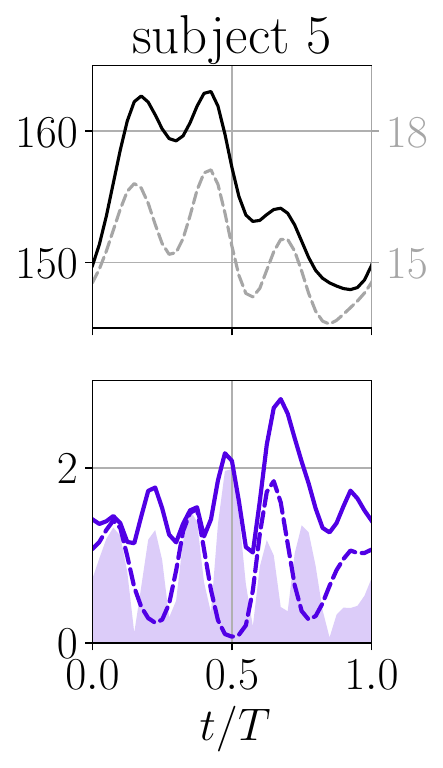}
    \includegraphics[scale=.5]{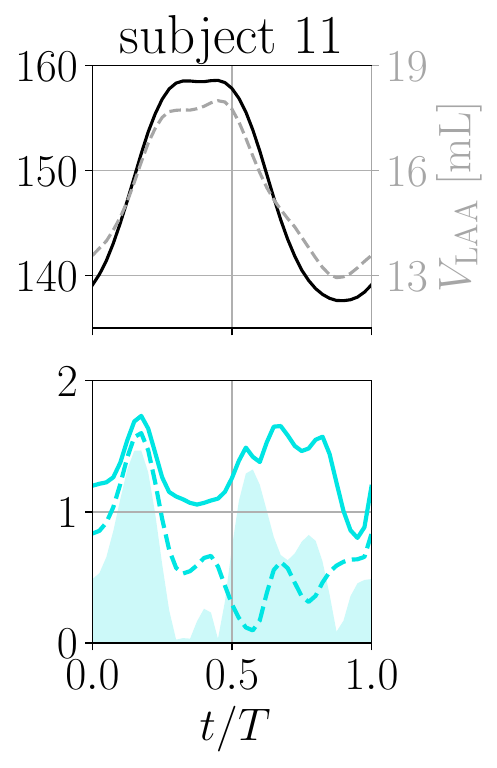}
    \caption{Temporal evolution of (top)~\line~LA and \linegray~LAA volumes and
    (bottom)~normalized phase-averaged \line~total ostial exchange rate~$\langle\gls{Qf;total;norm}\rangle$, \dashed~inflow~$\langle\gls{Qf;in;CFD;norm}\rangle$, and \filled~active flow rate~$\langle\gls{Qf;active;norm}\rangle$ for subjects~1, 12, 5, and~11.
    }
    \label{fig:flowrates}
\end{figure}

To examine how the differences in exchange arose within the cardiac cycle,
subjects~1, 5, 11, and~12 were selected for detailed comparison
(Fig.~\ref{fig:flowrates}).
These cases include the lowest and highest effective washout rates $k$
across the cohort (subjects~11 and~12), while subjects~1 and~5
have similar intermediate values.
In \autoref{fig:flowrates}, the volume histories in the first row
provide the context for the ostium flow rates
in the second row.
The solid curve represents the total ostial exchange rate, the dashed curve inflow,
and the shaded area the active, volume-driven contribution.
Since $\gls{Qf;total}=\gls{Qf;in;CFD}+\gls{Qf;out;CFD}
=\gls{Qf;active}+\gls{Qf;passive}$,
the difference between the solid and dashed curves gives outflow,
while the difference between the solid curves and the shaded areas
gives the passive contribution.

\newpage
Subjects~1 and~12, both with normal LAA EF and in SR, had comparatively 
large normalized total exchanged volumes, 2.85 and 3.67. 
Fluid exchange was predominantly volume-driven, with the active 
contribution accounting for 78.31\% and 65.84\% of total exchange, respectively.
In both subjects, the early total flow peaks were dominated by inflow
and accompanied LAA expansion, whereas the peaks near $t/T=0.4$
and $0.9$ were dominated by outflow during LAA contraction.
The passive contribution was small at these peaks. However, it
became more prominent prior to 
the E-wave and during ventricular diastasis, displaying 
variability between subjects. Its cycle-integrated contribution nevertheless 
remained smaller than volume-driven exchange.

Subjects~5 and~11, both with decreased LAA EF and in AF, 
exhibited a lower normalized total
exchanged volume (at 1.61 and 1.29, respectively) than subjects~1 and~12.
Both appendages still experienced appreciable volume-driven exchange,
which was similar to the passive
contribution (53.42\% and 50.34\% active contribution, respectively;
Tabs.~\ref{tab:cases} and~\ref{tab:results}).
Passive exchange contributed throughout the cycle in both cases,
but their flow histories differed.
Subject~5 showed several alternating LAA filling and emptying events, with the
largest peak in the total ostial exchange rate
near $t/T=0.675$, mostly attributable to passive exchange.
Subject~11 showed a more distinct early filling event, followed by
sustained outflow and appreciable passive exchange
during the decrease in LAA volume in the
second half of the cycle.

Together, these phase-resolved results show that volume-driven
and bidirectional exchange coexist, and that their relative importance
varies both between subjects and within the cardiac cycle.
In patients with impaired EF, the relevance of passive contributions 
to LAA flow exchange is consistent with the role of secondary swirling 
motions within the LAA proposed by \textcite{Duran2023}. Clinical evidence
of the association between LAA morphology and stroke risk in patients with AF 
supports this idea~\cite{DiBiase2012,Khurram2013}.

On the other hand, it is worth noting that volume-driven exchange remained 
substantial in subjects with AF, ranging from 24\% to 53\% in our study 
(see Tab.~\ref{tab:results}), which highlights the importance of resolving wall
motion even when modeling atrial and appendage flow during AF. Along these lines,
\textcite{DuenasPamplona2021}
compared patient-specific moving-wall and fixed-wall simulations in an AF patient with
Doppler-measured transmitral velocities, showing that the flow from 
moving-wall simulations followed the measurements more closely than the rigid model.
Clinical observations also demonstrate residual emptying during AF~\cite{iwama2012left}.
Other groups have compared moving-wall and fixed-wall 
simulations~\cite{Villalba2021,Kjeldsberg2024}, 
consistently finding more sustained flow across the
ostium and larger flow kinetic energy in the LAA of the moving-wall models, 
both in AF and SR.

\subsection{Similar Appendage Washout Rates Can Arise From Different Exchange Regimes}
\begin{figure}[h!tb]    
    \centering
    \setlength{\tabcolsep}{0pt} 
    \begin{tikzpicture}
        \node[inner sep=0pt,rotate=-90, anchor=east](cbar) at (0,0) {\includegraphics[width=2.7cm,height=0.2cm]{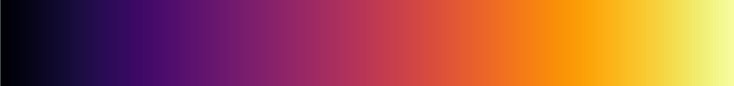}};
        \node(o)[anchor=west] at (-.15,0) {\small $-$};
        \node[anchor=west] at ([yshift=1.2cm]o.west) {\small $-$};
        \node[anchor=west] at ([yshift=2.7cm]o.west) {\small $-$};
        \node(o)[anchor=west] at (.25,0) {\small 0.1};
        \node[anchor=west] at ([yshift=1.2cm]o.west) {\small 0.5};
        \node[anchor=west] at ([yshift=2.7cm]o.west) {\small 1.0};
        \node[text centered] at (.1,-.5)
        {\small $c$ (-)};
        \node[anchor=north west](a) at ([xshift=1cm]cbar.west) {
             \includegraphics[scale=.15,trim = 150 250 150 80,clip]{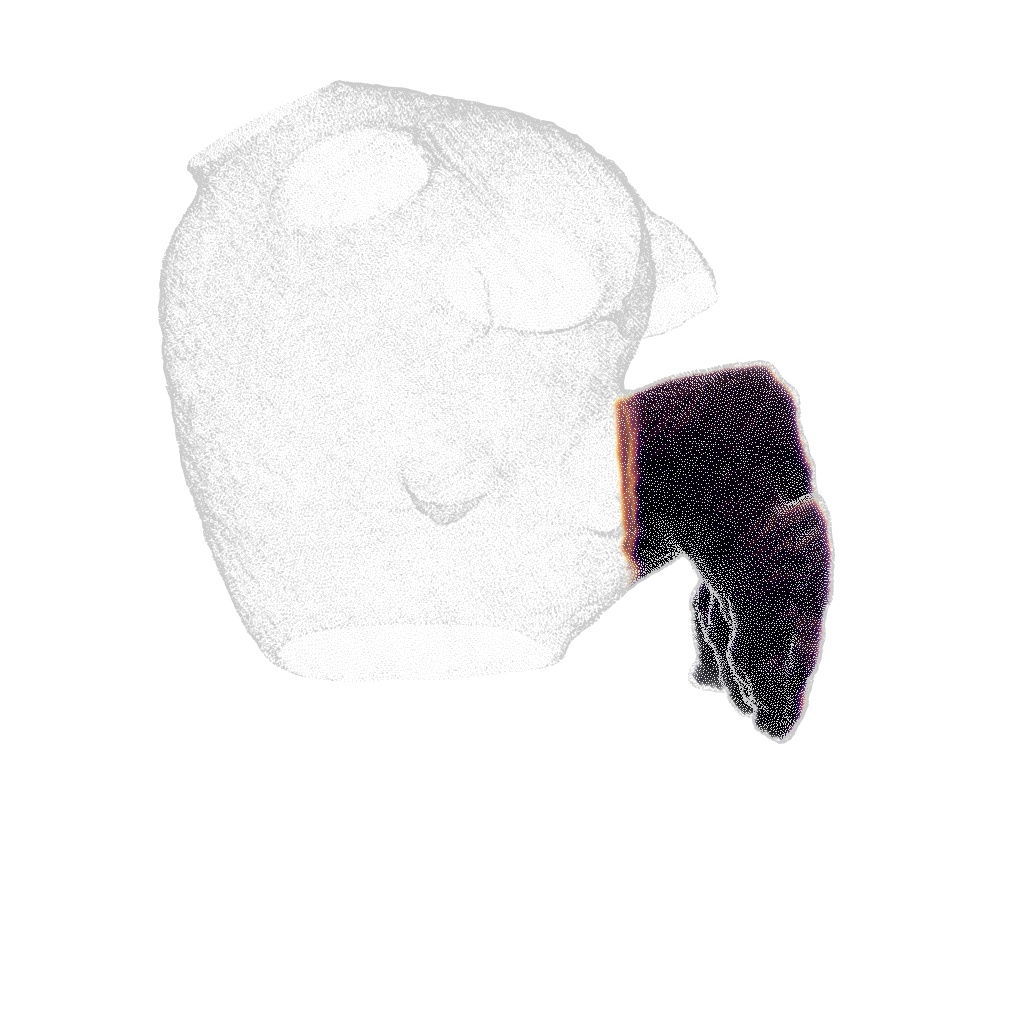}   };
        \node[anchor=north west](b) at (a.north east) {
        \includegraphics[scale=.15,trim = 150 250 150 80,clip]{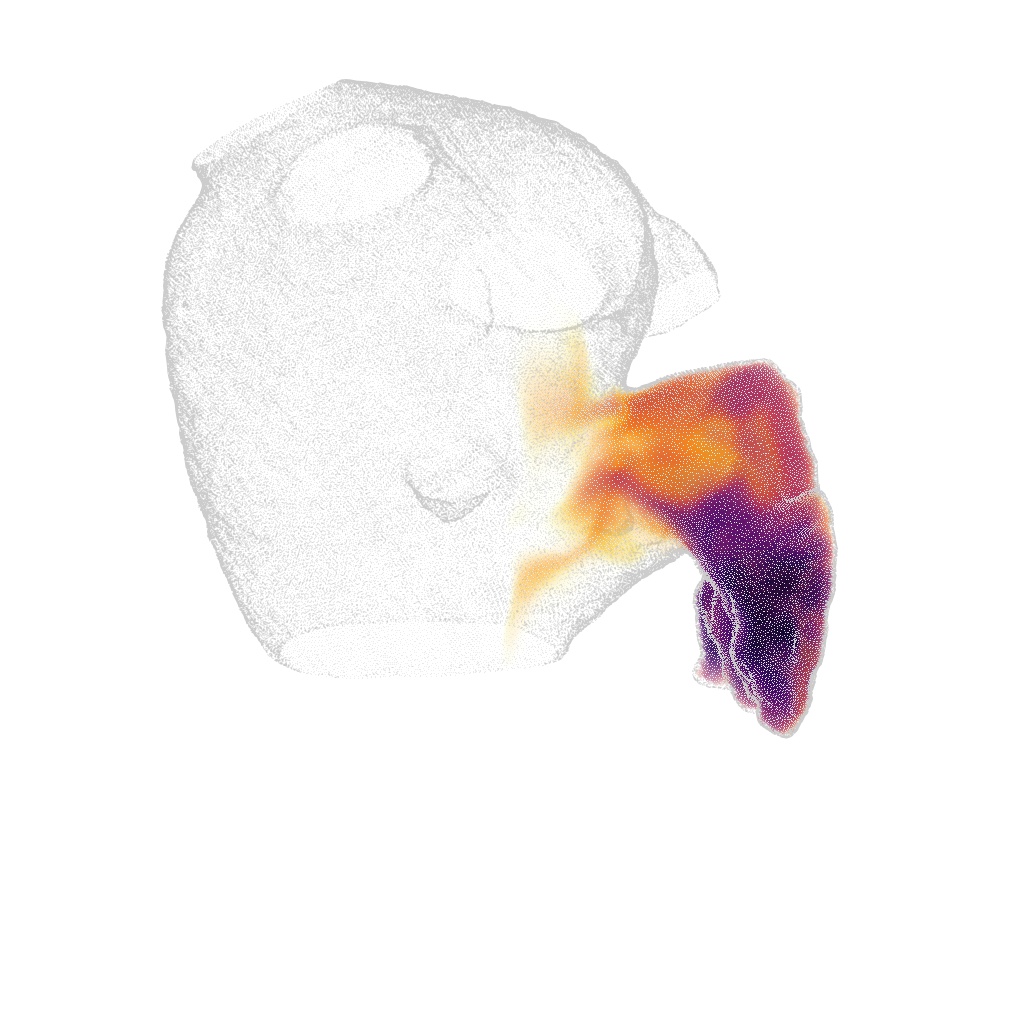}
        };
        \node[anchor=north west](c) at ([yshift=.5cm]b.north east) {
        \includegraphics[scale=.5]{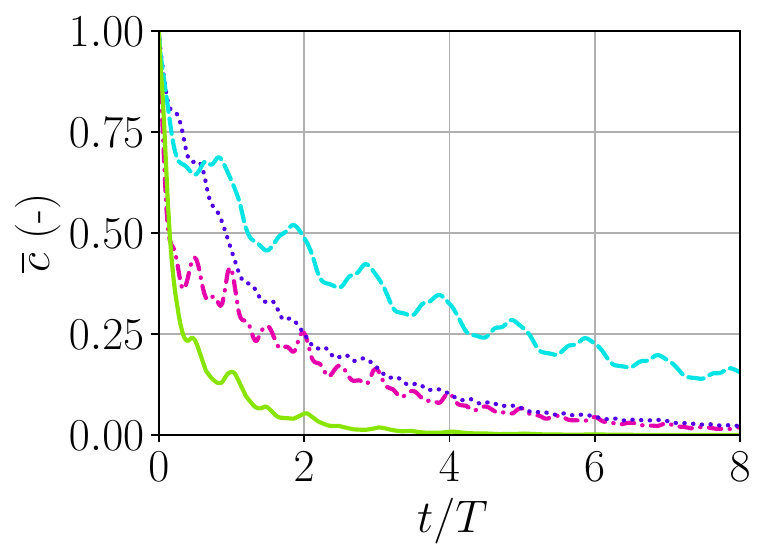}
        };
        \node[anchor=north west,text height=.4cm] at ([yshift=.5cm]a.north west) {(a)};
        \node[anchor=north west,text height=.4cm] at ([yshift=.5cm]b.north west) {(b)};
        \node[anchor=north west,text height=.4cm] at (c.north west) {(c)};
    \end{tikzpicture}
    \caption{Spatial and temporal evolution of LAA scalar washout.
    (a,~b)~Scalar concentration in subject~11 at initialization
    and after two cardiac cycles, respectively.
    (c)~Mean LAA scalar concentration from CFD for
    subjects~\lineOne~1, \lineFive~5, \lineEleven~11, and \lineTwelve~12.
    Time is measured from scalar initialization.
    }
    \label{fig:scalar_evolution}
\end{figure}

To quantify how closely fluid exchange through the ostium 
determines LAA clearance,
a passive scalar was initialized uniformly within the
LAA and its subsequent washout was followed over eight cycles.
The visualizations for subject~11 in \autoref{fig:scalar_evolution}a,b
illustrate scalar dilution within the appendage, 
with labeled fluid still present after two cycles.
The mean LAA concentration decreased over successive cycles in
all four subjects examined in detail, with fluctuations superimposed on
the overall decay (Fig.~\ref{fig:scalar_evolution}c).
These fluctuations can occur each cycle when previously ejected scalar-rich 
fluid returns to the LAA or scalar-poor fluid is exported.
These phenomena are analyzed in Section~\ref{sec:mixing_results}.
Across the cohort, the effective washout rate
$k$ ranged from \SI{0.200}{s^{-1}} in subject~11 to
\SI{0.866}{s^{-1}} in subject~12 (Tab.~\ref{tab:results}).

Subjects~1 and~5 had similar intermediate rates of 0.439 and
\SI{0.432}{s^{-1}}, respectively, despite their different exchange regimes, 
heart rhythms, and considerable differences in anatomy and function
(Tabs.~\ref{tab:cases} and~\ref{tab:results}).
Subject~1 showed a more rapid initial decrease in mean LAA concentration
and more pronounced within-cycle fluctuations than
subject~5 (see Fig.~\ref{fig:scalar_evolution}c),
consistent with its larger LAA EF (53.0 versus 20.5\%).
A complementary effect is observed for subjects~2, 3, and~13,
which have similar normalized total exchanged volumes (between 3.16 and 3.32),
but substantially different washout rates (between 0.405 and \SI{0.816}{s^{-1}}).

These data offer insight about why anatomical or functional
descriptors alone may not fully characterize appendage washout.
\textcite{Otani2016} reported a similar finding in two
patient-specific models with CT-derived wall motion.
They followed a scalar initialized within the LAA and observed
similar residual-fraction histories despite substantial
differences in atrial structure and function.
\textcite{sanatkhani2021} likewise reported substantial overlap
in washout-derived measures among traditional LAA shape groups,
with differences that could not be fully explained by size, location, and
appearance indices.
Complementary in-vivo evidence was recently provided by
\textcite{Severance2026}, who quantified LAA contrast transport
using dynamic contrast \gls*{4D-CT}.
The group of patients imaged during AF
exhibited a wide range of values of LAA residence time 
despite consistently having low LA EFs, and
LA EF explained only $<9\%$ of this variance. In the LV, simplified transit 
models have shown that EF uniquely determines chamber residence time as EF
approaches unity, but becomes an increasingly poor predictor 
as EF decreases~\cite{MartinezLegazpi2025}.  However, analogous models 
relating EF to residence time in the LA are lacking.
\textcite{sanatkhani2021} fitted a triple-exponential model
with a nonzero asymptote to concentration histories spanning
150~seconds.
They associated the three exponential terms
with rapid initial, intermediate, and slow late washout~\cite{sanatkhani2023}.
In the present study, $k$ was calculated from the mean of the
cycle-wise logarithmic concentration decreases,
as defined in Eq.~\ref{eq:cycle_decay}.
It therefore represents an effective rate over the analyzed
interval and does not establish a single underlying transport
timescale or exponential decay throughout that interval.
A further difference concerns the treatment of wall motion.
The simulations of \textcite{sanatkhani2021,sanatkhani2023}
used rigid walls and therefore omit volume-driven exchange
caused by appendage deformation.
Our results extend this analysis to moving atria and relate
washout to both volume-driven and passive exchange.

\subsection{Departures From Well-Mixed Models Distinguish Fluid Exchange From Washout}
\label{sec:mixing_results}

The CFD observations were compared
with the two washout-model predictions (Eq.~\ref{eq:mix})
to examine the relation between fluid
exchange and scalar washout.
The model-predicted concentration histories
for the four detailed cases are
provided in Appendix~\ref{app:model_histories}.

\begin{figure}[h!tb]
    \centering
        \begin{tikzpicture}
        \node[](a) at (0,0) {
    \includegraphics[scale=.5]{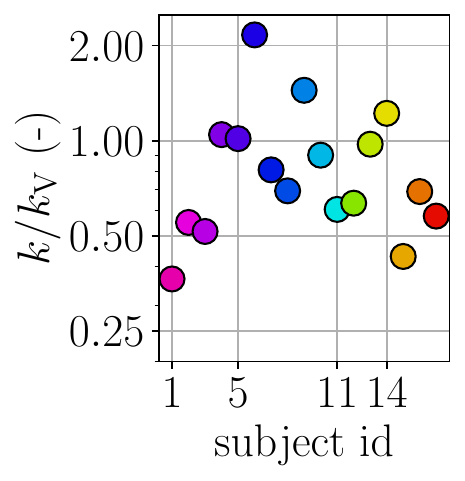}
                    };
        \node[anchor=north west](b) at (a.north east) {
    \includegraphics[scale=.5]{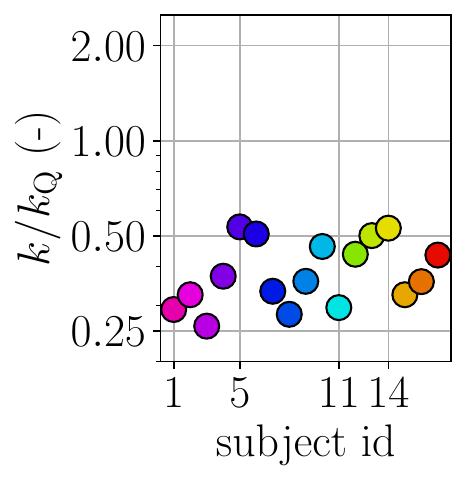}
        };
        \node[anchor=north west](c) at (b.north east) {
    \includegraphics[scale=.5]{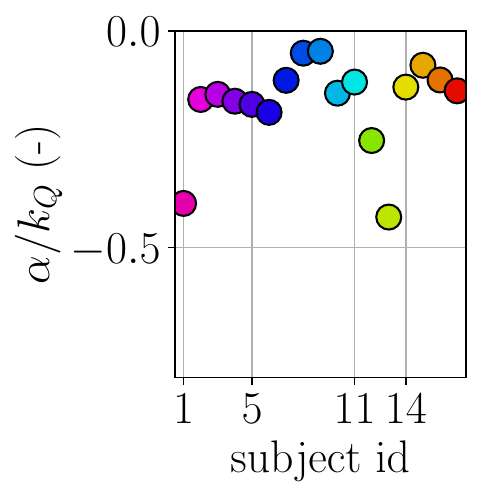}
                    };
        \node[anchor=north west](d) at (c.north east) {
    \includegraphics[scale=.5]{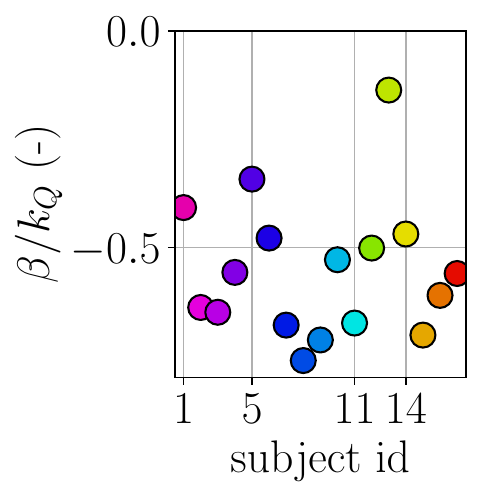}
        };
        \node[anchor=north west,text height=.4cm] at (a.north west) {(a)};
        \node[anchor=north west,text height=.4cm] at (b.north west) {(b)};
        \node[anchor=north west,text height=.4cm] at (c.north west) {(c)};
        \node[anchor=north west,text height=.4cm] at (d.north west) {(d)};
    \end{tikzpicture}
    \caption{Comparison of CFD-derived LAA washout with the two
    washout models and decomposition of the inflow-driven
    model discrepancy.
    Relative washout efficiencies
    (a)~$k/k_V$ and (b)~$k/k_Q$.
    Normalized contributions from
    (c)~scalar re-ingestion, $\gls{alpha}/k_Q$, and
    (d)~outflow concentration correction, $\gls{beta}/k_Q$
    (see Eq.~\ref{eq:washout_decomposition}).
    }
    \label{fig:scalar_models}
\end{figure}

In \autoref{fig:scalar_models}, we present the
relative washout efficiency $k/k_V$ for all cases.
The ratio $k/k_V$ fell on either side of unity across the cohort (Fig.~\ref{fig:scalar_models}a), indicating that the volume-driven model
underpredicted the effective washout rate
in some subjects and overpredicted it in others.
Including the passive contribution through the full CFD-derived
inflow accelerated the model-predicted washout.
Across the cohort, $k/k_Q$ ranged from approximately
0.26 to 0.53 (Fig.~\ref{fig:scalar_models}b), so the CFD-derived
rates were consistently below the inflow-driven well-mixed washout prediction.
The exchanged fluid therefore removed less scalar than predicted
under uniform mixing and scalar-free inflow.

The contributions to this discrepancy can be quantified by
the scalar-balance decomposition introduced in Eq.~\ref{eq:washout_decomposition}.
While the non-positive term \gls{alpha} measures the contribution from
re-ingestion of previously ejected scalar,
\gls{beta} indicates preferential export of
scalar-poor or scalar-rich fluid, depending on its sign.
Both contributions were negative in all subjects:
$\gls{alpha}/k_Q$ ranged approximately from $-0.05$ to $-0.43$, and
$\gls{beta}/k_Q$ from $-0.13$ to $-0.76$
(Fig.~\ref{fig:scalar_models}c,d).
The particularly pronounced scalar re-ingestion in subject~1
may partly explain the similar effective washout rates
of subjects~1 and~5 despite the smaller LAA volume
and higher LAA EF in subject~1.
The consistently negative \gls{beta} indicates
that the scalar concentration
in ejected fluid was lower than the mean LAA concentration
in every patient-specific simulation.
Subject~11 illustrates how this bias can arise.
After two washout cycles, scalar concentration was low near the
ostium and remained high at the appendage tip (Fig.~\ref{fig:scalar_evolution}b).
Because ostium outflow draws from the low-concentration
region near the ostium, the ejected fluid tends to
have a scalar concentration below the LAA mean.
The magnitude of \gls{beta} was larger than that of \gls{alpha} in most subjects,
identifying the selective export of scalar-poor fluid as a substantial
departure from perfect mixing. 

This pattern is consistent with a last-in, first-out (LIFO)
interpretation, in which recently entered fluid near the ostium is
more readily expelled while fluid deeper in the appendage is
retained.
In a queue model of LV transport,
\textcite{MartinezLegazpi2025} showed that, at a fixed EF,
LIFO transport represents the least favorable 
queuing pattern for avoiding regions of elevated
residence time.
More sophisticated models distinguish transport components within the whole LA~\cite{Back2024}, 
but lump ejection and re-ingestion across the LAA ostium into a single residual component. 
Like residence time, the size of this residual component can serve as an indicator of LAA stasis, 
but it does not reveal the underlying transport mechanisms. The present analysis provides a mechanistic basis for 
understanding how proposed~\cite{Martinez2014} or future therapies 
that alter LAA exchange patterns may improve fluid renewal and alleviate stasis.

\subsection{Residence-Time Distributions Reveal Long-Term Transport Heterogeneity Beyond Bulk Washout}

Instantaneous snapshots of the residence-time fields
for the four subjects examined in detail
are shown in \autoref{fig:rt_results}a--d.
Residence time and scalar washout describe transport in the same velocity field,
but differ in their initial and boundary conditions.
Washout follows fluid labeled within the LAA at $t=20T$,
whereas $T_R$ accumulates residence time since entry at the \glspl*{PV}.
LAA residence time therefore includes time spent in the atrial body
before entering the appendage and continues to grow during
repeated exchange between the two regions.

{\renewcommand{\arraystretch}{0} \setlength{\tabcolsep}{7pt} 
\begin{figure}[h!tb]
    \centering
\begin{tabular}{lllll}
    (a) & (b) & (c) & (d) & (e)
    \\
    \includegraphics[scale=0.1, trim = 150 200 270 150, clip]{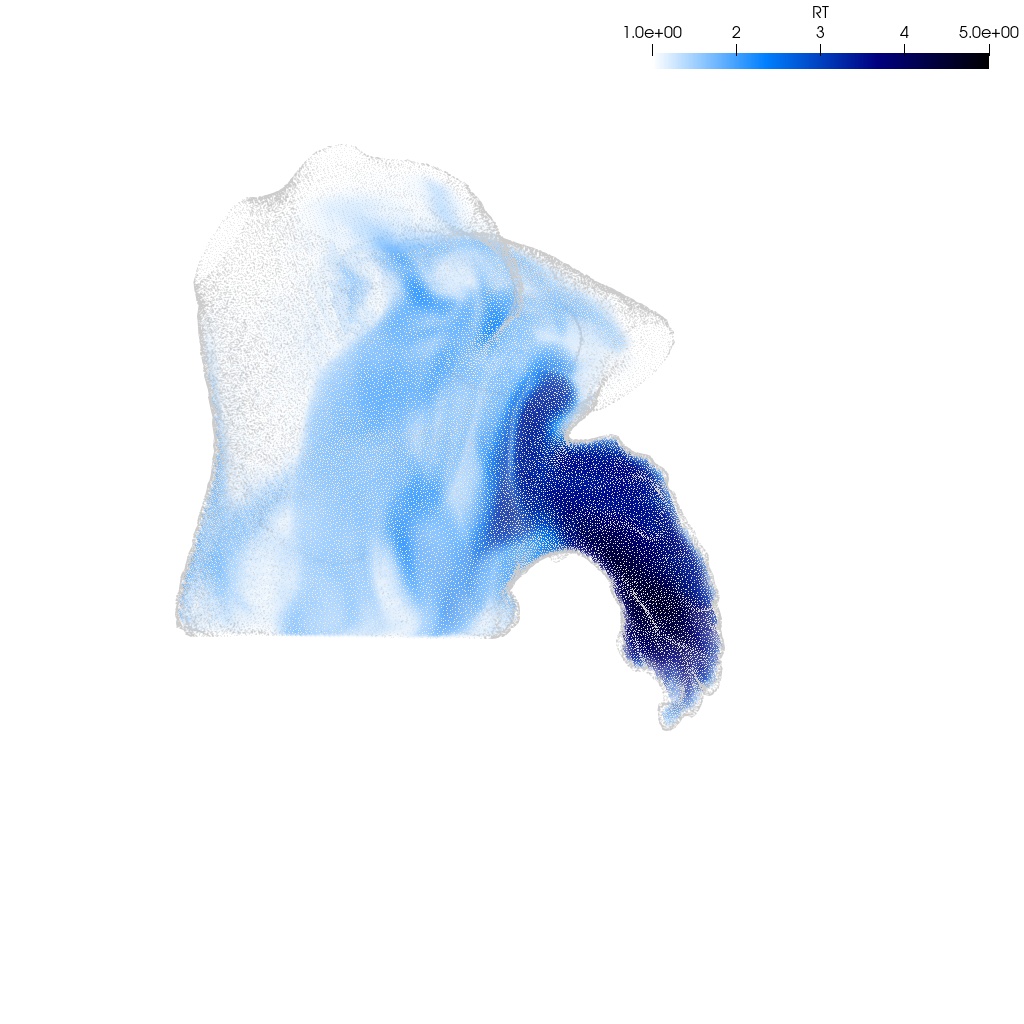} &
    \includegraphics[scale=0.1, trim = 90 100 200 200, clip]{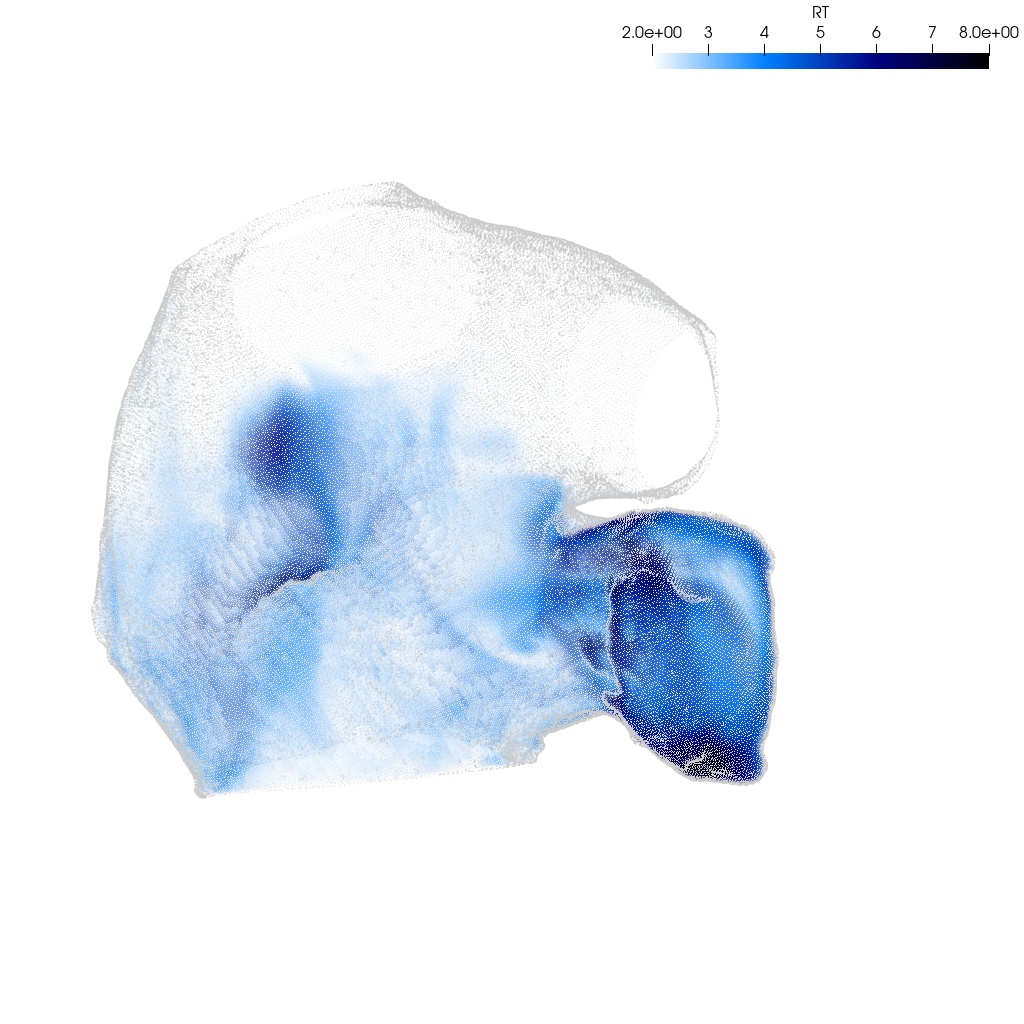} &
    \includegraphics[scale=0.1, trim = 150 200 180 80, clip]{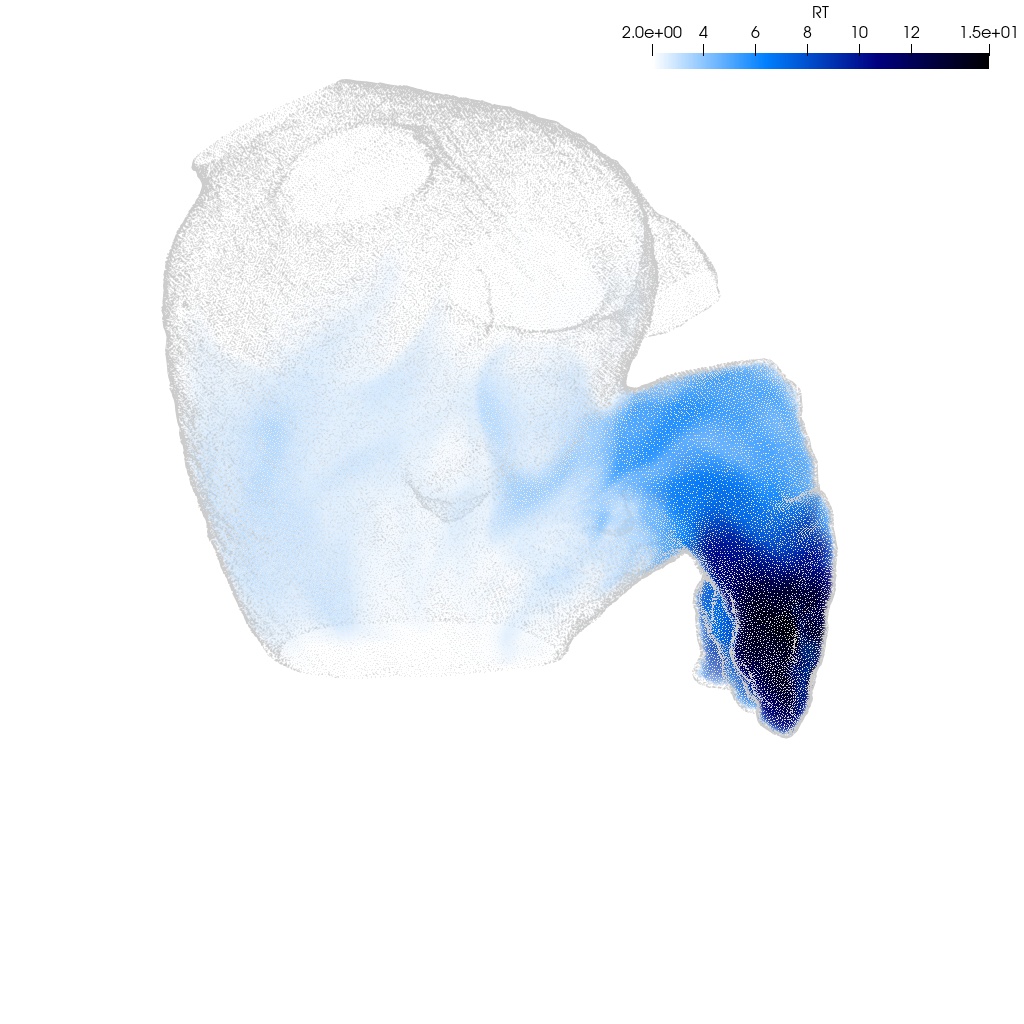}&
    \includegraphics[scale=0.1, trim = 120 200 125 0, clip]{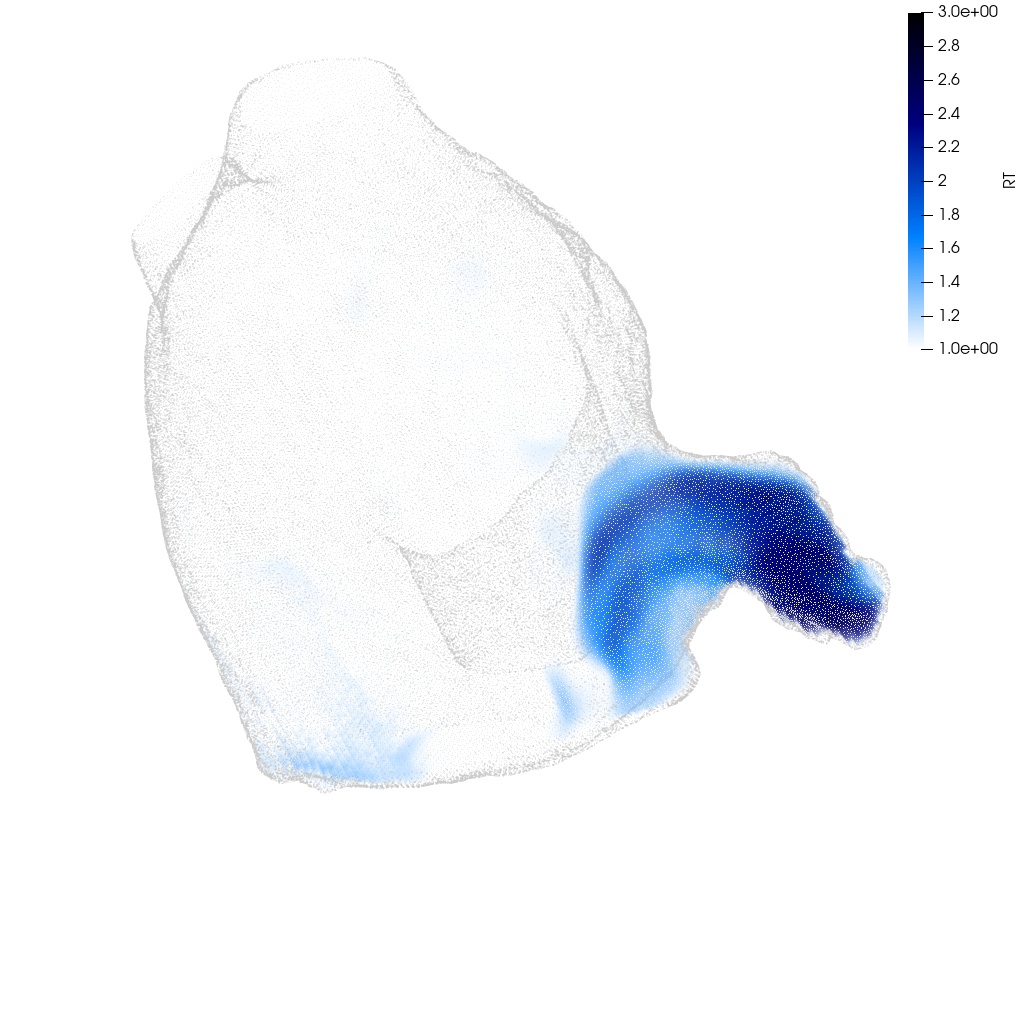}
    \\
     \begin{tikzpicture}
        \node[inner sep=0pt, anchor=west](cbar) at (0,0) {\includegraphics[width=1.75cm,height=0.2cm]{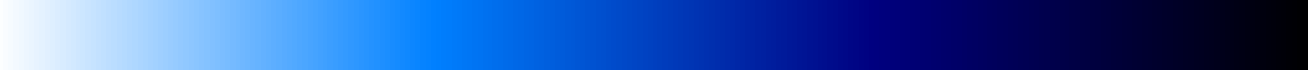}};
        \node(o)[anchor=north] at (0,.2) {\small $|$};
        \node at ([xshift=1.75cm]o) {\small $|$};
        \node(o)[anchor=north] at (0,-.25) {\small 1};
        \node[anchor=north] at ([xshift=1.75cm]o.north) {\small 5};
        \node[text centered,anchor=north] at (1.75/2,-.5)
        {\small $T_R$ (s)};
    \end{tikzpicture}&
    \begin{tikzpicture}
        \node[inner sep=0pt, anchor=west](cbar) at (0,0) {\includegraphics[width=1.75cm,height=0.2cm]{figures/vis/cbar_white_blue_black.png}};
        \node(o)[anchor=north] at (0,.2) {\small $|$};
        \node at ([xshift=1.75cm]o) {\small $|$};
        \node(o)[anchor=north] at (0,-.25) {\small 2};
        \node[anchor=north] at ([xshift=1.75cm]o.north) {\small 8};
        \node[text centered,anchor=north] at (1.75/2,-.5)
        {\small $T_R$ (s)};
    \end{tikzpicture}&
    \begin{tikzpicture}
        \node[inner sep=0pt, anchor=west](cbar) at (0,0) {\includegraphics[width=1.75cm,height=0.2cm]{figures/vis/cbar_white_blue_black.png}};
        \node(o)[anchor=north] at (0,.2) {\small $|$};
        \node at ([xshift=1.75cm]o) {\small $|$};
        \node(o)[anchor=north] at (0,-.25) {\small 2};
        \node[anchor=north] at ([xshift=1.75cm]o.north) {\small 15};
        \node[text centered,anchor=north] at (1.75/2,-.5)
        {\small $T_R$ (s)};
    \end{tikzpicture}&
    \begin{tikzpicture}
        \node[inner sep=0pt, anchor=west](cbar) at (0,0) {\includegraphics[width=1.75cm,height=0.2cm]{figures/vis/cbar_white_blue_black.png}};
        \node(o)[anchor=north] at (0,.2) {\small $|$};
        \node at ([xshift=1.75cm]o) {\small $|$};
        \node(o)[anchor=north] at (0,-.25) {\small 1};
        \node[anchor=north] at ([xshift=1.75cm]o.north) {\small 3};
        \node[text centered,anchor=north] at (1.75/2,-.5)
        {\small $T_R$ (s)};
    \end{tikzpicture}
    &
    \multirow[t]{2}{*}{
    \includegraphics[scale=.5]{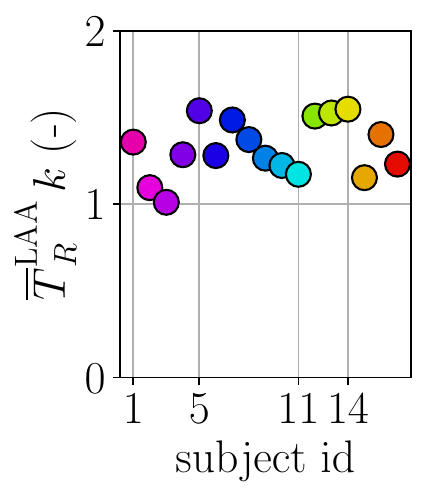}}
    \end{tabular}
    \caption{(a--d):~Spatial distribution of blood residence time $T_R$ in subjects~1, 5, 11,
    and~12.
    Subject-specific color limits were used to resolve 
    spatial variations in $T_R$ across the different residence-time ranges.
    (e)~Mean LAA residence time normalized by the washout timescale,
    $\overline{T}_R^{\mathrm{LAA}}k$.
    Each marker represents one subject.
    }
    \label{fig:rt_results}
\end{figure}}

Mean blood residence time was greater in the LAA than in the whole-LA
region for every configuration (Tab.~\ref{tab:results}).
Whole-LA means ranged from 0.73 to \SI{2.23}{s},
whereas LAA means ranged from 1.71 to \SI{6.44}{s}.
Whole-LA turnover scaling is discussed in Appendix~\ref{app:LA_turnover}.
Longer residence times have also been reported by
\textcite{Corti2022,DuenasPamplona2021,DuenasPamplona2022,Stoecker2025}.
Our previous CT-based analyses, which included some of the
geometries examined here, showed similar spatial
patterns~\cite{Villalba2021,Gonzalo2022,Duran2023}.
While our results show qualitative agreement, quantitative comparisons
are limited by differences in inlet conditions, rheological models,
and observation windows.
This localization is consistent with the LAA's position
outside the main flow path from the PVs to the MV.
While the atrial body conveys blood toward the MV,
renewal within the LAA requires exchange through its single
ostium.

Comparing mean LAA residence time with the washout rates
(Fig.~\ref{fig:rt_results}e),
the products $\overline{T}_R^\mathrm{LAA}k$ ranged
from approximately 1.00 in subject~3 to 1.54 in subject~14.
Mean LAA residence time generally
exceeded $1/k$.

Several differences from the reference problem
introduced in \autoref{eq:age_washout_reference} help interpret
these results.
In the present simulations,
residence time accumulated since entry through the PVs, so
$\overline{T}_R^\mathrm{LAA}$ includes time spent upstream in the LA body,
whereas the scalar was initialized at a single instant and only inside the LAA.
Its subsequent washout depended on the initialization phase
and the flow states encountered over the eight-cycle observation window.
Scalar re-ingestion additionally violates the scalar-free inflow condition.
Moreover, replacing  the washout integral with $1/k$ requires
exponential decay over the full washout history, which is not
established by the finite-window rate used here.
Once initialization transients have decayed,
averaging the residence-time field over several cycles can provide a description
of transport that is less dependent on a single labeling event.

\begin{figure}[h!tb]
    \centering
    \begin{tikzpicture}
        \node[](a) at (0,0) {
            \includegraphics[scale=.5]{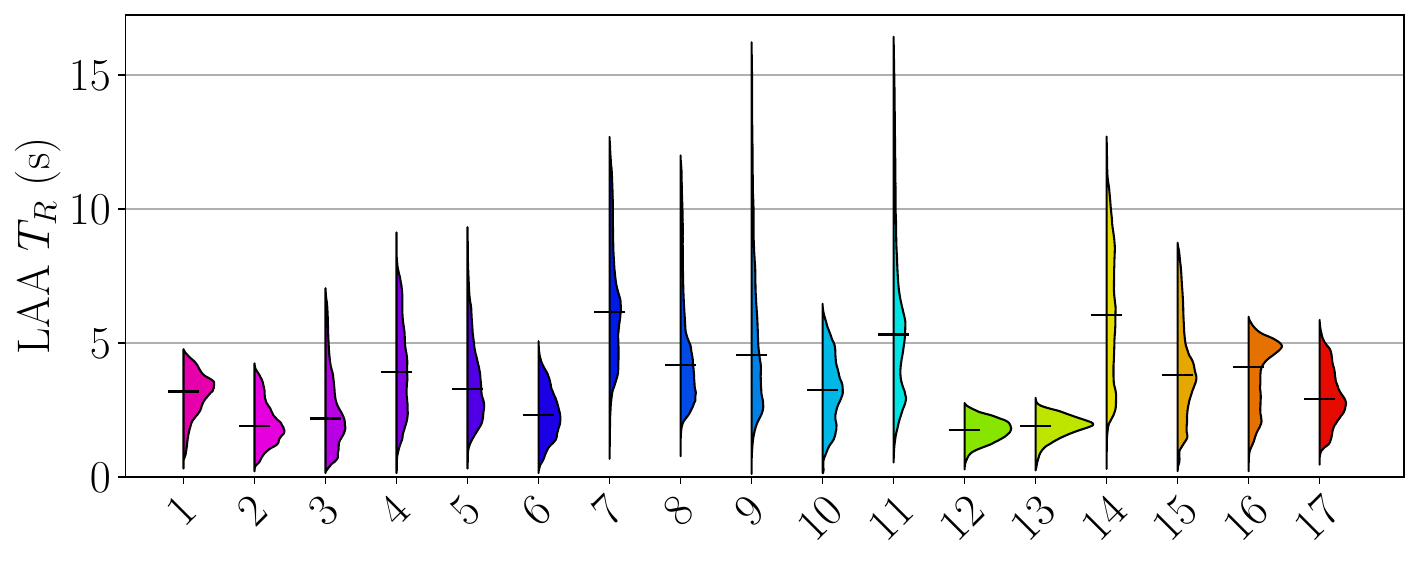}
        };
        \node[anchor=north west](b) at (a.north east) {
        \includegraphics[scale=.5]{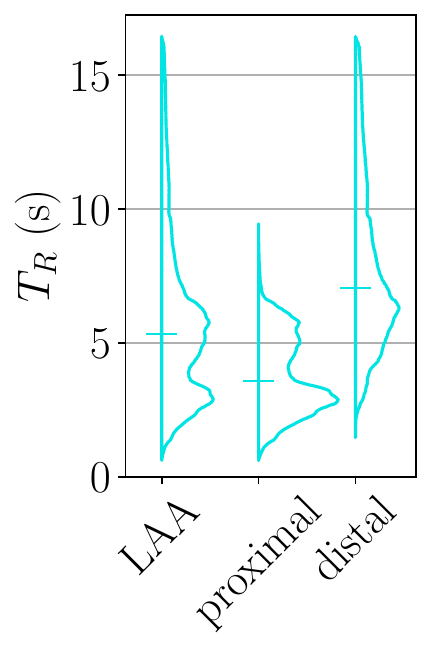}
        };
        \node[anchor=north west,text height=.4cm] at (a.north west) {(a)};
        \node(c)[anchor=north west,text height=.4cm] at (b.north west) {(b)};
    \end{tikzpicture}
    \caption{(a)~Pooled LAA residence-time distributions by subject.
    (b)~Corresponding distributions in the whole LAA and the
    proximal and distal regions in subject~11.
    Horizontal markers indicate medians.
    }
    \label{fig:violin}
\end{figure}

Beyond these mean values, the pooled probability
distributions revealed differences
in the range and prevalence of residence times in the LAA (Fig.~\ref{fig:violin}a).
Subjects~12 and~13 had comparatively compact LAA distributions,
whereas subjects~7, 8, 9, 11, and~14 had broader distributions
extending to residence times of \SI{10}{s} or more.
Several configurations, including subject~11, also exhibited
multiple apparent peaks,
which indicates that younger and older fluid populations
coexist within the appendage.

To relate the whole-LAA residence-time distribution to its spatial origin,
the appendage of subject~11 was divided into proximal and distal
regions of similar volume, as depicted in \autoref{fig:LAAcut}.
Blood residence time was generally lower in the proximal region near the ostium,
whereas the older peak and the extended high-$T_R$ tail of the
whole-LAA distribution originated predominantly in the distal region
(Fig.~\ref{fig:violin}b).
Although the regional distributions overlapped,
their different shapes show that the apparent
modes of the whole-LAA distribution
reflect spatial differences in renewal.

The regional residence-time distributions are consistent with the scalar
concentration field after two washout cycles
(Fig.~\ref{fig:scalar_evolution}b), where
concentration was lower near the ostium, while more scalar
remained in the distal appendage.
The residence-time field likewise showed elevated distal residence times
(Fig.~\ref{fig:rt_results}c).
Together, these observations indicate that fluid near the opening
was renewed more rapidly than fluid in the distal appendage.

Similar distal residence-time elevations were observed in our previous
studies~\cite{Villalba2021,Gonzalo2022,Duran2023,Stoecker2025}.
In particular, \textcite{Stoecker2025} associated the lower residence-time
mode of bimodal LAA distributions with fluid near the ostium
and the higher residence-time mode with fluid in the apex.
\textcite{DuenasPamplona2021} likewise found the longest residence
times at the LAA tip in both rigid and moving-wall models.
Complementary evidence comes from rigid-wall scalar-washout
studies \cite{Bosi2018,sanatkhani2021} that
found persistent tip concentration despite washout near the
ostium.
Spatially heterogeneous LAA transport has also been observed
in vivo.
\textcite{Severance2026} used \gls*{4D-CT} to follow the
transport of an injected contrast agent within the LAA.
They demonstrated delayed distal contrast
enhancement and prolonged contrast-derived residence times
relative to the LAA ostium in a patient with AF.

Across our cohort, lower peak bulk ostium velocities generally accompanied
higher mean LAA residence times (Tabs.~\ref{tab:results} and~\ref{tab:ostium_flow_scales}).
Subjects~7 and~14
had the lowest peak bulk velocities (2.8 and \SI{4.2}{cm/s}) and the longest mean LAA
residence times (6.44 and \SI{6.13}{s}), respectively.
This pattern is in line with 
the clinical interpretation of low Doppler LAA emptying velocity as
an indirect marker of stasis in AF~\cite{goldman1999,hautmann2023}.
However, a single peak velocity may
obscure substantial differences in mean LAA residence time.
For example, subjects~5 and~8 had similar simulated peaks (5.9 and
\SI{5.8}{cm/s}), yet their mean LAA residence times differed markedly
(3.56 and \SI{4.81}{s}, respectively). 

\subsection{Limitations}
\label{sec:limitations}

The cohort comprised only 17 subjects, but covered
a broad range of anatomies, functions, and rhythms during imaging.
The comparisons therefore describe the combined effects of these
characteristics and do not isolate the influence of any single factor.

Wall motion was prescribed from a periodic reconstruction of the
imaging data. The temporal resolution of these images
and the applied Fourier interpolation can
affect the reconstructed deformation and associated flow,
as shown in our previous wall-motion study~\cite{Stoecker2025}.
Equal plug velocities were prescribed at the PVs, and the MV
was represented by an open or closed boundary without resolving
the valve leaflets.
The findings consequently depend on these inflow, outflow,
and motion assumptions, which can affect atrial circulation and
appendage renewal~\cite{Duran2023,Rodriguez2024}.

All simulations were performed using a cardiac period of \SI{1}{s},
corresponding to a heart rate of \SI{60}{bpm}, and blood was
modeled as a Newtonian fluid
with a constant kinematic viscosity
of $\nu=\SI{0.04}{cm^2/s}$.
While this approach facilitated comparisons across the cohort,
it neglects the patient-specificity of these values.
Of note, heart rate can also vary within an individual with physical
activity, stress, or rhythm disorders.
The periodic motion prescription additionally excluded irregular
beat-to-beat changes during AF.
The effects of these physiological variations were outside the scope
of this study.
Some earlier analyses of overlapping geometries used
non-Newtonian constitutive laws, which altered secondary flow
and further increased distal residence times~\cite{Gonzalo2022}.

We did not model coagulation or thrombus formation and evolution, 
which will be addressed in future work.

Scalar washout was observed for eight cycles following a single
initialization phase, and the reported rates summarize the final
seven cycles of that interval.
The effective washout rate may therefore depend
on the initialization phase and the flow history during the observation window.
Repeating the analysis at different initialization phases
and following washout for longer would help assess
this sensitivity and better characterize slow late washout.

Dynamic contrast \gls*{4D-CT} provides a route for future validation
of the simulated transport.
Following the approach proposed by \textcite{Severance2026},
contrast transport could be simulated under conditions matched
to the imaging acquisition.
\section{Conclusion}
\label{sec:conclusions}

Patient-specific moving-wall simulations driven by \gls*{4D-CT} imaging 
showed that fluid exchange across the LAA ostium
combines volume-driven filling and emptying with
simultaneous bidirectional transport.
The cohort included patients imaged during SR and AF and
covered a range of exchange regimes. Across our cohort,
volume-driven transport
accounted for approximately 24--78\% of total exchange.
Among the nine subjects imaged during AF,
volume-driven transport accounted for
approximately 24--53\% of total ostial exchange.
This finding reinforces the importance of
accounting for wall motion even when modeling LAA flow during AF, as
emphasized in previous comparisons of rigid and moving-wall
models~\cite{DuenasPamplona2021,Villalba2021,Kjeldsberg2024}.
However, fluid exchange through the LAA ostium did not by itself determine
how effectively the appendage was renewed.
Across the cohort, effective washout rates were only approximately
26--53\% of those predicted by a well-mixed washout model using
the actual ostium inflow and assuming scalar-free incoming fluid.

Decomposition of the scalar balance identified two contributions
to this discrepancy: re-ingestion of previously expelled scalar and
preferential export of scalar-poor fluid.
Preferential export of scalar-poor fluid
reduced washout more than scalar re-ingestion in most subjects.
Thus, the volume of fluid crossing the ostium was insufficient
to characterize washout; the scalar content of the incoming
and outgoing fluid also mattered.
Consequently, subjects with different anatomy, rhythm, and exchange regimes
could have similar effective washout rates.

Residence-time distributions complemented the ostium transport analysis
by revealing the coexistence of younger and older fluid populations.
In the regional analysis of subject~11, the older population
originated predominantly in the distal appendage, consistent
with the scalar retained there after two washout cycles.
Together, these findings show how appreciable exchange across
the opening of a deforming cavity can coexist with prolonged
retention within it.
Characterizing LAA renewal therefore requires both the scalar
transport across the ostium and the spatial distribution of
fluid retained in the appendage.

\newpage
\appendix
\renewcommand{\thefigure}{A~\Roman{figure}}
\renewcommand{\theHfigure}{A~\Roman{figure}}
\setcounter{figure}{0}
\renewcommand{\thetable}{A~\Roman{table}}
\renewcommand{\theHtable}{A~\Roman{table}}
\setcounter{table}{0}

\section{Grid Refinement Study}
\label{app:grid_refinement}

\begin{figure}[h!tb]
    \centering
        \begin{tikzpicture}
        \node[](a) at (0,0) {
    \includegraphics[scale=0.45]{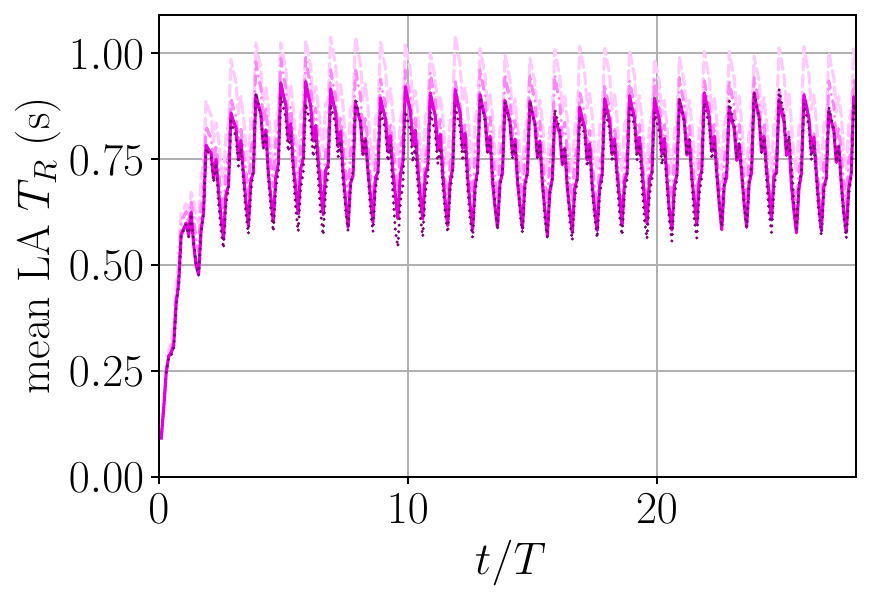}
                    };
        \node[anchor=south west](b) at ([xshift=.5cm]a.south east) {
    \includegraphics[scale=0.45]{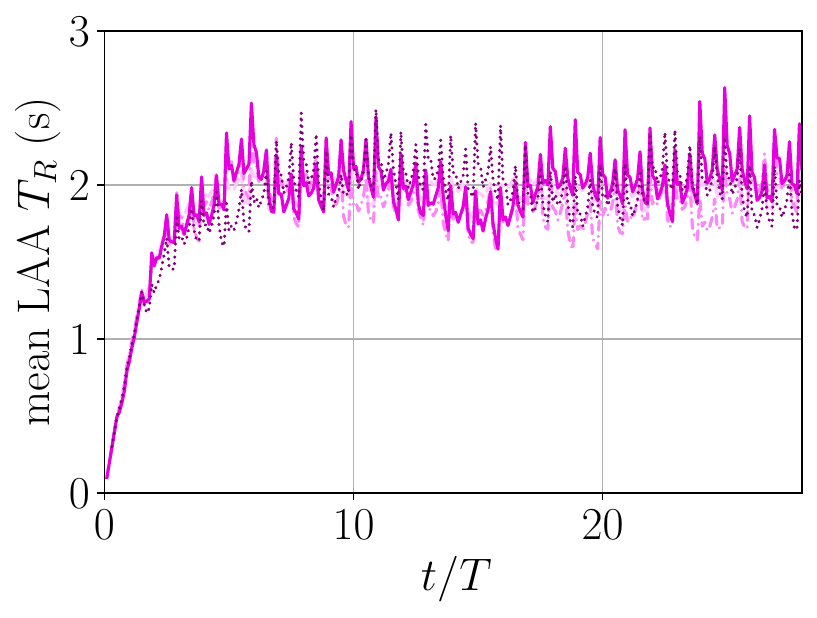}
        };
        \node[anchor=north east](c) at (a.south east) {
    \includegraphics[scale=0.45]{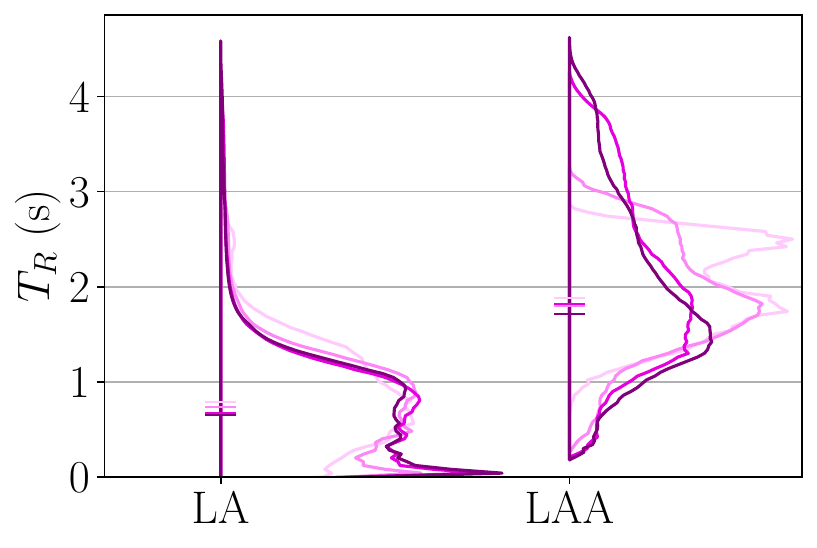}
        };
        \node[anchor=north east](d) at (b.south east) {
        \includegraphics[scale=0.45]{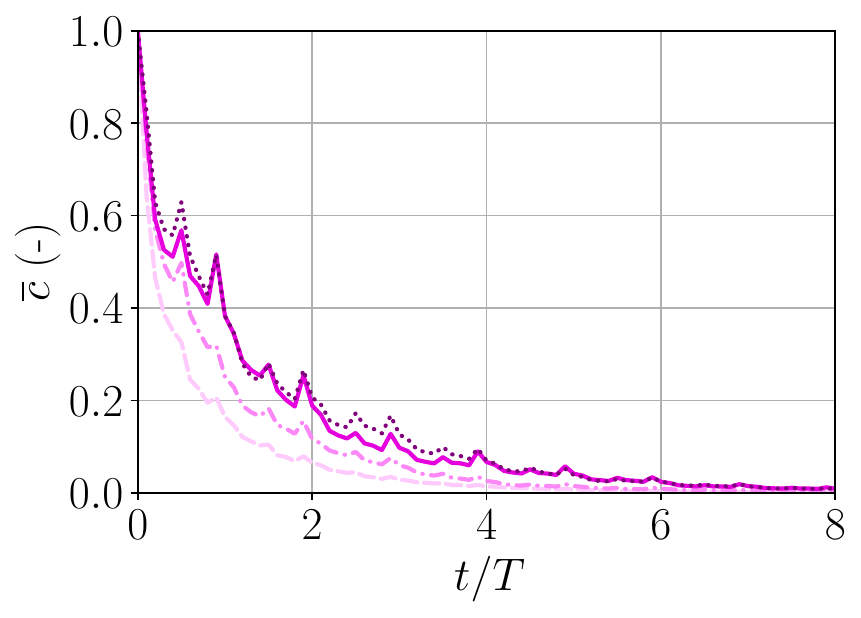}
        };
        \node[anchor=north west,text height=.4cm] at ([xshift=-.1cm,yshift=.1cm]a.north west) {(a)};
        \node[anchor=north west,text height=.4cm] at ([xshift=-.1cm,yshift=.1cm]b.north west) {(b)};
        \node[anchor=north west,text height=.4cm] at ([xshift=-.1cm,yshift=.1cm]c.north west) {(c)};
        \node[anchor=north west,text height=.4cm] at ([xshift=-.1cm,yshift=.1cm]d.north west) {(d)};
    \end{tikzpicture}
    \caption{Grid refinement for subject~2:
    (a)~Mean whole-LA residence time and (b)~mean LAA residence time.
    (c)~Pooled whole-LA and LAA residence-time distributions,
    and (d)~evolution of mean scalar concentration in the LAA.
    Horizontal markers in (c) indicate the medians.
    Grids: \lineVeryCoarse~$64^3$, \lineCoarse~$144^3$, \lineNom~$256^3$,
    and \lineFine~$384^3$.
    }
    \label{fig:grid_refinement}
\end{figure}

Sensitivity to the numerical resolution was assessed for subject~2 using
uniform Cartesian grids of $64^3$, $144^3$, $256^3$, and $384^3$ points.
To satisfy the \gls*{CFL} condition, the corresponding time steps
were set to $\Delta t=\SI{20e-5}{s}$, \SI{8e-5}{s},
\SI{5e-5}{s}, and \SI{2.5e-5}{s}, respectively.
To remove sensitivity to the Lagrangian mesh of marker points,
all simulations used the mesh corresponding to the finest
grid step size.
Simulations were run for 28 cycles, and the final 8 cycles
were used for analysis.
The residence-time histories and pooled probability distributions
are shown in Fig.~\ref{fig:grid_refinement}, alongside the
temporal evolution of mean scalar concentration in the LAA.
Summary statistics are reported
in Tab.~\ref{tab:grid_refinement}.

Mean residence time changed little between the intermediate and
finest grids.
The whole-LA and LAA means varied from 0.74 to \SI{0.87}{s},
and 1.89 to \SI{2.06}{s}, respectively.
The probability distributions retained some resolution
dependence, particularly at the high-$T_R$ tails in the LAA
(Fig.~\ref{fig:grid_refinement}c).
\autoref{tab:grid_refinement} reports the median residence time for both
the whole LA and the LAA, along with the 90th percentile for the LAA.
Additionally, we computed the \gls*{EMD} between the three coarser grids
and the finest grid.
The \gls*{EMD} is a metric to assess the similarity between two
probability distributions~\cite{Rubner1998} and, in one dimension,
can be computed as
$\mathrm{EMD}(F,G) = \int_{-\infty}^{\infty}\left| F(x) - G(x) \right| \mathrm{d}x$,
where $F$ and $G$ are the corresponding cumulative distribution functions~\cite{Vallender1974}.
An \gls*{EMD} of zero indicates a perfect match.
The resulting values are reported in \autoref{tab:grid_refinement}
and indicate convergence of the results with
grid refinement.

Scalar washout showed similar convergence upon refinement.
The mean washout rate was 0.67, 0.63, 0.54, and
\SI{0.56}{s^{-1}} for the $64^3$, $144^3$, $256^3$, and $384^3$
grids, respectively.
Relative to the finest grid, the differences were 19.6, 12.5,
and 3.6\% for the coarse and intermediate grids.
Similar convergence was observed for the normalized total
exchanged volume $\widehat{V}_\mathrm{total}$.

\begin{table}[h!tb]
    \setlength{\tabcolsep}{5pt} 
    \centering
    \begin{tabular}{ccc|rrrr}
    \hline
    \multicolumn{3}{c|}{quantity} & 64$^3$ & 144$^3$ & 256$^3$ & 384$^3$ \\
    \hline
    \multirow{7}{*}{$T_R$ (s) }& \multirow{4}{*}{ LA }
      & mean   & 0.87 & 0.79 & 0.75 & 0.74 \\  \cline{3-7}
     && median & 0.81 & 0.75 & 0.68 & 0.68 \\ \cline{3-7}
     && std    & 0.53 & 0.53 & 0.58 & 0.58 \\ \cline{3-7}
     && EMD    & 0.15 & 0.07 & 0.02 & --~~ \\ \cline{3-7}
    \cline{2-6}
    & \multirow{4}{*}{ LAA }
        & mean   & 1.96 & 1.89 & 2.06 & 1.97 \\ \cline{3-7}
     &  & median & 1.95 & 1.87 & 1.95 & 1.81 \\ \cline{3-7}
     &&  \makecell{90th percentile} & 2.55 & 2.68 & 3.35 & 3.28 \\ \cline{3-7}
     &  & EMD    & 0.35 & 0.24 & 0.11 & --~~ \\ \cline{3-7}
    \hline
    \multicolumn{3}{c|}{$k$ (\si{s^{-1}})}
    & 0.67 & 0.63 & 0.54 & 0.56 \\ \hline
    \multicolumn{3}{c|}{$\widehat{V}_\mathrm{total}$ {(-)} }
    & 2.41 & 3.04 & 3.21 & 3.45 \\ \hline
    \end{tabular}
    \caption{Residence-time statistics, effective washout rates, and total exchanged volume
    for subject~2 at the four numerical resolutions. 
    }
    \label{tab:grid_refinement}
\end{table}

\newpage
\section{Well-Mixed Washout Models}
\label{app:perfect_mixing}

Consider an appendage with time-dependent volume $V(t)$
and scalar inventory $M(t)$.
Under perfect mixing, the scalar concentration $\gls{c;model}$ is spatially
uniform, such that $M=V\gls{c;model}$, and outgoing fluid has the
perfectly mixed concentration $\gls{c;model}$, while
incoming fluid is assumed to be scalar-free.
The scalar and volume balances then give
\begin{equation}
    \frac{\mathrm{d}(V\gls{c;model})}{\mathrm{d}t}
    =-Q_{\mathrm{out}}\gls{c;model},
    \qquad
    \frac{\mathrm{d}V}{\mathrm{d}t}
    =Q_{\mathrm{in}}-Q_{\mathrm{out}},
\end{equation}
where the nonnegative rates $Q_{\mathrm{in}}$ and
$Q_{\mathrm{out}}$ denote inflow and outflow.
Expanding the scalar balance and substituting the volume
balance to eliminate the outflow term yields
\begin{equation}
    V\frac{\mathrm{d}\gls{c;model}}{\mathrm{d}t}
    =-Q_{\mathrm{in}}\gls{c;model}.
\end{equation}
For an initial concentration $\gls{c;model}(t_0)=1$, integration gives
\begin{equation}
    \gls{c;model}(t)=
    \exp\left[
        -\int_{t_0}^{t}
        \frac{Q_{\mathrm{in}}(\tau)}{V(\tau)}
        \,\mathrm{d}\tau
    \right].
\end{equation}

\newpage
\section{Decomposition of Washout}
\label{app:washout_decomposition}
The well-mixed washout prediction can be generalized by allowing
incoming and outgoing fluid to have concentrations that differ
from the instantaneous mean $\overline{c}$.

Let $\overline{c}_{\mathrm{in}}$ and
$\overline{c}_{\mathrm{out}}$ denote the mean
concentrations of incoming and outgoing fluid, respectively.
For purely advective transport through an incompressible
domain, the scalar balance is
\begin{equation}
    \frac{\mathrm{d}(V\overline{c})}{\mathrm{d}t}
    =
    Q_{\mathrm{in}}\overline{c}_{\mathrm{in}}
    -Q_{\mathrm{out}}\overline{c}_{\mathrm{out}}.
\end{equation}
Using $\mathrm{d}V/\mathrm{d}t
=Q_{\mathrm{in}}-Q_{\mathrm{out}}$ and dividing by
$V\overline{c}$ gives, for $\overline{c}>0$,
\begin{equation}
    -\frac{\mathrm{d}\ln\overline{c}}{\mathrm{d}t}
    =
    \frac{Q_{\mathrm{in}}}{V}
    -\frac{Q_{\mathrm{in}}}
    {V}
    \frac{\overline{c}_{\mathrm{in}}}
    {\overline{c}}
    +\frac{Q_{\mathrm{out}}}{V}
    \left(
        \frac{\overline{c}_{\mathrm{out}}}{\overline{c}}-1
    \right).
\end{equation}

Integrating over a cardiac cycle beginning at $t_n$ and
dividing by its period $T$ yields
\begin{equation}
    k_n=k_{Q,n}+\alpha_n+\beta_n,
\end{equation}
with
\begin{align}
    k_{Q,n}
    &=\frac{1}{T}\int_{t_n}^{t_n+T}
    \frac{Q_{\mathrm{in}}}{V}\,\mathrm{d}t,\\
    \gls{alpha}_n
    &=-\frac{1}{T}\int_{t_n}^{t_n+T}
   \frac{Q_{\mathrm{in}}}
    {V}
    \frac{\overline{c}_{\mathrm{in}}}
    {\overline{c}}\,\mathrm{d}t,\\
    \gls{beta}_n
    &=\frac{1}{T}\int_{t_n}^{t_n+T}
    \frac{Q_{\mathrm{out}}}{V}
    \left(
        \frac{\overline{c}_{\mathrm{out}}}{\overline{c}}-1
    \right)\,\mathrm{d}t.
\end{align}

The first term is the inflow-driven washout prediction.
The non-positive term $\gls{alpha}_n$ accounts for incoming scalar,
which reduces washout relative to scalar-free inflow.
The term $\gls{beta}_n$ describes the effect of the outgoing
concentration relative to the appendage mean.
Preferential export of scalar-poor fluid gives $\gls{beta}_n<0$
and reduces washout, whereas preferential export of
scalar-rich fluid gives $\gls{beta}_n>0$.

The mean rates $k$, $k_Q$, $\gls{alpha}$, and $\gls{beta}$ are obtained
by averaging the corresponding cycle-specific terms over $n=1,\ldots,7$,
as in the main analysis.

\newpage
\section{Ostium Flow Scales}
\label{app:ostium_flow_scales}
The ostium flow scales were evaluated over $21\leq t/T<28$
(Tab.~\ref{tab:ostium_flow_scales}).
The peak bulk absolute relative velocity was determined as
\begin{equation}
    w_\mathrm{bulk} =\max_{t}
           \frac{Q_{\mathrm{total}}(t)}{A(t)}, \\
\end{equation}
over the sampled time instants in this interval.
The mean ostium area $\bar{A}$ was obtained
by averaging the area of the moving ostium surface over time.
An area-equivalent diameter
$D=2\sqrt{A/\pi}$ was used to calculate
the LAA aspect ratio
and 
Reynolds number 
\begin{equation}
    \frac{L}{D} = \frac{\bar{V}_\mathrm{LAA}}{\bar{A}\, \bar{D}}, \qquad
    Re=\frac{w_{\mathrm{bulk}}D}{\nu},
\end{equation}
with $\nu=\SI{0.04}{cm^2/s}$.
As summarized in \autoref{tab:ostium_flow_scales},
the aspect ratio was between 0.42 and 1.15,
peak bulk
velocities ranged from 2.8 to
\SI{11.7}{cm/s}, giving Reynolds numbers between 185 and 957.

\begin{table}[h!tbp]
\centering
\setlength{\tabcolsep}{5pt} 
\begin{tabular}{|c|r|r|r|r|}
\hline
    \multirow{2}{*}{\makecell{subject}}
    &  \multirow{2}{*}{\makecell{$\bar{\gls{area}}$ \\ {(cm$^2$)}}}
    & \multirow{2}{*}{\makecell{$L/D$ \\{(-)}}} 
    & \multirow{2}{*}{\makecell{$w_{\mathrm{bulk}}$ \\  {(cm/s)}}}
    & \multirow{2}{*}{\makecell{ $Re$ \\ {(-)} }}\\
     &&&& \\
    \hline
    1 
    & 3.86 & 0.75 & 11.0 & 559 \\
    2 
    & 4.45 & 0.46 & 6.4 & 373 \\
    3 
    & 8.07 & 0.55 & 11.7 & 957 \\
    4 
    & 7.84 & 0.43 & 4.8 & 380 \\
    5 
    & 7.39 & 0.68 & 5.9 & 452 \\
    6 
    & 8.96 & 0.73 & 9.4 & 799 \\
    7 
    & 5.14 & 0.42 & 2.8 & 185 \\
    8 
    & 4.66 & 0.54 & 5.8 & 352 \\
    9 
    & 6.75 & 0.71 & 5.4 & 401 \\
    10 
    & 6.62 & 0.92 & 8.3 & 628 \\
    11 
    & 7.15 & 0.73 & 4.4 & 330 \\
    12 
    & 2.64 & 0.66 & 9.3 & 386 \\
    13 
    & 3.58 & 0.50 & 6.7 & 378 \\
    14 
    & 7.16 & 0.80 & 4.2 & 308 \\
    15 
    & 4.0 & 0.80 & 5.1 & 308 \\
    16 
    & 4.14 & 0.82 & 6.2 & 362 \\
    17 
    & 4.29 & 1.15 & 10.2 & 628 \\
\hline
\end{tabular}
\caption{Ostium flow scales over $21\leq t/T<28$:
mean ostium area $\bar{\gls{area}}$,
aspect ratio $L/D$ of the LAA, peak
bulk velocity $w_{\mathrm{bulk}}$,
the corresponding Reynolds numbers,
$Re$.}
\label{tab:ostium_flow_scales}
\end{table}

\newpage
\section{Well-Mixed Concentration Histories}
\label{app:model_histories}
\begin{figure}[h!tb]
    \centering
        \begin{tikzpicture}
        \node[](a) at (0,0) {
        \includegraphics[scale=.5]{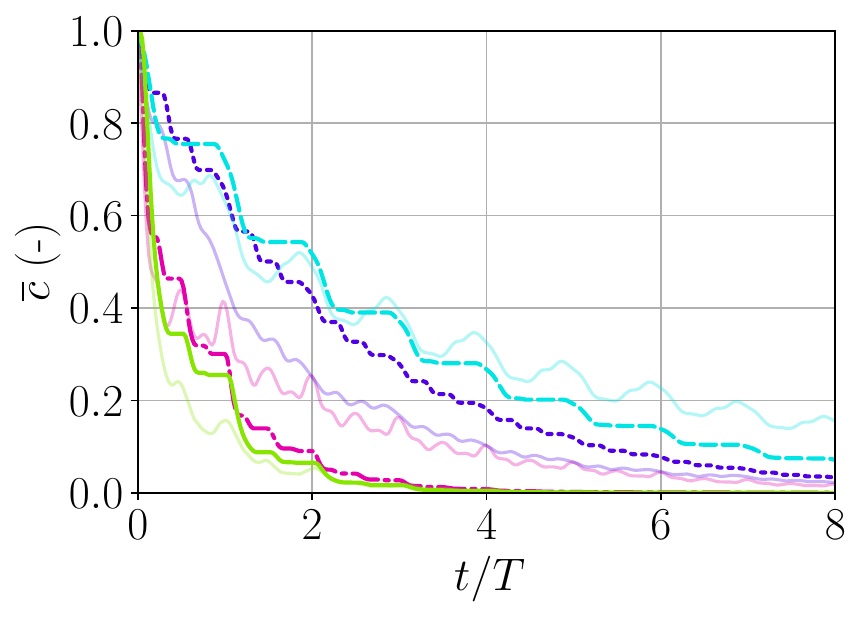}
        };
        \node[anchor=north west](b) at (a.north east) {
        \includegraphics[scale=.5]{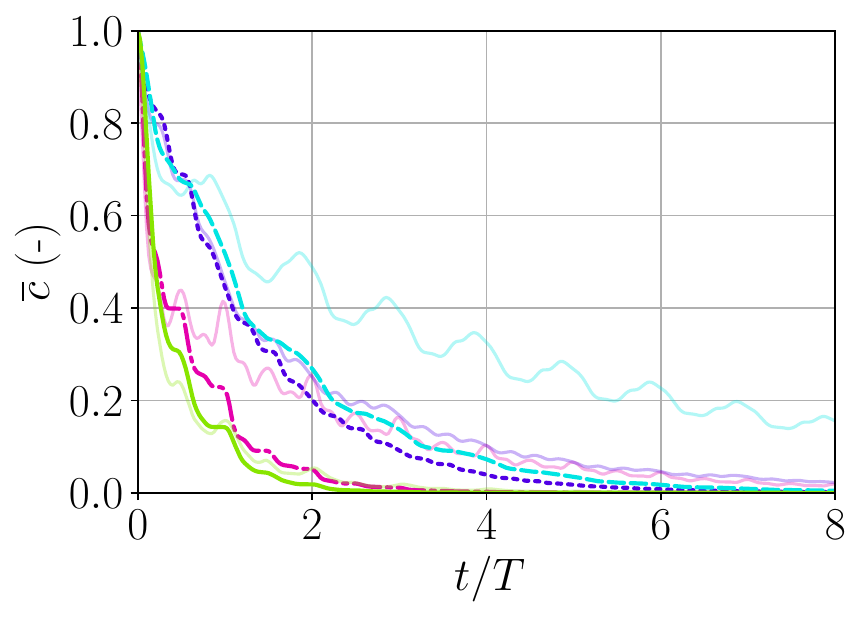}
        };
        \node[anchor=north west,text height=.4cm] at (a.north west) {(a)};
        \node[anchor=north west,text height=.4cm] at (b.north west) {(b)};
    \end{tikzpicture}
    \caption{
    Mean LAA scalar concentration for subjects~\lineOne~1, \lineFive~5, \lineEleven~11,
    and \lineTwelve~12.
    Faint curves show the CFD results in both panels. Prominent curves
    show predictions from the (a)~volume-driven and (b)~inflow-driven
    well-mixed washout models. Time is measured from scalar initialization.
    }
    \label{fig:scalar_models_time}
\end{figure}

The model-predicted concentration histories depicted in \autoref{fig:scalar_models_time}
complement the cohort-level rate
comparisons in Section~\ref{sec:mixing_results} by resolving
the initial transient and subsequent within-cycle variations.
In subjects~1 and~11, the volume-driven model (Fig.~\ref{fig:scalar_models_time}a)
initially predicted
higher concentrations than CFD, but lower concentrations at later
times.
In subject~5, the CFD concentrations consistently remained below the model prediction.
For subject~12, in turn, the decrease in concentration during the first cycle was
underpredicted, but the model showed close agreement over the subsequent
cycles.

The inflow-driven model predictions (Fig.~\ref{fig:scalar_models_time}b) 
matched the CFD results well during the first cycle for subjects 5 and 12, but
overpredicted the washout in all cases in the subsequent cycles.

\newpage
\section{Whole-LA Residence Time and Turnover Time}
\label{app:LA_turnover}
For a reservoir in steady state, the mean residence time of its contents can be 
compared with its turnover time $\tau_0=V/Q$,
defined as the ratio of its volume to the 
through-flow rate~\cite{BolinRodhe1973}.
For zero-age inflow, piston flow gives a mean residence time of $\tau_0/2$,
whereas perfect mixing gives $\tau_0$.
Mean residence time can also exceed $\tau_0$ when rapid passage coexists
with prolonged retention.

\begin{figure}[h!tb]
    \centering
    \includegraphics[scale=.5]{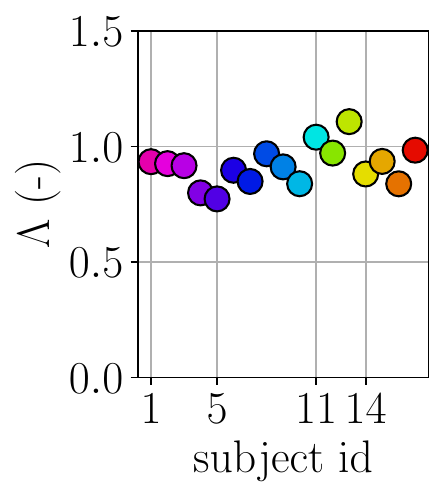}
    \caption{Whole-LA mean residence time normalized by the nominal
    turnover timescale,
    $\Lambda=\overline{T}_R^{\mathrm{LA}}\mathrm{LVSV}/
    (\overline{V}_{\mathrm{LA}}T)$.
    Each marker represents one subject.}
    \label{fig:LA_turnover}
\end{figure}

For the LA, 
$\overline{V}_{\mathrm{LA}}T/\mathrm{LVSV}$ defines a
nominal LA turnover timescale, as used in our previous
studies~\cite{Gonzalo2022,Duran2023,Stoecker2025}.
Mean whole-LA residence time normalized by this timescale clustered
near unity (Fig.~\ref{fig:LA_turnover}), which supports it as a relevant
residence-time scale, but does not necessarily establish a particular mixing regime.

\bibliography{bibliography}

\end{document}